\documentclass[11pt, oneside]{Thesis} 

\graphicspath{{Pictures/}} 

\usepackage[square, numbers, comma, sort&compress]{natbib} 
\usepackage{slashed}
\hypersetup{urlcolor=blue, colorlinks=true} 
\title{\ttitle} 

\begin{document}

\frontmatter

\setstretch{1.3} 

\fancyhead{} 
\rhead{\thepage}
\lhead{}

\pagestyle{fancy} 

\newcommand{\HRule}{\rule{\linewidth}{0.5mm}} 

\hypersetup{pdftitle={\ttitle}}
\hypersetup{pdfsubject=\subjectname}
\hypersetup{pdfauthor=\authornames}
\hypersetup{pdfkeywords=\keywordnames}


\begin{titlepage}
{
\setlength{\parindent}{0pt}

\thispagestyle{empty}

\begin{center}
{ {\Large \bf Universidade de São Paulo \\ \vskip 2mm Instituto de F\'isica} }
\end{center}

\vskip 2.0cm

\begin{center}
{ \LARGE \bf
The three point function in Liouville and $\mathcal{N}=1$ Super Liouville Theory
}

\vskip 1.3cm {\normalsize Mart\'in Dionisio Arteaga Tupia \\ Orientador: Prof. Dr. Elcio Abdalla}
\end{center}

\vskip 2.5cm

\hfill \parbox{5.5cm}{\normalsize Dissertação de mestrado apresentada ao Instituto de Física para a obtenção do título de Mestre em Ciências.}
\vfill

\vskip 2cm

\textbf{Banca Examinadora:}

Prof. Dr. Elcio Abdalla (IFUSP) \\
Prof. Dr. João Carlos Alves Barata (IFUSP) \\
Prof. Dr. Antonio Lima Santos (UFSCar)

\vskip 3.0cm

\begin{center}
\textbf{São Paulo \\ 2015}
\end{center}

}

\end{titlepage}

%

\newpage{\ }
\thispagestyle{empty} 

\begin{titlepage}
{
\setlength{\parindent}{0pt}

\thispagestyle{empty}

\begin{center}
{ {\Large \bf Universidade de São Paulo \\ \vskip 2mm Instituto de F\'isica} }
\end{center}

\vskip 2.0cm

\begin{center}
{ \LARGE \bf
A Func\~{a}o de Tr\^{e}s pontos nas Teorias de Liouville\\ e $\mathcal{N}=1$ Super Liouville 
}

\vskip 1.3cm {\normalsize Mart\'in Dionisio Arteaga Tupia \\ Orientador: Prof. Dr. Elcio Abdalla}
\end{center}

\vskip 2.5cm

\hfill \parbox{5.5cm}{\normalsize Dissertação de mestrado apresentada ao Instituto de Física para a obtenção do título de Mestre em Ciências.}
\vfill

\vskip 2cm

\textbf{Banca Examinadora:}

Prof. Dr. Elcio Abdalla (IFUSP) \\
Prof. Dr. João Carlos Alves Barata (IFUSP) \\
Prof. Dr. Antonio Lima Santos (UFSCar)

\vskip 3.0cm

\begin{center}
\textbf{São Paulo \\ 2015}
\end{center}

}

\end{titlepage}

\thispagestyle{empty}

\textcolor[rgb]{1.00,1.00,1.00}{palabra} 

\newpage 




%




\setstretch{1.3} 

\acknowledgements{\addtocontents{toc}{\vspace{1em}} 


   The research included in this dissertation could not have been performed if not for the assistance, patience, and support of many individuals.  I would like to extend my gratitude first and foremost to my thesis advisor Dr. Elcio Abdalla for mentoring me over the course of my graduate studies.  
He has helped me through extremely difficult times over the course of the analysis and the writing of the dissertation and for that I sincerely thank him for his confidence in me.

I am deeply grateful to the members of the jury professors Jo\~{a}o Barata and Antonio Lima Santos for agreeing to read the manuscript, to participate in the defense of this thesis and to help me to improve this dissertation with their valuable and insightful suggestions. 

I would additionally like to thank Lyn Chia, Rene Negron, Ricardo Landim and Antonio Sanchez for their support in both the research and especially the revision process that has lead to this document.  Their knowledge and understanding of the written word has allowed me to fully express the concepts behind this research.

I would also like to extend my appreciation to professors Josif Frenkel, Adilson da Silva and Enrico Bertuzzo who have served as a voice of quiet wisdom in matters ranging from the most basic aspects of physics, to the paths that my career has eventually taken.


I would like to extend my deepest gratitude to my parents Dionisio Arteaga and Juana Tupia, my girlfriend Homeira Kh. and my friends Fernando A., Daniela M., Rosio I., Gustavo C., Eder R., Fernando C., Carlos G., Rocio C., Miguel G., Mariela A., Luciane M., Roberta G. without whose love, support and understanding I could never have completed this master degree.

Finally I would like to thank the Brazilian Ministry of Education (Coordenação de Aperfeiçoamento de Pessoal de Nível Superior - CAPES) for a Master Scholarship, and the University of S\~{a}o Paulo for the amazing research environment.

}
\clearpage 


\pagestyle{fancy} 
\lhead{\emph{Contents}} 
\tableofcontents 

\lhead{\emph{List of Figures}} 
\listoffigures 


\setstretch{1.3} 

\pagestyle{empty} 

\dedicatory{Dedicado a Juana, Dionisio,\\ Fiorella, Isidora y Homeira, que con su amor me gu\'ian.\\ Y para Roberto, Mila y Nicho \\ quienes con su recuerdo me impulsan.} 

\addtocontents{toc}{\vspace{2em}} 


\mainmatter 

\pagestyle{fancy}

\chapter*{Resumo}

\label{Resumo} 

Neste trabalho s\~{a}o apresentados alguns conceitos b\'asicos da Teoria de Liouville e
$\mathcal{N}=1$ Super Liouville, enfatizando o c\'alculo das fun\c c\~{o}es de tr\^{e}s pontos dessas teorias. Uma introdu\c c\~{a}o a Teoria de Campos Conformes (CFT) e a Supersimetria tamb\'em s\~{a}o
inclu\'{\i}das, as quais constituem ferramentas b\'asicas da presente pesquisa.
\chapter*{Abstract} 

\label{Abstract} 

In this dissertation we present some basic features about Liouville and $\mathcal{N}=1$ Super Liouville Theory, and focus in the computation of their three point functions. Additionally, we include an introduction to Conformal Field Theories (CFT) and Supersymmetry, which are the basic tools of the present research.

\chapter{Introduction} 

\label{Chapter1} 

\lhead{Chapter 1. \emph{Introduction}} 

Liouville theory has been extensively studied for several years in the context of two dimensional quantum gravity. This theory is understood as a toy model in order to try to understand some technical and conceptual problems arising within this field \cite{zamo}, without the inconvenience or difficulty of facing the mathematics of quantum gravity in higher dimensions.
 
Its origin was in the highly regarded work by Poincar\'e \cite{Poincare}. However, for physicists the main source was the seminal paper written by Polyakov \cite{bosonicstring}. That paper gave rise to a revolution in the dark days of string theory. With the fresh point of view it was possible to have a bosonic string theory in a dimension which would not be necessarily $26$. Such a theory has been called non-critical string theory. In this way, non-critical string theory appears on the scene, and a vital part of this field is Liouville theory. Liouville theory could be also understood as the quantum theory of the world-sheet.

An important cornerstone in the development of Liouville theory was the discovery of the Dorn-Otto-Zamolodchikov-Zamolodchikov proposal also named as the DOZZ formula. It was first derived by H. Dorn, J. Otto \cite{dotto1} and A. Zamolodchikov, Al. Zamolodchikov \cite{zamo2}. Such formula is an exact expression for the structure constant of the Liouville three point function.

In recent years with the advent of the $AdS/CFT$ correspondence, the DOZZ formula gains more importance. Because, as it was proposed by O. Coussaert, M. Henneaux and P. van Driel  in \cite{AdS3/CFT2}, there exists a particular case of the correspondence called $AdS_{3}/CFT_{2}$ correspondence, where the $CFT_{2}$ is identified as the Liouville theory. In this sense, Liouville theory, viewed as a $CFT$ in two dimensions, should have a gravitational dual theory living in an three dimensional Anti - de Sitter space-time $AdS_{3}$. Hence, if we were able to find such a gravitational theory and then compute the minimal surface formed by fixing three points at the boundary, it would be possible to compare that result with the DOZZ formula. A comparison of an exact expression, in both sides of the duality, would be an important step in the difficult task for proving the $AdS/CFT$ correspondence. Nonetheless, the comparison between the minimal surface cited above and the DOZZ formula is currently an unsolved question.
  
Therefore, due to its important role the DOZZ formula needs a careful and detailed examination. In order to study such an important result, in Chapter \ref{Chapter2} we will develop the necessary mathematical tools, which come from Conformal Field Theory (CFT). From there, bosonic Liouville theory will be studied in Chapter \ref{Chapter3}.

The usual way to compute the DOZZ correlator is by applying the so-called Coulomb Gas method. Nonetheless, we will perform this computation by applying an alternative method initially given in the papers of O'Raifeartaigh, J. Pawlowski and V. Sreedhar \cite{dual} \cite{dual2}. Such a method will be named the OPS method. The main importance of the OPS method lies in the fact that it provides explicitly the lattice of poles, which are located in the denominator of the DOZZ formula. Whilst the Coulomb gas method maintains such lattice of poles hidden within its Taylor expansion of the exponential potential $\mu e^{b\phi}$. We will discuss these points further in Chapter \ref{Chapter4}.

Natural questions arise when we want to incorporate fermionic fields into Liouville theory. What does conformal invariance look like?, Is it still possible to compute precisely the new tree point function? If this is the case, what does it look like? In order to answer the first question, it will be necessary to introduce the machinery of Supersymmetry, which we will perform in Chapter \ref{Chapter5}. The main objective of Chapter \ref{Chapter5} lies in the second question. We will also perform the computation of the Super Liouville three point function in this chapter. Our result follows previous research by E. Abdalla et al. \cite{3}, L. Alvarez-Gaume and Ph. Zaugg \cite{Alvarez1} \cite{Alvarez2} and R. Rashkov, M.Stanishkov \cite{Rashkov}. In contrast to our previous computation, we used the usual Coulomb Gas method for the super Liouville three point function.


\chapter{Conformal Field Theory} 

\label{Chapter2} 

\lhead{Chapter 2. \emph{Conformal Field Theory}} 


\section{The beginning of Conformal field theory: Ising model at critical phase transition}

\begin{figure}[htbp]
  \centering
    \includegraphics[scale=0.65]{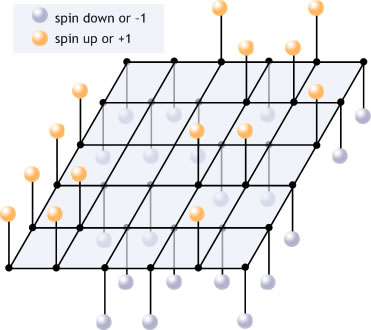}
    \rule{26em}{0.3pt}
  \caption[Ising's square lattice]{Spins $\sigma_{i}=\pm 1$ are distributed on the square lattice.}
  \label{fig:Ising}
\end{figure}

The critical behaviour of some systems at second order phase transitions is described by Conformally invariant Quantum Field Theory. The emblematic example is, of course, the Ising model in two dimensions. This model, with a constant interaction $\epsilon$, is described by the Hamiltonian,
\begin{equation}
H_{I} = -\epsilon \sum_{\langle ij\rangle } \sigma_{i} \sigma_{j},
\end{equation}
where $\sigma_{i}$ represents the spin which takes the values $\sigma_{i}=\pm 1$ and the notation ${\langle ij\rangle}$ indicates summation only over pairs of nearest neighbour sites. The spin distribution is on the sites of a square lattice, as shown in Fig. \ref{fig:Ising}.

In addition, when the system is subject to a constant external  magnetic field $h$, we have a total hamiltonian given by
\begin{equation}
H=-\epsilon\sum_{\langle ij\rangle}\sigma_{i}\sigma_{j}-h\sum_{i}\sigma_{i}.
\end{equation}
One of the most important quantities in this model is the magnetization $M$. As usual we define it as the expectation value of the spin $M=\langle\sigma_{i}\rangle$. Because the magnetic field $h$ is acting on our system, we define another important statistical quantity in the Ising model, the magnetic susceptibility $\chi$. In this way, it is not difficult to imagine that magnetic susceptibility measures the response of our system when a magnetic field is acting on it. Thus, we have  
\begin{equation}
\chi=\left.\frac{\partial M}{\partial h}\right|_{h=0}. 
\end{equation}
After a small calculation, it is found that magnetic susceptibility is proportional to the variance of the total spin $\sigma_{\text{Tot}}=\sum_{i}\sigma_{i}$. In this context it is also usual to define the connected pair correlation function as $\Gamma_{c}(i-j)=\langle\sigma_{i}\sigma_{j}\rangle-\langle\sigma_{i}\rangle\langle\sigma_{j}\rangle$. Hence, using such a function our magnetic susceptibility is rewritten as \cite{Molignini}
\begin{equation}
\chi=\beta\sum_{i}\Gamma_{c}(i)
\end{equation}
where $\beta=1/T$, using the Boltzmann constant $k_{B}=1$.

Closer to the critical point, characterized by the critical temperature $T_{c}$, the functions defined above have a very special form. Around criticality there exists another important parameter, the correlation length $\xi$. For instance,  the correlation function $\Gamma_{i}$ is expressed as
\begin{equation}
\Gamma(i-j)\sim e^{-\frac{|i-j|}{\xi(T)}}.
\end{equation}
However, the correlation length closer to the critical point is given by an expression proportional to $|T-T_{c}|^{-1}$, diverging at $T_{c}$. At criticality $T_{c}$, one obtains that $\Gamma(i-j)$ becomes proportional to $|i-j|^{2-d-\eta}$ (for a $d$ dimensional space). Hence, for $d=2$ we have $|i-j|^{-\eta}$, and for the two dimensional Ising model $\eta=1/4$.

Finally, in this regimen it is usual to use the so called ``Scaling hypothesis''. This hypothesis establishes that the free energy $F$ near the critical point is a homogeneous function of its parameters, the external field $h$ and the reduced parameter $\tau=\frac{T}{T_{c}}-1$. In this case,  parameters $a$, $b$ and a scaling parameter $\lambda$ exists such that
\begin{equation}
F(\lambda^{a}\tau,\lambda^{b}h)=\lambda F(\tau,h).
\end{equation}   
This scaling property around criticality gives rise to the idea of Conformal Field Theory (CFT). The study of CFT is the main objective of this chapter and is the basis of String theory, in particular, Liouville theory which is the central issue of the work presented.

\section{Conformal Theories in {\it d} dimensions}\label{CTDdimensions}

\subsection{Conformal group in {\it d} dimensions}   

We started by considering the space $\mathbb{R}^{p,q}$ (where $d=p+q$ is the total dimension) with flat metric $g_{\mu\nu} = \eta_{\mu\nu}$. In this space we construct the line element $ds^{2}=g_{\mu\nu}dx^{\mu}dx^{\nu}$ and analyse how the metric $g_{\mu\nu}$ transforms under a change of coordinates $x \mapsto x'$. We have
\begin{equation}
g^{'}_{\mu\nu}(x')=\frac{\partial x^{\alpha}}{\partial x'^{\mu}} \frac{\partial x^{\beta}}{\partial x'^{\nu}} g_{\alpha\beta}(x).
\label{metrictransformation}
\end{equation} 
On the other hand, we define the {\it conformal} group as the subgroup of coordinate transformations that leaves the metric invariant, up to a {\it scale} change such as
\begin{equation}
g^{'}_{\mu\nu}(x)=\Omega(x) g_{\mu\nu}(x).
\label{conformaldefinition}
\end{equation}
These transformations preserve the angle between two vectors. For instance, in the case of two vectors $v$ and $w$ which belong to a finite dimensional Euclidean Vector Space $\mathbb{V}$ we know from Calculus that
\begin{equation}
\theta=\arccos (v.w/(v^{2}w^{2})^{1/2}).
\label{angle}
\end{equation}
From this equation it is easy to see that the angle $\theta$ given by  eq.(\ref{angle}) is preserved. According to the definition of the scalar product $v.w = g_{\mu\nu}v^{\mu}w^{\nu}$, we have
\begin{eqnarray*}
\theta' &=& \arccos \left( \frac{(v.w)'}{{(v^{2}w^{2})'}^{1/2}}\right) \\
&=&\arccos \left( \frac{g^{'}_{\mu\nu}v^{\mu}w^{\nu}}{(\Omega(x)^{2}v^{2}w^{2})^{1/2}}\right) \\
&=&\arccos \left( \frac{\Omega(x)g_{\mu\nu}v^{\mu}w^{\nu}}{\Omega(x)(v^{2}w^{2})^{1/2}}\right)\\ 
&=&\theta.
\end{eqnarray*}
Now using, $\Omega(x)=1$, in eq.(\ref{conformaldefinition}), we are led to the invariant metric $g^{'}_{\mu\nu}(x')=g_{\mu\nu}(x)$. In this way, we find that the {\it Poincaré} group \footnote{Poincaré Group$=$Translations$\otimes$ 
Lorentz transformations.} is a subgroup of the conformal group.\\
In order to find a closed expression for generators of the conformal group, we analyse first the infinitesimal coordinate transformation $x^{\mu} \longrightarrow x'^{\mu} = x^{\mu} + \varepsilon^{\mu}$. If it is substituted into eq.(\ref{metrictransformation}) along with the definition of line element $ds^{2}$, we get
\begin{equation}
ds^{2} \longrightarrow ds'^{2} = ds^{2} + (\partial_{\mu}\varepsilon_{\nu} + \partial_{\nu}\varepsilon_{\mu})dx^{\mu}dx^{\nu}.
\label{dstransformation}
\end{equation}
Therefore, if we want an expression obeying conformal invariance as given by eq.(\ref{conformaldefinition}), we need the second term in eq.(\ref{dstransformation}) to be proportional to the line element $ds^{2}$. This is achieved if we have
\begin{equation}
\partial_{\mu}\varepsilon_{\nu} + \partial_{\nu}\varepsilon_{\mu} = k \eta_{\mu\nu}, 
\label{eq1}
\end{equation}
where $k$ is an scalar function. In order to obtain such a constant, we multiply eq.(\ref{eq1}) by $\eta^{\mu\nu}$ as follows,
\begin{eqnarray*}
\eta^{\mu\nu}[\partial_{\mu}\varepsilon_{\nu} + \partial_{\nu}\varepsilon_{\mu}]&=&k \eta^{\mu\nu} \eta_{\mu\nu}\\
2(\partial.\varepsilon)&=&k(d)\\
k&=&\frac{2 (\partial.\varepsilon)}{d}.
\end{eqnarray*}
Hence, eq.(\ref{eq1}) can be written as,
\begin{equation}\label{eq2}
\partial_{\mu}\varepsilon_{\nu} + \partial_{\nu}\varepsilon_{\mu}=\frac{2 (\partial.\varepsilon)}{d}\eta_{\mu\nu}.
\end{equation}
For our purposes eq.(\ref{eq2}) should be rewritten in a more convenient form. We explicitly compute it below,
\begin{eqnarray*}
\partial_{\mu}\partial^{\mu}\left[ d\partial_{\mu}\varepsilon_{\nu} + d\partial_{\nu}\varepsilon_{\mu}\right] &=&\partial_{\mu}\partial^{\mu}\left[ (\partial.\varepsilon)\eta_{\mu\nu}+ (\partial.\varepsilon)\eta_{\mu\nu}\right], \nonumber\\
2\partial_{\mu}\partial^{\mu}n_{\mu\nu}(\partial.\varepsilon)&=&n_{\mu\nu}\square(\partial.\varepsilon)+d\partial_{\mu}\partial_{\nu}(\partial.\varepsilon),\nonumber
\end{eqnarray*} 
where we used $\eta_{\mu\nu}n^{\mu\nu}=d$. Finally, we obtain
\begin{equation}\label{eq3}
(n_{\mu\nu}\square + (d-2)\partial_{\mu}\partial_{\nu})(\partial.\varepsilon)=0.
\end{equation}

Eqs.(\ref{eq3}) or (\ref{eq2}) are called the condition for {\it conformal locality}.

Let us first analyse the general case $d>2$ in the condition of conformal locality. In the next section, we will analyse the very special case $d=2$ in eq.(\ref{eq2}).

It is not difficult to see that if $d>2$, there are third derivatives applied on $\epsilon$. Hence, in order to maintain eq.(\ref{eq3}), we require that $\epsilon$ should be at most a quadratic expression on $x^{\mu}$. In this way, it is usual to classify these different expressions of $\epsilon$ in four parts. By defining two arbitrary constant vectors $a^{\mu}$, $b^{\mu}$ and a constant $\lambda$, we write this classification as:

\begin{enumerate}
\item Translations:\,\, $\epsilon^{\mu}=a^{\mu}$, zero order in $x^{\mu}$,
\item Rotations:\,\,$\epsilon^{\mu}= \omega_{\nu}^{\mu}x^{\nu}$, first order in $x^{\mu}$,
\item Scale transformations:\,\,$\epsilon^{\mu}=\lambda x^{\mu}$, first order in $x^{\mu}$,
\item Special conformal transformations:\,\,$\epsilon^{\mu}=b^{\mu}x^{2}-2x^{\mu}(b\cdot x)$, second order in $x^{\mu}$.
\end{enumerate}

These four transformations give rise to the so called conformal group. We return to this point later. In the next section, as mentioned above, the special case of two dimensions will be analysed.
  
\subsection{Conformal algebra in 2 dimensions}

Here we study the conformal algebra in two dimensions $d=2$ arising from eq.(\ref{eq2}). Taking into account the euclidean metric given by $g_{\mu\nu}=\delta_{\mu\nu}$, this equation becomes the Riemann Cauchy equations: 
\begin{equation}\label{riemanncauchy}
\frac{\partial \varepsilon_{1}}{\partial x^{1}}=\frac{\partial \varepsilon_{2}}{\partial x^{2}},\,\,\,\,\,\,\,\, \frac{\partial \varepsilon_{2}}{\partial x^{1}}=-\frac{\partial \varepsilon_{1}}{\partial x^{2}}
\end{equation}
Now, let us define the following complex relations $z=x^{1}+ix^{2}$, $\overline{z}=x^{1}-ix^{2}$ and $\varepsilon(z)=\varepsilon^{1}+i\varepsilon^{2}$, $\overline{\varepsilon}(\overline{z})=\varepsilon^{1}-i\varepsilon^{2}$. From eq.(\ref{riemanncauchy}) it follows that our two dimensional conformal transformations are nothing less than the analytic complex coordinate transformations eq.(\ref{complextransf}),
\begin{equation}\label{complextransf}
z\longrightarrow f(z), \,\,\,\,\,\,\,\,\,\,\,\, \overline{z}\longrightarrow f(\overline{z}).
\end{equation}
On the other hand, our differential line element, which in $\mathbb{R}^{2}$ is defined as $ds^{2}=(dx^{1})^{2}+(dx^{2})^{2}$ turns into $ds^{2}=dzd\overline{z}$. Under these analytic transformations we analyse how the line element $ds^{2}$ transforms,
\begin{equation}\label{lineC}
ds^{2}=dzd\overline{z} \longrightarrow ds'^{2}=\left| \frac{\partial f}{\partial z} \right|^{2} dzd\overline{z}.
\end{equation}
This relation is nothing more than our definition of conformal transformations by replacing the conformal factor by $\Omega = \left|\partial f/\partial z\right|^{2}$. 
Now, we can derive the commutation relations of generators associated with infinitesimal transformations. The infinitesimal (anti-) holomorphic transformations are
\begin{equation}
z\longrightarrow z' = z + \varepsilon_{n}(z), \,\,\,\,\,\,\,\,\overline{z}\longrightarrow \overline{z}' = \overline{z} + \overline{\varepsilon}_{n}(z),
\end{equation}
where $n\in \mathbb{Z}$. Explicitly, we find that the functions $\varepsilon_{n}(z)$ and $\overline{\varepsilon}_{n}(z)$ are given by
\begin{equation}\label{inftransf1}
\varepsilon_{n}(z)=-z^{n+1}, \,\,\,\,\,\,\,\, \overline{\varepsilon}_{m}(z)=-\overline{z}^{m+1}.
\end{equation}
Hence, the following infinitesimal generators are obtained,
\begin{equation}\label{infgenerators1}
l_{n}=-z^{n+1}\partial_{z} \,\,\,\,\,\,\,\,  \overline{l}_{m}=-\overline{z}^{m+1}\partial_{\overline{z}}. 
\end{equation}
The commutator between two of these generators $[l_{m},l_{n}]$ is obtained by applying it over an arbitrary function $\Psi(z,\overline{z})$. We have,
\begin{eqnarray*}
[l_{m},l_{n}]\Psi(z,\overline{z})&=&(l_{m}l_{n}-l_{n}l_{m})\Psi(z,\overline{z})\\
\\
&=&l_{m}l_{n}\Psi(z,\overline{z})-l_{n}l_{m}\Psi(z,\overline{z})\\
\\
&=&(-z^{m+1}\partial_{z})(-z^{n+1}\partial_{z})\Psi(z,\overline{z})-(-z^{n+1}\partial_{z})(-z^{m+1}\partial_{z})\Psi(z,\overline{z})\\
\\
&=&(n-m)z^{m+n+1}\partial_{z}\Psi(z,\overline{z}).
\end{eqnarray*}
However, $-z^{(m+n)+1}\partial_{z}$ is the generator $l_{m+n}$. Thus, in operator form we get
\begin{equation}\label{llcommutation}
[l_{m},l_{n}]=(m-n)l_{m+n}.
\end{equation}
For the other commutators $[\overline{l}_{m},\overline{l}_{n}]$ and $[l_{m},\overline{l}_{n}]$ we proceed analogously. In this way, we have the next two relations,
\begin{eqnarray}
[\overline{l}_{m},\overline{l}_{n}] = (m-n)\overline{l}_{m+n}\label{lolo}\,,\hspace{0.3cm}\text{and} \hspace{0.7cm}[l_{m},\overline{l}_{n}] = 0.
\end{eqnarray}
The latter equation shows that the local conformal algebra is the direct sum of two independent (perhaps isomorphic) sub-algebras $\mathbb{H}\bigoplus\mathbb{\overline{H}}$. Due to this relation, we can regard $z$ and $\overline{z}$ as two independent coordinates. Each of these sub-algebras, $\mathbb{H}$ and $\mathbb{\overline{H}}$ are defined by $[l_{m},l_{n}]=(m-n)l_{m+n}$ and $[\overline{l}_{m},\overline{l}_{n}] = (m-n)\overline{l}_{m+n}$, respectively.
By using $l_{n}$, we can define the vector field $\mathbb{W}(z)$. According to this, we could then say that holomorphic conformal transformations are generated by the vector field $\mathbb{W}(z)$, as is clear from eq.(\ref{vectorialfield}):
\begin{eqnarray}
\mathbb{W}(z)&=&-\Sigma_{n}a_{n}l_{n}\label{vectorialfield}\\
\mathbb{W}(z)&=&\Sigma_{n}a_{n}z^{n+1}\partial_{z}\label{az}
\end{eqnarray}
If we want non-singularity as $z\rightarrow 0$ for $a_{n}\neq0$, the requirement is $n+1\geq0$. As a result, our first restriction for $n$ is
\begin{equation}\label{1restriction}
n\geq-1.
\end{equation} 
We can now proceed analogously as $z$ goes to $\infty$. In this order, let us define the following transformation by $z=-1/w$. The next step is to build the vectorial field $\mathbb{W}(z)$. Therefore, we have:
\begin{eqnarray}
\mathbb{W}(z)&=&\Sigma_{n}a_{n}\left(-\frac{1}{w}\right)^{n+1}\left(\frac{dz}{dw}\right) ^{-1}\partial_{w},\\
\mathbb{W}(z)&=&\Sigma_{n}a_{n}\left(-w\right)^{1-n}\partial_{w}.\label{a/w} 
\end{eqnarray}
Applying the same analysis as before, eq.(\ref{a/w}) gives us another restriction, $1-n\geq0$. Hence, we have also the restriction for $n$,
\begin{equation}\label{2restriction}
n\leq 1. 
\end{equation}
Due to (\ref{1restriction}) and (\ref{2restriction}), $\mathbb{W}(z)$ is globally \footnote{Here, we mean by global in the sense $\mathbb{W}(z)$ is defined over all Riemann sphere.} defined only for a certain range of values for $n$ ($n\in \mathbb{Z}$):
\begin{equation}\label{range}
-1\leq n \leq 1, \,\,\,\, n \in \mathbb{Z}
\end{equation}
As $n$ is an integer, it only takes three values: $n=-1,0,1$. From this limited number of allowed values for $n$, we can conclude that in two dimensions we only have three generators, for each sub-algebra $\mathbb{H}$, $\mathbb{\overline{H}}$, of the global conformal group. Explicitly, they are labelled by $l_{-1}, l_{0}, l_{1} \in \mathbb{H}$ and $\overline{l}_{-1}, \overline{l}_{0}, \overline{l}_{1} \in \mathbb{\overline{H}}$. For instance, it is possible to identify $l_{-1}$ and $\overline{l}_{-1}$ as the translation generators. In the same way, dilatations are identified with $l_{0}+\overline{l}_{0}$ and rotations with $i(l_{0}-\overline{l}_{0})$. Finally, $l_{1}$ and $\overline{l}_{1}$ are the generators of special conformal transformations.

\subsection{Constraints imposed by conformal invariance in {\it d} dimensions}

In this section we analyse some constraints imposed by conformal invariance, in the general {\it d}-dimensional case, to the $N$-point functions of a quantum theory. In the next section, we will deal with the very special case of a $d=2$ conformal theory.

We start by defining some general properties about a theory which possesses conformal invariance, then we define the {\it quasi-primary fields}, along with certain properties about these fields. Finally, these definitions and properties are used to find the general form of the two, three and four point functions, up to their structure constants. 

It is important to point out that it is not possible to find these structure constants by using conformal field methods only. In fact, they depend on the particular kind of Lagrangian we use.
 
Now, fields in a conformal theory have the following properties:

\begin{enumerate}
\item In a conformal field theory, we have a set $\mathfrak{C_{F}}$ of fields $\mathfrak{F}_{i}$ ,which are labelled by the index $i$ (to denote different fields). Hence, we have the set $\{ \mathfrak{F}_{i} \}$. This set of fields is, in general, infinite and in particular includes derivatives of fields $\mathfrak{{F}}_{i}$.
\begin{equation}\label{setfields}
\mathfrak{C_{F}}=\{ \mathfrak{F}_{i} \}
\end{equation} 
\item There exists a subset of $\mathfrak{C_{F}}$ called the set of {\it quasi-primary fields} $\{\varphi_{j}\}$,
\begin{equation}
\{\varphi_{j}\}\subset\mathfrak{C_{F}}.
\end{equation}
Quasi-primary fields transform under a {\it global} conformal transformation $x \rightarrow x'$ according to the rule
\begin{equation}\label{fieldstransf}
\varphi_{j}(x)= \left|\frac{\partial x'}{\partial x}\right|^{\Delta_{j}/d}\varphi_{j}(x'),
\end{equation}  
where $\Delta_{j}$ is the dimension associated to $\varphi_{j}$. Eq. (\ref{fieldstransf}) imposes the next \\
transformation for correlation functions:
\begin{equation}\label{covariantproperty}
\langle\varphi_{1}(x_{1})\dots\varphi_{n}(x_{n})\rangle=\left|\frac{\partial x'}{\partial x}\right|^{\Delta_{1}/d}_{x_{1}}\dots\left|\frac{\partial x'}{\partial x}\right|^{\Delta_{n}/d}_{x_{n}}\langle\varphi_{1}(x'_{1})\dots\varphi_{n}(x'_{n})\rangle
\end{equation} 
indexes $x_{j=1,\dots,n}$ on Jacobians mean that they are evaluated for these points.
\item Any other element of $\mathfrak{C_{F}}$ can be expressed as a linear combination of the quasi-primary fields and their derivatives. Thus, if $\psi_{k}\in\mathfrak{C_{F}}$ we have,
\begin{equation}
\psi_{k}=\sum_{i \in \mathbb{N}}a^{i}_{k}\varphi_{i}
\end{equation}
where $\varphi_{i}$'s are the quasi-primary fields and $a^{i}_{k}$'s are the coefficients of this linear expansion.
\item There is a vacuum $|0\rangle$ invariant under the global conformal group.
\end{enumerate}

From eq.(\ref{covariantproperty}) we impose some restrictions over the two and three point functions, under the global conformal group. Remember that the global conformal group is made up by the Poincaré group, scale transformations and special conformal transformations, they are listed below:

\begin{enumerate}
\item[a.] Translations, $\varepsilon^{\mu}=a^{\mu}$.
\item[b.] Rotations, $\varepsilon^{\mu}=\Theta^{\mu}_{\nu} x^{\nu}$.
\item[c.] Scale transformations, $\varepsilon^{\mu}=\lambda x^{\mu}$.
\item[d.] Special conformal transformations, $\varepsilon^{\mu}=b^{\mu} x^{2} - 2 x^{mu} b.x$. 
\end{enumerate}

In order to obtain the desired general form for the $N$ - point functions, we should analyse how they are restricted by the cited transformations.

\subsubsection*{a. Translation invariance:}

From the usual properties of translations, it is easy to conclude that a general $N$-point function depends not on $N$ independent coordinates $x_{i}$, but rather only on their difference $x_{i}-x_{j}$. A simple combinatorial analysis shows us that the total independent quantities are $N(N-1)$.

\subsubsection*{b. Rotation invariance:}
For simplicity we consider spinless objects. Rotational invariance tell us that for a dimension $d$ large enough \footnote{In lower dimensions the $N$ - point functions have linear relations among their coordinates which reduce the number of independent quantities.}, $N$ - point functions only depend on the $N(N-1)/2$ distances $r_{ij}=|x_{i}-x_{j}|$. Note that here we label the difference between $x_{i,j}$ only by $r_{ij}$.

\subsubsection*{c. Scale invariance:} 
A scale transformation is defined by:
\begin{equation}
x_{k} \rightarrow \lambda x_{k}
\end{equation}
Therefore, it is also valid for the following relation:
\begin{equation*}
r_{ij} \rightarrow \lambda r_{ij}.
\end{equation*}
From this last equation, scale invariance allows only dependence on the ratios $r_{ij}/r_{kl}$, because the scale factor is cancelled.

\subsubsection*{d. Special conformal transformation:} 

From the above transformation, we have
\begin{equation}
{r'}_{ij}^{2}=\frac{r_{ij}^{2}}{(1+2b.x_{i}+b^{2}x_{i}^{2})(1+2b.x_{j}+b^{2}x_{j}^{2})}.
\end{equation}
According to this, we see that the so called cross-ratios,
\begin{equation}
\frac{r_{ij}r_{kl}}{r_{ik}r_{jl}}
\end{equation}
are invariant not only under special conformal transformations, but also under the full conformal group. 

Having defined these restrictions, we find out how they affect the two and three point functions.
From eq.(\ref{fieldstransf}), the 2-point function of two quasi-primary fields $\varphi_{1}$ and $\varphi_{2}$ has a covariant transformation law given by eq.(\ref{covariantproperty}), thus we have:
\begin{equation}
\langle\varphi(x_{1})\varphi(x_{2})\rangle = \left|\frac{\partial x'}{\partial x}\right|^{\Delta_{1}/d}_{x=x_{1}}\left|\frac{\partial x'}{\partial x}\right|^{\Delta_{2}/d}_{x=x_{2}}\langle\varphi(x'_{1})\varphi(x'_{2})\rangle.
\end{equation}
As stated above, invariance under translation and rotation fix our two point function to be dependent on $r_{12}=|x_{1}-x_{2}|$. On the other hand, invariance under dilatation, say $x\rightarrow \lambda x$, impose that $r_{12}$ is raised to $\Delta_{1}+\Delta_{2}$. Finally, special conformal transformation implies that $\Delta_{1}=\Delta_{2}=\Delta$ or the two point function vanishes otherwise. In this way, our two point function has the following expression
\begin{equation*}
\quad\langle\varphi(x_{1})\varphi(x_{2})\rangle = 
\left\{
\begin{aligned}
&\frac{C_{12}}{r^{2\Delta}}\hspace{0.3cm}\Delta_{1}=\Delta_{2}=\Delta\\
&0\hspace{0.8cm}\Delta_{1}\neq\Delta_{2}
\end{aligned}
\right.
\end{equation*}
On the other hand, for the three point function by using similar arguments about invariance under the conformal group we obtain:
\begin{equation}
\langle\varphi(x_{1})\varphi(x_{2})\varphi(x_{3})\rangle=\sum_{ijk}\frac{C_{ijk}}{r^{i}_{12}r^{j}_{23}r^{k}_{13}}
\end{equation}
Here, due to special conformal transformations, we have:
\begin{eqnarray}
 i=\Delta_{1}+\Delta_{2}-\Delta_{3}\\
 j=\Delta_{2}+\Delta_{3}-\Delta_{1}\\
 k=\Delta_{3}+\Delta_{1}-\Delta_{2}
\end{eqnarray}

\section{Conformal Theories in 2 dimensions}\label{CFT2D}

Here, we focus on the special case of conformal theory in two dimensions. One of the most important notions is the concept of primary fields, thus it will be the next point to focus on.  
  
\subsection{Primary fields}

In order to generalize eq.(\ref{lineC}), we start by defining the {\it primary fields}. A primary field $\varphi(z,\overline{z})$ of conformal weight $(h,\overline{h})$ in a $2d$ conformal field theory is a field which has a transformation rule, under the analytical maps $z\rightarrow f(z)$ and $\overline{z}\rightarrow f(\overline{z})$, defined by
\begin{equation}\label{primarytransformation}
\varphi(z,\overline{z})\rightarrow \left(\frac{\partial f(z)}{\partial z}\right)^{h}\left(\frac{\partial f(\overline{z})}{\partial \overline{z}}\right)^{\overline{h}}\varphi(f(z),f(\overline{z})),
\end{equation}    
where $h,\overline{h} \in \mathbb{R}$. It is important to point out that $\overline{h}$ is not the complex conjugate of $h$. An equivalent way to define eq.(\ref{primarytransformation}) is by the quantity $\varphi(z,\overline{z})dz^{h}d\overline{z}^{\overline{h}}$. By imposing that this expression is invariant under the analytical transformations described above, we obtain:
\begin{equation}\label{primayfieldtransf}
\varphi(z,\overline{z}) dz^{h} d\overline{z}^{\overline{h}} = \varphi(f(z),f(\overline{z})) df(z)^{h} df(\overline{z})^{\overline{h}}
\end{equation}
Not all fields are primaries, this is a problem because in this case we do not know their transformations rule. Fortunately, these non-primary fields in a conformal field theory can be expressed in terms of primary fields only, this is why primaries are so important! In addition, these  non-primary fields are called {\it secondary fields}.

In virtue of eq. (\ref{primarytransformation}), all primary fields are automatically quasi-primaries. Therefore, we can conclude that primary fields satisfying eq.(\ref{fieldstransf}) under the global conformal group.

As usual in quantum field theory, it is useful to analyse how such a transformation behaves for infinitesimal transformations $z \longrightarrow z + \epsilon(z)$ and its anti-holomorphic counterpart $\overline{z} \longrightarrow \overline{z} + \overline{\epsilon}(\overline{z})$. By inserting these transformations into eq.(\ref{primarytransformation}), we obtain
\begin{eqnarray}
\left(\frac{\partial (z + \epsilon_{(z)})}{\partial z}\right)^{h} &=& (1 + \partial_{z}\epsilon_{(z)})^{h} = 1 + h \partial\, \epsilon_{(z)},\label{Eq1a}\\
\left(\frac{\partial (\overline{z} + \overline{\epsilon}_{(\overline{z})})}{\partial \overline{z}}\right)^{\overline{h}} &=& (1 + \partial_{\overline{z}}\overline{\epsilon}_{(\overline{z})})^{\overline{h}} = 1 + \overline{h}\,\, \overline{\partial}\overline{\epsilon}_{(\overline{z})}\label{Eq1b},,
\end{eqnarray}
where $\partial_{z} = \partial$ and $\partial_{\,\overline{z}} = \overline{\partial}$ are a short notation for holomorphic and anti-holomorphic partial derivatives, respectively. The Taylor expansion of $\varphi(z+\epsilon,\overline{z}+\overline{\epsilon})$, around $(z_{0},\overline{z}_{0})=(0,0)$, at first order yields
\begin{equation}\label{Taylorexpansion}
\varphi(z+\epsilon,\overline{z}+\overline{\epsilon}) = \varphi(z,\overline{z}) + \epsilon\partial \varphi(z,\overline{z}) + \overline{\epsilon}\overline{\partial} \varphi(z,\overline{z}).
\end{equation} 
Substituting eqs.(\ref{Eq1a}), (\ref{Eq1b}) and (\ref{Taylorexpansion}) into eq.(\ref{primarytransformation}), we have
\begin{eqnarray}\label{QUASI}
\varphi(z',\overline{z}') &=& (1 + h \partial\, \epsilon)(1 + \overline{h}\,\, \overline{\partial}\overline{\epsilon})(\varphi(z,\overline{z}) + \epsilon\partial \varphi(z,\overline{z}) + \overline{\epsilon}\overline{\partial} \varphi(z,\overline{z}))\nonumber\\
&=& (1 + h\partial \epsilon + \overline{h}\,\, \overline{\partial}\overline{\epsilon})(\varphi(z,\overline{z}) + \epsilon\partial \varphi(z,\overline{z}) + \overline{\epsilon}\overline{\partial} \varphi(z,\overline{z}))\nonumber\\
&=& \varphi + (h\partial \epsilon + \overline{h}\,\, \overline{\partial}\overline{\epsilon} + \epsilon\partial + \overline{\epsilon}\overline{\partial} )\varphi(z,\overline{z}).
\end{eqnarray}
According to the above equation, an expression for transformation of a primary field with conformal weight $(h,\overline{h})$, under the infinitesimal conformal transformation described before is obtained. From eq.(\ref{QUASI}) the infinitesimal variation of a primary field is given by the expression,
\begin{equation}\label{deltaepsilon1}
\delta_{\epsilon\overline{\epsilon}}\varphi(z,\overline{z}) = \left[(h\partial\epsilon + \epsilon\partial) + (\overline{h}\partial_{\overline{z}}\overline{\epsilon} + \overline{\epsilon}\partial_{\overline{z}})\right]\varphi(z,\overline{z}).
\end{equation}
Following the same lines as in quantum field theory, we assume that this infinitesimal transformation arise from a Taylor expansion at first order on its generator. Such a generator is understood as a conserved charge. Hence, our next step is to examine what role the conserved charge plays in this context.  
 
\section{Conserved charges and Radial ordering}\label{RadialOrdering}

According to quantum theory, to each symmetry we can associate a conserved current $j^{\mu}$, which is conserved, 
\begin{equation}\label{conservedcurrent}
\partial_{\mu}j^{\mu}=0.
\end{equation}
Along with this current, due to Noether's theorem, we can associate a conserved charge $Q$. In a $d$ dimensional quantum field theory, with $d-1$ spacial dimensions, we obtain such a conserved charge by performing the integral,
\begin{equation}\label{chargecurrentcondition}
Q=\int d^{d-1}xj_{0}.
\end{equation}
From quantum theory we also know that for any field $\Phi$, its infinitesimal variation can be expressed as
\begin{equation}\label{relacioncargafield}
\delta_{\epsilon}\Phi = \epsilon\left[Q,\Phi\right]. 
\end{equation}
This charge $Q$, generating the local coordinates transformation, can be constructed from the energy-momentum tensor $T_{\mu\nu}$ by using eq.(\ref{chargecurrentcondition}). Because the current $j_{\mu}$ could be constructed as
\begin{eqnarray}
j_{\mu} &=& T_{\mu\nu}x^{\nu},\,\, \text{dilatation current},\label{dilatationcurrent}\\
j_{\mu} &=& T_{\mu\nu}\epsilon^{\nu}, \,\, \text{other currents}. \label{othercurrents}
\end{eqnarray}
In a conformal theory $T_{\mu\nu}$ is {\it traceless}, this restriction is known as the scale invariance condition. From the dilatation current (due to scale invariance) eq.(\ref{dilatationcurrent}) and the conserved current eq.(\ref{conservedcurrent}), we directly obtain $\partial_{\mu}(T^{\mu}_{\nu}x^{\nu})= T_{\mu}^{\mu}=0$. Thus, these expressions are given by:
\begin{eqnarray}
\partial_{\mu}j^{\mu}&=&T_{\mu}^{\mu} \hspace{1cm}= 0,\,\,\, \text{for dilatation current},\\
\partial_{\mu}j^{\mu}&=&\frac{1}{2}T_{\mu}^{\mu}(\partial.\epsilon) = 0,\,\,\, \text{other currents}.
\end{eqnarray}
Now, we want to explore the scale invariance condition for conformal theories. We use a general metric $ds^{2}=g_{\alpha\beta}dx^{\alpha}dx^{\beta}$, obtaining
\begin{equation}\label{generalizeconservedcurrent}
g^{\mu\beta}\partial_{\mu}T_{\alpha\beta}=0.
\end{equation}
In a flat Euclidean space $g_{\mu\nu}=\delta_{\mu\nu}$ and its line element can be expressed as $ds^{2}=dzd\overline{z}$ ($g_{z\,z} =  g_{\overline{z}\,\overline{z}} = 0$). The traceless condition takes the form
\begin{equation}
T_{z\overline{z}}=0.
\end{equation}
Along with eq.(\ref{generalizeconservedcurrent}) and $g_{z\,z} = g_{\overline{z}\,\overline{z}} = 0$, it implies
\begin{equation}
\partial_{\overline{z}}T_{zz}=0, \,\,\,\,\text{and}\,\,\,\,\partial_{z}T_{\overline{z}\overline{z}}=0.
\end{equation} 
To find the charge $Q$ in terms of the energy-momentum tensor $T_{\mu\nu}$, we replace eq.(\ref{othercurrents}) into eq.(\ref{chargecurrentcondition}), and write
\begin{equation}\label{Qoint}
Q=\frac{1}{2\pi i}\oint\left\lbrace dz T_{z z}\epsilon + d\overline{z} T_{\overline{z}\overline{z}}\overline{\epsilon}\right\rbrace.
\end{equation}
The constant factor in front of the contour integral is a suitable factor for a posteriori calculation. Hence, eqs.(\ref{relacioncargafield}) and (\ref{Qoint}) yields
\begin{equation}\label{deltaphi1}
\delta\varphi(w,\overline{w})_{\epsilon,\overline{\epsilon}} = \frac{1}{2\pi i}\oint_{\mathcal{C}}dz\left[T_{z z}\epsilon,\varphi(w,\overline{w})\right] + \frac{1}{2\pi i}\oint_{\mathcal{C}}d\overline{z}\left[T_{\overline{z}\overline{z}}\overline{\epsilon},\varphi(w,\overline{w})\right]  .
\end{equation}
We should be careful with the commutators $\left[T_{z z}\epsilon,\varphi(w,\overline{w})\right]$ and $\left[T_{\overline{z}\overline{z}}\overline{\epsilon},\varphi(w,\overline{w})\right]$ inside the latter integrals. Because they are thought to be working into path integrals later. Hence, these commutators have to be time ordered. 

The well known time ordering in quantum field theory becomes {\it radial ordering} in conformal field theory. Holomorphic (resp. antiholomorphic) transformations $z \rightarrow f(z)$ are complex mappings of the form $e^{\tau + i\sigma}$, where $\tau$ can be understood as the physical time. Therefore, it is possible to identify this transformation over the complex plane as
\begin{equation}
r\,e^{i\theta}=e^{\tau}\,e^{i\sigma}.
\end{equation}

Thus, radial displacements are related to time variations. From this statement we are able to define the ordering of operator products as $A(z)B(w)$. For instance, it is easy to see that
\begin{equation}
A(z)B(w),\,\,\, \text{for}\,\,\,|z|<|w|\nonumber
\end{equation}
does not make sense, since this is not equivalent to the time-ordering in quantum field theory (QFT). Because in QFT we place operators evaluated at later times to the left. Therefore, we find that a product of operators $A(z)B(w)$ is defined as $|z|>|w|$. In this way, we denote radial ordering for two operators as $\mathcal{R}(A(z)B(w))$ and define it as follows:
\begin{equation}
  \mathcal{R}\left\lbrace  A(z)B(w)\right\rbrace  =\left\lbrace
  \begin{array}{l}
     A(z)B(w)\,\,\, \text{for}\,\,\,|z|>|w|\\
     B(w)A(z)\,\,\, \text{for}\,\,\,|w|>|z|\\
  \end{array}
  \right.
  \label{radialordering1}
\end{equation}
By taking the contour integral of the commutator expression $[A,B]=AB-BA$. Where $A(z)B(w)$ and $B(w)A(z)$ are defined by eq.(\ref{radialordering1}), we have:   
\begin{equation}
\oint_{|z|>|w|}A(z)B(w)dz - \oint_{|w|>|z|}B(w)A(z)dz  =  \oint_{\mathcal{C}}\mathcal{R}(A(z)B(w))dz\nonumber
\end{equation} 
The difference between contour integrals in the complex plane $\mathbb{C}$ is best illustrated in Fig.\ref{fig:radialordering}:
\begin{figure}[htbp]
  \centering
    \includegraphics[scale=0.5]{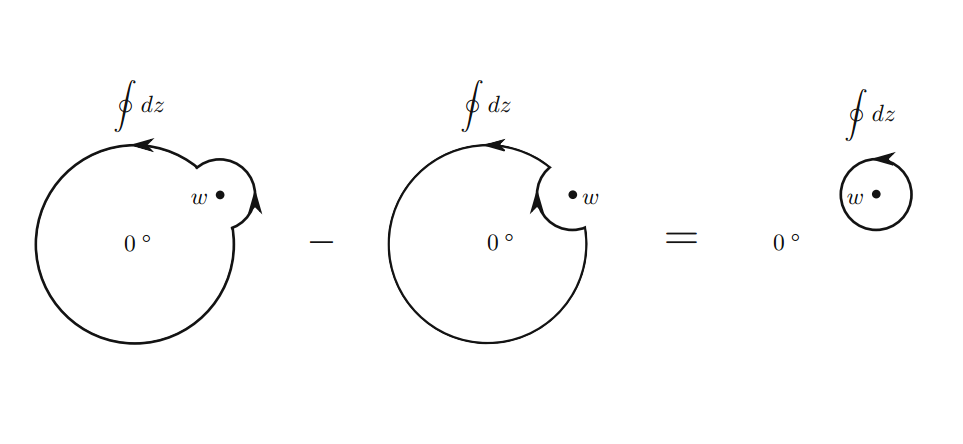}
    \rule{32em}{0.5pt}
  \caption[Radial ordering]{Contour integration used to define radial ordering, $\mathcal{R}(A(z)B(w))$.}
  \label{fig:radialordering}
\end{figure}

Note that if the operators involved are fermionic we have to be careful, because we get a minus sign from each permutation. All above definitions allow us to promote the equal time commutator with the spatial integral $[\int dx A,B]_{E.T.}$ to the contour integral of the radial ordering product $\oint dz \mathcal{R}\left\lbrace  A(z)B(w)\right\rbrace  $. Now, from eq.(\ref{deltaphi1}) it is easy to obtain
\begin{equation}\label{deltaepsilon2}
\delta\varphi(w,\overline{w})_{\epsilon,\overline{\epsilon}}= \frac{1}{2\pi i}\oint_{\mathcal{C}}dz\,\epsilon\,\mathcal{R}\left\lbrace T_{z z}\varphi(w,\overline{w})\right\rbrace  + \frac{1}{2\pi i}\oint_{\mathcal{C}}d\overline{z}\,\overline{\epsilon}\,\mathcal{R}\left\lbrace T_{\overline{z}\overline{z}}\varphi(w,\overline{w})\right\rbrace . 
\end{equation}
Comparing (\ref{deltaepsilon1}) and (\ref{deltaepsilon2}), we see that
\begin{eqnarray}
\frac{1}{2\pi i}\oint_{\mathcal{C}}dz\,\epsilon\,\mathcal{R}\left\lbrace T_{z z}\varphi(w,\overline{w})\right\rbrace &=&\left[ h\partial_{w}\epsilon(w) + \epsilon(w)\partial_{w}\right] \varphi(w,\overline{w}), \label{cauchyholomorphicF1}\\
\frac{1}{2\pi i}\oint_{\mathcal{C}}d\overline{z}\,\overline{\epsilon}\,\mathcal{R}\left\lbrace T_{\overline{z}\overline{z}}\varphi(w,\overline{w})\right\rbrace  &=&\left[ \overline{h}\partial_{\overline{w}}\overline{\epsilon}(\overline{w}) + \overline{\epsilon}(\overline{w})\partial_{\overline{w}}\right]  \varphi(w,\overline{w}).\label{cauchyholomorphicF2}
\end{eqnarray} 
Now, taking into account the Cauchy's residue theorem, 
\begin{equation}
\oint_{\mathcal{C}} f(z)dz = 2\pi i \sum_{k}Res (f, a_{k}),
\end{equation}  
where $Res (f, a_{k})$ is the residue of $f(z)$ around each one of its singularities $a_{k}$. Separating eq.(\ref{cauchyholomorphicF1}) in two holomorphic terms, we obtain
\begin{eqnarray}
h\partial_{w}\epsilon(w)\varphi(w,\overline{w})&=&\frac{1}{2\pi}\oint_{\mathcal{C}(w)}dz \,h\frac{\epsilon(z)}{(z-w)^{2}}\varphi(w,\overline{w}),\label{equationTphi1}\\
\epsilon(w)\partial_{w}\varphi(w,\overline{w})\varphi(w,\overline{w})&=&\frac{1}{2\pi}\oint_{\mathcal{C}(w)}dz\,\frac{\epsilon(z)}{(z-w)}\partial_{w}\varphi(w,\overline{w}),\label{equationTphi2}
\end{eqnarray} 
where $\mathcal{C}(w)$ is the contour around $w$. From eqs.(\ref{cauchyholomorphicF1}), (\ref{cauchyholomorphicF2}), (\ref{equationTphi1}) and (\ref{equationTphi2}), we find
\begin{equation}
\mathcal{R}\left\lbrace T_{z z}\varphi(w,\overline{w})\right\rbrace =\frac{h}{(z-w)^{2}}\varphi(w,\overline{w})+\frac{1}{(z-w)}\partial_{w}\varphi(w,\overline{w}). 
\end{equation} 
And similarly for the anti-holomorphic part $\mathcal{R}\left\lbrace T_{\overline{z}\overline{z}}\varphi(w,\overline{w})\right\rbrace $. We adopt the usual convention of dropping the symbol $\mathcal{R}$ in all expressions involving product of operators, from now on. Thus, we have for the holomorphic and the anti-holomorphic operator product between the stress-energy tensor and the field $\varphi$
\begin{eqnarray}
T_{z z}\varphi(w,\overline{w})&=&\frac{h}{(z-w)^{2}}\varphi(w,\overline{w})+\frac{1}{(z-w)}\partial_{w}\varphi(w,\overline{w}),\label{FutureNORMALORDER} \\
T_{\overline{z}\overline{z}}\varphi(w,\overline{w})&=&\frac{\overline{h}}{(\overline{z}-\overline{w})^{2}}\varphi(w,\overline{w})+\frac{1}{(\overline{z}-\overline{w})}\partial_{\overline{w}}\varphi(w,\overline{w}). 
\end{eqnarray}
We always have to keep in mind that the left-hand side is understood as into a radial ordering $\mathcal{R}$. 

In the next section we start by working with operator products, such as $T_{z z}\varphi(w,\overline{w})$. In this context a natural problem is the divergence which arises when we wish to define such an operator product in the same space-time point. This problem leads us to the concept of Operator Product Expansion (OPE). 

\section{Operator Product Expansion OPE}\label{OPE}

By analysing primary fields and their constraints for a Conformal Field Theory, we have arrived at the concept of radial ordering and it has led to the concept of operator products. Now, in this section we examine in detail the concept of {\it Operator Product Expansion} which arises to deal with the problem of defining operator products at the same point (in our context a point on the complex plane $\mathbb{C}$). For this purpose let us first start by describing what is understood by a {\it local} operator in a CFT.

In quantum field theory it is usual to reserve the term `` field '' for objects $\phi$ which are placed in the path integral. On the other hand, in a CFT a field relates to any local expression. For instance, the field $\phi$ as in QFT, but it also includes other local expressions as $\partial^{n}\phi$ or $\mathit{e}^{i\phi}$.

Now, let us explain in the next paragraphs, what is called an {\it operator product expansion} (OPE). When we study the behaviour of local operators as they approach each other. We start by giving ourself the idea of two local operators at nearby points. But what happens when they are infinitesimally approximated? This is a natural question in this context and to answer it, we first expose where the concept of OPE arises.

As was mentioned in Section \ref{CTDdimensions}, the Ising model is the birthplace of all these tools and concepts. Hence, to understand how the concept of OPE arised, it is useful to look at this model.

We recall that the Ising hamiltonian is described by $H=-\epsilon\sum_{ij}\sigma_{i}\sigma_{j}$, where $\left\lbrace i,j\right\rbrace $ are labelling sites on a square lattice. Now, if each node of the lattice is labelled by $r$, then $\sigma_{i}$ means $\sigma(r_{i})$. As we know, each node is associated to a binary-valued spin $\sigma_{i}=\pm 1$. In addition, we promote the constant coupling interaction $\epsilon$ to a variable one, and also to depend on the distance between the nodes,
\begin{equation}
\epsilon \rightarrow \epsilon_{ij}=\epsilon_{(r_{i}-r_{j})}.
\end{equation}
In this way, the Ising hamiltonian is given by:
\begin{equation}\label{IsingEij}
H=-\sum_{ij}\epsilon_{ij}\sigma_{i}\sigma_{j}
\end{equation}
Each configuration $\lbrace i,j\rbrace$ allows us to define a statistical weight $\omega(\lbrace i,j\rbrace) \propto \exp{(\sum_{ij}\epsilon_{ij}\sigma_{i}\sigma_{j})}$, where $\epsilon_{ij}>0$. A local observable $\phi_{n}(r)$ in the lattice model is the sum of products of nearby spins over some region.

Defining the partition function as $Z=\sum_{ij}\omega(\lbrace i,j\rbrace)$. Furthermore, correlation functions are defined as the expectation values of local observables and are given by
\begin{equation}
\langle \phi_{1}(r_{1})\dots\phi_{n}(r_{n})\rangle=Z^{-1}\sum_{ij}\phi_{1}(r_{1})\dots\phi_{n}(r_{n})\,\omega(\lbrace i,j\rbrace).
\end{equation}   
In general the connected pieces of these correlations functions fall off over the same distance scale as the interaction $\epsilon_{ij}$. However, close to the critical point the correlation length can become larger than the lattice length. We label such a lattice length by $L$.

The scaling limit is obtained as $L \rightarrow 0$, while we keep the correlation length and the domain fixed. Correlation functions, as defined above, do not posses a finite scaling limit in general. Nonetheless, renormalization theory suggests that there is a combination of local lattice observables which is multiplicatively renormalizable. This means that the limit
\begin{equation}\label{LIMITAAA1}
\lim_{L\rightarrow 0}L^{-\sum_{k=1}^{n}x_{k}}\langle \phi_{1}(r_{1})\dots\phi_{n}(r_{n})\rangle
\end{equation}
exists for certain values of $\lbrace x_{k}\rbrace$. It is customary to denote this limit by the simpler expression
\begin{equation}\label{CORRELATION}
\langle \phi_{1}(r_{1})\dots\phi_{n}(r_{n})\rangle .
\end{equation}
The numbers $\lbrace x_{k}\rbrace$ can be identified with our conformal weights. One important reason why eq.(\ref{CORRELATION}) is not true in general is because of the limit in eq.(\ref{LIMITAAA1}) which in fact only exists if the points $\lbrace r_{i}\rbrace$ are non-coincident. Thus, correlation functions given by eq.(\ref{CORRELATION}) are singular as $r_{i}\rightarrow r_{j}$. Nevertheless, the nature of these singularities is prescribed by the OPE 
\begin{equation}\label{DOTS}
\langle \phi_{i}(r_{i})\phi_{j}(r_{j})\dots\rangle=\sum_{k}C_{ijk}(r_{i}-r_{j})\langle \phi_{k}(\frac{r_{i}+r_{j}}{2})\dots\rangle .
\end{equation}  
The main point here is that, in the limit when $|r_{i}-r_{j}|$ is much less than the separation between $r_{i}$ and all the other arguments, which are represented by dots, the coefficients $C_{ijk}$ are independent of these. As a result, eq.(\ref{DOTS}) is usually written as
\begin{equation}
\phi_{i}(r_{i})\phi_{j}(r_{j})=\sum_{k}C_{ijk}(r_{i}-r_{j})\phi_{k}(\frac{r_{i}+r_{j}}{2}).
\end{equation}
It must be stressed that this is merely a short-hand for eq.(\ref{DOTS}).

Coming back to CFT, we consider the set of all local operators and denote it by $ \{\mathcal{O}_{i} \} $. For $ \mathcal{O}_{i}$ and $ \mathcal{O}_{j}$ belong to $ \{\mathcal{O}_{i} \} $, the OPE is defined as has been stated above,
\begin{equation}\label{OPE}
 \mathcal{O}_{i}(z,\overline{z}) \mathcal{O}_{j}(w,\overline{w}) = \sum_{k}C^{k}_{ij}(z-w,\overline{z}-\overline{w})\mathcal{O}_{k}(w,\overline{w}).
\end{equation}      
The structure coefficients $C^{k}_{ij}(z-w,\overline{z}-\overline{w})$ are (singular) functions which only depend on the distance between the operators. We have to recall here that these products are always understood inside time-ordered correlation functions (or in our case inside radial ordering) as follows,
\begin{equation}
\left\langle \mathcal{O}_{i}(z,\overline{z}) \mathcal{O}_{j}(w,\overline{w})\dots\right\rangle  = \sum_{k}C^{k}_{ij}(z-w,\overline{z}-\overline{w})\left\langle \mathcal{O}_{k}(w,\overline{w})\dots\right\rangle  .
\end{equation}
Other operator insertions are represented by dots as before. Correlations functions are always assumed to be radial ordered, it means that everything commutes since the ordering of operators is determined inside the correlation function anyway. Then we would have expressions, such as $ \mathcal{O}_{i}(z,\overline{z}) \mathcal{O}_{j}(w,\overline{w}) = \mathcal{O}_{j}(w,\overline{w})\mathcal{O}_{i}(z,\overline{z})$.

If operators are Grassmannian, we recall that they get a minus sign from each commutation, even inside radial ordering products.

Operator product expansions  have singular behaviour as $z \rightarrow w$, and this singular behaviour will turn out to contain the same information as commutation relations. It was shown by using the Cauchy theorem from complex analysis. In this way, singular terms also tells us how operators transform under symmetries. Thus, we will pay special attention to singular terms in the OPEs, as we did in a previous section for $T_{\overline{z}\,\overline{z}}\varphi(w,\overline{w})$. 

The eq.(\ref{OPE}) is an asymptotic expansion, however in a conformal theory it has been argued to converge \cite{Tong} \cite{CFToh}. The coordinate dependence of $C^{k}_{ij}(z-w,\overline{z}-\overline{w})$ can be determined by dimensional analysis for operators of fixed scaling dimension $\Delta$ and $\overline{\Delta}$. Thus we have, from (\ref{OPE}),
\begin{equation*}
C^{k}_{ij}(z-w,\overline{z}-\overline{w}) \sim \frac{1}{|z-w|^{\Delta_{\mathcal{O}_{i}}+\Delta_{\mathcal{O}_{j}}-\Delta_{\mathcal{O}_{k}}}|\overline{z}-\overline{w}|^{\overline{\Delta}_{\mathcal{O}_{i}}+\overline{\Delta}_{\mathcal{O}_{j}}-\overline{\Delta}_{\mathcal{O}_{k}}}} .
\end{equation*}

\section{Bosonic Field, the first example}

Let us illustrate how such a formalism works for the free scalar field. We start by using the string theory action, the so called Polyakov action \cite{bosonicstring},
\begin{equation}\label{Polyakovaction}
S_{B}=\frac{1}{4\pi\alpha'}\int\,d^{2}\sigma\,\sqrt{g}\,g^{\alpha\beta} \partial_{\alpha}X^{\mu}\partial_{\beta}X_{\mu} .
\end{equation}
Taking into account the canonical choice $g^{\alpha\beta}=\mathit{e}^{2\phi}\delta^{\alpha\beta}$ by using the reparametrization invariance to fix the metric on the right hand side to be proportional to the flat Euclidean metric. Thus, eq.(\ref{Polyakovaction}) is reduced to
\begin{equation}\label{Polyakovaction1}
S_{B}=\frac{1}{4\pi\alpha'}\int\,d^{2}\sigma\,\partial_{\alpha}X^{\mu}\partial^{\alpha}X_{\mu} .
\end{equation}
From equation eq.(\ref{Polyakovaction1}), the classical equation of motion is $\partial^{2}X=0$. Now, we will start by familiarizing ourselves with some techniques using path integrals, in order to derive an analogous statement at the quantum level.

As we recall from ordinary calculus that the integral of a total derivative vanishes. We promote this statement to the path integral language
\begin{equation}
0 = \int\,\mathcal{D}X\,\frac{\delta}{\delta\,X(\delta)} \mathit{e}^{-S} = \int\,\mathcal{D}X\, \mathit{e}^{-S}\left[ \frac{1}{2\pi\alpha'}\partial^{2}X(\sigma)\right] ,
\end{equation} 
but this is nothing more than the Ehrenfest theorem. This theorem states that expectation values of operators obey classical equations of motion,
\begin{equation}\label{quantummotion}
\left\langle \partial^{2}X(\sigma)\right\rangle = 0.
\end{equation} 

\subsection{Propagator of the $X$ field.}

Deriving the propagator of the $X$ field by using the path integral technique which has been developed above. We apply the same argument of taking the integral of a total derivative that vanishes. However, the integrand is taken to be $e^{-S}X$. The computation follows as
\begin{equation}
0 = \int\mathcal{D}X\frac{\delta}{\delta \, X(\sigma)} \left[ \mathit{e}^{-S}X(\sigma')\right]  = \int\mathcal{D}X \mathit{e}^{-S}\left[\frac{1}{2\pi\alpha'} \partial^{2} X(\sigma)X(\sigma') + \delta^{2}(\sigma - \sigma')\right],\nonumber 
\end{equation}
\begin{equation}\label{XXdifferentialeq}
\left\langle \partial^{2} X(\sigma)X(\sigma')\right\rangle = -2\pi\alpha'\delta^{2}(\sigma-\sigma'). 
\end{equation}
Eq.(\ref{XXdifferentialeq}) is the differential equation for the propagator $ \left\langle X(\sigma)X(\sigma')\right\rangle $. Thus, we can write
\begin{equation}\label{Propagator1}
\partial^{2}\left\langle X(\sigma)X(\sigma')\right\rangle = -2\pi\alpha'\delta^{2}(\sigma-\sigma').
\end{equation}
To solve this equation, we checked the standard result from complex calculus,
\begin{equation}\label{Propagator2}
\partial^{2}\ln(\sigma-\sigma')^{2} = 4\pi\delta^{2}(\sigma-\sigma').
\end{equation}
From eqs.(\ref{Propagator1}) and (\ref{Propagator2}), we conclude that the expression of the propagator of a free scalar field in two-dimensions is given by
\begin{equation}\label{infrareddiverg}
\left\langle X(\sigma)X(\sigma')\right\rangle = -\frac{\alpha'}{2}\ln(\sigma-\sigma')^{2}.
\end{equation}
We realize that this propagator has an ultra-violet divergence which is common to all field theories. In addition, it also has an infra-red divergence as $|\sigma-\sigma'|\rightarrow \infty$.

After a change of coordinates $z=\sigma_{1}+i\sigma_{2}$, the classical equation of motion $\partial^{2}X(\sigma)=0$ can be written in the convenient form $\partial\overline{\partial}X(z,\overline{z})=0$. It allows us to split $X$ as left-moving (holomorphic) and right-moving (anti-holomorphic) pieces, 
\begin{equation}
X(z,\overline{z}) = X(z) + \overline{X}(\overline{z}).
\end{equation}
They have propagators
\begin{eqnarray}
\left\langle X(z)X(w)\right\rangle &=& -\frac{\alpha'}{2}\ln(z-w),\\
\left\langle \overline{X}(\overline{z})\overline{X}(\overline{w})\right\rangle &=& -\frac{\alpha'}{2}\ln(\overline{z}-\overline{w}).
\end{eqnarray}
From these relations it follows that the field $X(z)$ is not itself a conformal field. However its derivative has a rather nice looking short distance expansion,
\begin{equation}
\left\langle \partial X(z)\partial X(w)\right\rangle  = -\frac{\alpha'}{2}\frac{1}{(z-w)^{2}} + \dots
\end{equation} 
where dots mean non-divergent terms.

\subsection{The Energy-Momentum Tensor and Primary Operators}

To compute the OPE of the energy-momentum tensor $T_{zz}$ with other operators plays a vital role in a CFT. For instance, the classical theory of the free scalar field has an energy-momentum tensor which is given by
\begin{equation}\label{TZ}
T(z) = -\frac{1}{\alpha'}\partial X(z)\partial X(z),
\end{equation} 
where $T(z)=T_{zz}$. It involves a product of two operators defined at the same point $z$. This fact reminds us of an OPE expansion. In addition, as we have seen from eq.(\ref{infrareddiverg}), it has an infra-red and ultra-violet divergence due to its logarithmic behaviour. 

In the same way as it usually performed in QFT, we define the normal ordering. It is well known that normal ordering naively means to put all annihilation operator modes to the right. Another equivalent way to do this, which is more convenient in CFT, is by defining normal ordering, usually denoted by $:\dots :$. We adopt this notation as follows,  
\begin{equation}\label{NORMALORDER}
T(z)=-\frac{1}{\alpha'}:\partial X(z) \partial X(z):\,\, \doteq -\frac{1}{\alpha'}\lim_{z \rightarrow w} (\partial X(z) \partial X(w) - \langle\partial X(z) \partial X(w)\rangle).
\end{equation}
It is not difficult to see that $\langle T(z) \rangle=0$.

From eq.(\ref{FutureNORMALORDER}) we already know what is the general form of an OPE between the energy-momentum operator and one primary field. Thus, such a form is expected for the energy-momentum operator described by eq.(\ref{TZ}) and a suitable primary field. 

By using eq.(\ref{infrareddiverg}), we will compute two illustrative examples of OPEs. The first one is $T(z)\partial X(w)$ and the other $T(z)e^{ik X_{(w)}}$. In this calculation we need to be aware of the fact that both OPEs are inside radial ordering. We recall that radial ordering is an analogous of time ordering in QFT. Hence, radial ordering and normal ordering should be also connected in CFT. 

Fortunately, the well known Wick's theorem still holds. It states that a normal ordered product can be split into several pieces which only involve time ordered products. Formally, we have for arbitrary operators $\mathcal{O}_{k}$ \footnote{Here, for pedagogical purposes we write explicitly the radial ordering by $\mathcal{R}[ \dots ]$.},
\begin{eqnarray}
\mathcal{R}[\prod_{k}\mathcal{O}_{k}]\,=\, :\prod_{k}\mathcal{O}_{k}:+\sum_{k=\mu,\nu}\langle \mathcal{O}_{\mu}\mathcal{O}_{\nu}\rangle \times :\prod_{k\neq \mu, \nu}\mathcal{O}_{k}: +\sum \text{cross-contractions}.
\end{eqnarray}
Using Wick's theorem and eq.(\ref{infrareddiverg}) we find $T(z)\partial X(w)$, 
\begin{equation}\label{TX1}
T(z)\partial X(w)=\frac{\partial X(w)}{(z-w)^{2}} + \frac{\partial^{2} X(w)}{z-w} + \dots 
\end{equation}
By comparing this last expression  with eq.(\ref{TZ}), we become aware that $\partial X(w)$ is a primary field of conformal weight $h=1$.

The other OPE is computed in two parts \cite{Tong}. First, we compute the OPE 
\begin{eqnarray}
\partial X(z):e^{ikX_{(w)}}:&=&\sum_{j=0}^{\infty}\frac{(ik)^{j}}{j!}\partial X(z):X^{j}(w):\nonumber\\
&=&-\frac{i\alpha' k}{2}\frac{:e^{ikX_{(w)}}:}{z-w}+\dots
\end{eqnarray} 
Now, we can compute 
\begin{eqnarray}
T(z)e^{ikX_{(w)}}&=&-\frac{1}{\alpha'}:\partial X(z)\partial X(z):\,:e^{ikX_{(w)}}:\nonumber\\
&=&\frac{\alpha'k^{2}}{4}\frac{e^{ikX_{(w)}}}{(z-w)^{2}}+\frac{\partial_{w}e^{ikX_{(w)}}}{z-w}+\dots
\end{eqnarray}
Finally, we also write the OPE between two energy-momentum tensors. Using eq.(\ref{TX1}), we have
\begin{equation}
T(z)T(w)=\frac{1/2}{(z-w)^{4}} +\frac{2T(w)}{(z-w)^{2}}+\frac{\partial T(w)}{z-w}\dots
\end{equation}
We will come back to study this OPE later.

\section{The second example, the Fermionic Field}

Another important example of a two dimensional conformal field theory is provided by  the theory of a single massless Majorana-Weyl fermion. We recall that Majorana condition is equivalent to the requirement of reality condition \cite{K}. On the other hand, Weyl condition  defines a chirality for our spinor, that is, a two dimensional analogue of the well known four-dimensional $\gamma_{5}$. Once both conditions are established, this theory is described by the massless fermion action, 
\begin{equation}\label{Faction}
S_{F}=-\frac{1}{2\pi}\int d\sigma^{2}\psi\overline{\partial}\psi + \overline{\psi}\partial\overline{\psi}.
\end{equation}
In two dimensions the usual Dirac action takes the form given in eq.(\ref{Faction}), because we can decompose $\sigma^{\mu}\partial_{\mu}$ as
\begin{eqnarray}
\widehat{\partial}=\sigma_{x}\partial_{x}+\sigma_{y}\partial_{y}=
\left( 
\begin{matrix}
0&&\partial_{x}-i\partial_{y}\\
\partial_{x}+i\partial_{y}&&0
\end{matrix}
\right)=\frac{1}{2} 
\left( 
\begin{matrix}
0&&\partial\\
\overline{\partial} &&0
\end{matrix}
\right)
\end{eqnarray}
and also
\begin{eqnarray}\label{Faction2}
\left( 
\begin{matrix}
\psi\\
\overline{\psi}
\end{matrix}
\right)=\frac{1}{2}(1+\sigma_{z})
\left( 
\begin{matrix}
\psi\\
0
\end{matrix}
\right)+\frac{1}{2}(1-\sigma_{z})
\left( 
\begin{matrix}
0\\
\overline{\psi}
\end{matrix}
\right).
\end{eqnarray}
In eq.(\ref{Faction}) as much as eq.(\ref{Faction2}), $\sigma_{x}, \sigma_{y}$ and $\sigma_{z}$ are the Pauli matrices. Here, $\sigma_{z}$ plays a similar role of $\gamma_{5}$. 

From the fermionic action the energy-momentum tensor is obtained as is usual in QFT. As in the bosonic case, the fermionic energy-momentum tensor has a holomorphic and an anti-holomorphic part. We will use only the holomorphic part, thus we only give such an expression. The holomorphic part is given by
\begin{eqnarray}
T_{\psi}(z)=\frac{1}{2}\psi(z)\partial\,\psi(w),\\
\overline{T}_{\overline{\psi}}(\overline{z})=\frac{1}{2}\overline{\psi}(\overline{z})\overline{\partial}\,\,\overline{\psi}(\overline{w}).
\end{eqnarray}
By applying the same technique as in the bosonic action, we mean that the total functional derivative of a path integral vanishes and we obtain
\begin{equation}
0 = \int\mathcal{D}\psi\mathcal{D}\overline{\psi}\frac{\delta}{\delta \, \psi(z)} \left[ \psi(w)\mathit{e}^{-S_{F}}\right]  = \int\mathcal{D}\psi\mathcal{D}\overline{\psi} \mathit{e}^{-S_{F}}\left[\frac{1}{\pi} \overline{\partial}\psi(z)\psi(w) + \delta^{2}(z - w)\right] 
\end{equation}
thus its propagator is given by 
\begin{equation}
\langle\psi(z)\psi(w)\rangle=-\frac{1}{z-w}.
\end{equation}
Following the same lines as before (for the bosonic OPE) the operator product expansion between the energy momentum operator and the fermionic field is
\begin{equation}\label{FermionicOPE}
T_{\psi}(z)\psi(w)=\frac{1/2}{(z-w)^{2}}\psi(w)+\frac{1}{z-w}\partial\psi(w)+\dots
\end{equation}
By comparing this last equation and our general OPE expression between the energy-momentum $T(z)$ and a primary field eq.(\ref{FutureNORMALORDER}), we become aware that this fermionic OPE eq.(\ref{FermionicOPE}) is describing a primary field of conformal weight $h=1/2$. 

In the present fermionic case, it is important to point out that on the cylinder we can define two kinds of boundary conditions for Majorana-Weyl fermions. They can be either periodic or anti-periodic boundary conditions. Under periodic boundary conditions they are called the Neveu-Schwarz type (NS), and under anti-periodic ones they are called the Ramond type (R), we explicitly write these conditions on the plane as: \footnote{We realize that a periodic fermion on the cylinder becomes an anti-periodic fermion on the plane and \it{vice versa} \cite{K}.}
\begin{eqnarray}
\text{Neveu-Schwarz:}\hspace{0.5cm} \psi(e^{2\pi i}z)=+\psi(z),\nonumber\\
\text{Ramond:}\hspace{0.5cm} \psi(e^{2\pi i}z)=-\psi(z).
\end{eqnarray}


\section{Central Charge and Virasoro algebra}

An important topic to observe in CFT is with regard to the OPE between two energy-momentum tensors. The first question that we could ask is if this operator is a primary field. To answer this question we analyse its OPE and compare with eq.(\ref{FutureNORMALORDER}), which is our general OPE expression for a primary field. Hence, by dimensional and analyticity reasons this OPE is described as
\begin{equation}\label{TTOPE}
T(z)T(w)=\frac{c/2}{(z-w)^{4}}+\frac{2}{(z-w)^{2}}T(w)+\frac{1}{z-w}\partial_{w}T(w)+\dots
\end{equation}
Because the first term $(c/2)/(z-w)^{4}$ is present there, we realize that $T(Z)$ is not a primary field. However, this OPE is still an important result, because from this expression is possible to derive an equivalent representation which is called the Virasoro algebra.

\begin{figure}[htbp]
  \centering
    \includegraphics[scale=0.4]{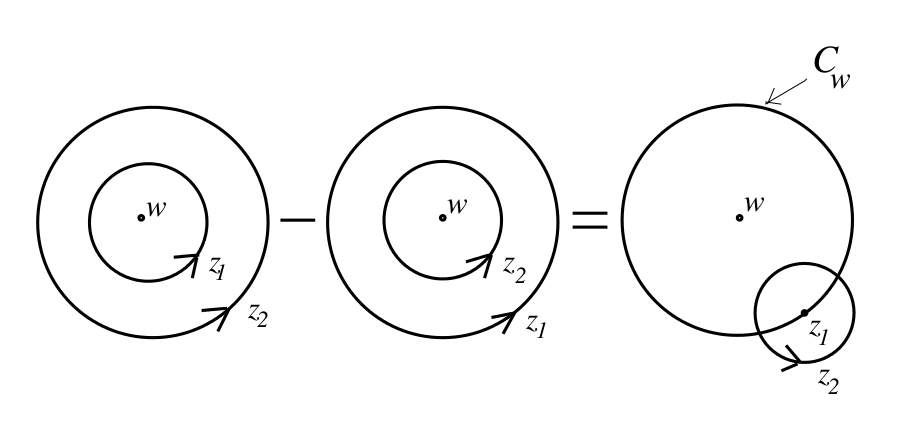}
    \rule{20em}{0.3pt}
  \caption[Contour]{Contours involved in eq.(\ref{contour1}).}
  \label{fig:amor1}
\end{figure}

In order to arrive at this important result, we recall the general expression of an OPE for a primary field $\varphi$, 
\begin{equation}
T(z)\varphi(w,\overline{w})=\frac{h}{(z-w)^{2}}\varphi(w,\overline{w})+\frac{1}{(z-w)}\partial_{w}\varphi(w,\overline{w}),
\end{equation}
where we do not care about non-singular terms. We also recall that by taking into account the Laurent expansion of $T(z)$, where $\widehat{L}_{n}$'s are the operational modes. When these operator modes act on $\varphi$, we obtain the next contour integral, 
\begin{eqnarray}
\widehat{L}_{m}\varphi(w)=\oint \frac{dz}{2\pi i}(z-w)^{m+1}T(z)\varphi(w).
\end{eqnarray}
Now, we want to build a commutator relating two of these operator modes. To get this commutator relation we compute the action of $\widehat{L}_{n}\widehat{L}_{m}$ and $\widehat{L}_{m}\widehat{L}_{n}$ on a primary field $\varphi$. Explicitly, we have the next double contour integrals,
\begin{eqnarray}
\widehat{L}_{n}\widehat{L}_{m}\varphi(w)=\oint_{C_{2}}\frac{dz_{2}}{2\pi i}\oint_{C_{1}}\frac{dz_{1}}{2\pi i}(z_{2}-w)^{n+1}(z_{1}-w)^{m+1}T(z_{2})T(z_{1})\varphi(w),\\
\widehat{L}_{m}\widehat{L}_{n}\varphi(w)=\oint_{C_{1}}\frac{dz_{1}}{2\pi i}\oint_{C_{2}}\frac{dz_{2}}{2\pi i}(z_{2}-w)^{m+1}(z_{2}-w)^{n+1}T(z_{1})T(z_{2})\varphi(w).
\end{eqnarray}
Subtracting these equations in order to build the commutator $\left[ \widehat{L}_{n},\widehat{L}_{m}\right]$, we have 
\begin{eqnarray}\label{contour1}
\hspace{-0.5cm}\left[ \widehat{L}_{n},\widehat{L}_{m}\right] \varphi(w)=\oint_{C_{w}}\frac{dz_{1}}{2\pi i}(z_{1}-w)^{m+1}\oint_{C_{z_{1}}}\frac{dz_{2}}{2\pi i}(z_{2}-w)^{n+1}T(z_{2})T(z_{1})\varphi(w),
\end{eqnarray}
where the contour integrals are graphically explained in Fig.(\ref{fig:amor1}).

After computing this contour integrals, we find the so-called Virasoro algebra. Thus, this algebra is defined by the commutator
\begin{equation}\label{VirasoroAlgebra}
\left[ \widehat{L}_{n},\widehat{L}_{m}\right]=(n-m)\widehat{L}_{n+m}+ \frac{c}{12}n(n^{2}-1)\delta_{n+m,0}.
\end{equation}

\section{{\it In} and {\it Out} states}

We can recall that the complex coordinates, which are a natural way to describe the infinite conformal algebra arising in $d=2$, was made up by using the coordinates $\xi_{1}$ and $\xi_{2}$. Explicitly, we have for the holomorphic (anti-holomorphic) coordinate $z=\xi_{1}+i\xi_{2}$ ($\overline{z}=\xi_{1}-i\xi_{2}$). We will see in the next chapter that Liouville theory arises in the context of string theory. When we take into account Liouville contributions such a theory is called non-critical string theory. Therefore, it is natural to ask ourselves how this complex coordinates arise in the string theory context. 

In String Theory instead of a world line, which could be parametrized by a proper-time $\tau$, we describe the theory on the {\it world sheet}. Thus, it is parametrized by to coordinates, say $\tau$ and $\sigma$. For our purposes, we take into account the closed string. In this context, those parameters have the ranges $\tau\in \mathbb{R}$ and $0<\sigma<2\pi$. For instance if we consider only a section of $\mathbb{R}$, these parameters are describing a cylinder. Nonetheless, if we want to describe this theory in a complex plane $\mathbb{C}$, it is necessary to map the cylinder to the plane. This is performed by the holomorphic transformation
\begin{equation}\label{CylynderPlaneTransf}
 z=e^{\tau+i\sigma}
\end{equation} 
where the anti-holomorphic coordinate is $\overline{z}=e^{\tau - i\sigma}$. From eq.(\ref{CylynderPlaneTransf}) it is easy to see that we can identify the origin and the infinity of the complex plane as
\begin{eqnarray}
z&=&0,\hspace{0.3cm}\text{as}\hspace{0.3cm} \tau \rightarrow -\infty,\label{A1}\\
z&=&\infty ,\hspace{0.3cm} \text{as} \hspace{0.3cm}\tau \rightarrow \infty.
\end{eqnarray} 
Additionally, from Chapter \ref{Chapter2} we learned that if the point $z=0$ is well defined, the $\widehat{L}_{n}$ operators are constrained to $n \geq -1$. Combining the last statement and eq.(\ref{A1}), it is possible to define the {\it in}-vacuum state $|0\rangle$ in a CFT (which correspond to $\tau \rightarrow -\infty$) to be invariant under the global conformal group. Such a vacuum state vanishes when $\widehat{L}_{n}$ acts on it, thus we have 
\begin{equation}\label{Nmenor1}
\widehat{L}_{n}|0\rangle = 0,\hspace{0.3cm}\text{as}\hspace{0.2cm} n \geq -1.
\end{equation}
In the same way, we can obtain the so called {\it in}-states. These states are determined by the limit
\begin{equation}\label{INstates}
|\mathcal{O}_{\text{in}} \rangle\equiv \lim_{z,\overline{z}\rightarrow 0} \mathcal{O}(z,\overline{z})|0\rangle,
\end{equation}
where, as we mentioned before, they are defined at the past infinite $\tau\rightarrow -\infty$. In CFT we also define other states, these ones are called {\it out}-states. To define such states, we first start by using eq.(\ref{Nmenor1}) and get 
\begin{equation}\label{Nmenor2}
\langle 0|\widehat{L}_{-n}=0,\hspace{0.3cm} n \geq -1.
\end{equation}
Therefore, in a similar way as we defined in-states, it is defined the out-states. Before performing this, we make the next change of variables $z\rightarrow w=1/z$. We recall that such a transformation was made in Chapter \ref{Chapter2}, to define correctly the $\widehat{L}_{n}$ operators at infinity on the Riemann sphere. Thus, as $z\rightarrow \infty$ we have $w\rightarrow 0$, and a similar expression for the anti-holomorphic coordinate. Now, we are prepared to give the expression for the out states,
\begin{equation}\label{OUTstates}
\langle\mathcal{O}_{\text{out}}|\equiv \lim_{w,\overline{w}\rightarrow 0} \langle 0| \mathcal{O}(w,\overline{w}),
\end{equation}
where, in contrast to in-states, eq.(\ref{OUTstates}) is determined at $\tau\rightarrow \infty$. Considering the operators which give rise to the in- and out-states as primaries of conformal dimensions $h$ and $\overline{h}$, they have a well defined transformation rule. By using this transformation for primaries and the mappings $z\rightarrow w=1/z$, $\overline{z}\rightarrow \overline{w}=1/\overline{z}$,\, we obtain 
\begin{equation}\label{EQHome1}
\mathcal{O}(w,\overline{w})=\mathcal{O}(1/w,1/\overline{w})(-w^{-2})^{h}(-\overline{w}^{-2})^{\overline{h}}.
\end{equation}
From this equation, it is easy to see that the out-states could also be defined as
\begin{equation}\label{EQHome2}
\langle\mathcal{O}_{\text{out}}|=\lim_{z,\overline{z} \rightarrow \infty}\langle 0|\mathcal{O}(z,\overline{z})z^{2h}\overline{z}^{2\overline{z}}.
\end{equation}
From eq.(\ref{EQHome1}) it is natural to define the adjoint state $|\dots\rangle^{\dagger}$ by:
\begin{equation}\label{EQHome3}
\left[\mathcal{O}(z,\overline{z})\right]^{\dagger}\equiv \mathcal{O}\left(1/z,1/\overline{z}\right)\frac{1}{z^{2h}}\frac{1}{\overline{z}^{2\overline{h}}}  
\end{equation}
This is a natural definition indeed, by using eqs.(\ref{EQHome1}), (\ref{EQHome2}) and (\ref{EQHome3}), we have
\begin{eqnarray}
\langle \mathcal{O}_{\text{out}}|=\lim_{w,\overline{w}\rightarrow 0}\mathcal{O}(w,\overline{w})=\langle 0|\mathcal{O}\left(1/z,1/\overline{z} \right)\frac{1}{z^{2h}}\frac{1}{\overline{z}^{2\overline{h}}}, \\
=\lim_{z,\overline{z}\rightarrow 0}\langle 0 |\left[\mathcal{O}(z,\overline{z})\right]^{\dagger}=\left[ \lim_{z,\overline{z}\rightarrow 0}\mathcal{O}(z,\overline{z})|0\rangle\right]^{\dagger}=|\mathcal{O}\rangle_{\text{in}}^{\dagger}.   
\end{eqnarray}
Therefore, we obtain the usual expression $\langle \mathcal{O}_{\text{out}}|=|\mathcal{O}\rangle_{\text{in}}^{\dagger}$.

\subsection{Positive norm of primary states in CFT}

In the same way that we defined primary operators, we are now able to define the so called {\it primary} states. They are defined similarly as we did for the in- and out-states
\begin{equation}\label{PrimaryStates}
|h,\overline{h}\rangle=\lim_{z,\overline{z}\rightarrow 0}\varphi_{h,\overline{h}}(z,\overline{z})|0\rangle,
\end{equation}
where $\varphi_{h,\overline{h}}$ is a primary field. Once we define these states, we use eq.(\ref{FutureNORMALORDER}) to compute the commutator 
\begin{eqnarray}\label{OO}
\left[\widehat{L}_{n},\varphi_{h,\overline{h}}(z,\overline{z}) \right]&=&\oint \frac{dw}{2\pi i}w^{n+1}T(w)\varphi_{h,\overline{h}}(z,\overline{z})\\
&=&\left(z^{n+1}\frac{d}{dz}+(n+1)z^{n}h \right) \varphi_{h,\overline{h}}(z,\overline{z}).
\end{eqnarray}
After applying the vacuum state $|0\rangle$, we compute $\left[\widehat{L}_{n},\varphi_{h,\overline{h}}(z,\overline{z}) \right]|0\rangle$ and realize that for $n>0$ the limit in eq.(\ref{PrimaryStates}) vanishes
\begin{eqnarray}
(\widehat{L}_{n}\varphi_{h,\overline{h}}-\varphi_{h,\overline{h}}\widehat{L}_{n})|0\rangle &=&(z^{n+1}\frac{d}{dz}+(n+1)z^{n}h ) \varphi_{h,\overline{h}}|0\rangle\nonumber\\
\widehat{L}_{n}\lim_{z,\overline{z}\rightarrow 0}\varphi_{h,\overline{h}}|0\rangle-0&=&\lim_{z,\overline{z}\rightarrow 0}(z^{n+1}\frac{d}{dz}+(n+1)z^{n}h ) \varphi_{h,\overline{h}}|0\rangle\nonumber\\
\widehat{L}_{n}|h,\overline{h}\rangle &=& 0. 
\end{eqnarray}
For $n=0$ we obtain
\begin{eqnarray}
\widehat{L}_{0}|h,\overline{h}\rangle &=& h|h,\overline{h}\rangle,\\
\widehat{\overline{L}}_{0}|h,\overline{h}\rangle &=& \overline{h}|h,\overline{h}\rangle.
\end{eqnarray}
These equations define what is called highest-weight states. In addition, by using eq.(\ref{VirasoroAlgebra}) and our definition of primary states, it is easy to see that the central charge $c$ and the conformal dimension should be both larger than zero, in order to define a positive norm in the Hilbert space of CFT. By applying the Virasoro algebra on primary states $|h\rangle$, we get
\begin{equation}
\langle h|L_{-n}^{\dagger}L_{-n}|h\rangle=\left(2nh+\frac{c}{12}n(n^{2}-1) \right) \langle h|h\rangle>0
\end{equation}
and finally, this result implies $c>0$ and $h>0$.

\chapter{Liouville Theory} 

\label{Chapter3} 

\lhead{Chapter 3. \emph{Liouville Theory}} 

\section{Introduction}

Liouville theory or Liouville Quantum Gravity emerges as a 2D toy model which tries to understand and answer some technical and conceptual problems about the difficult issue of the quantization of gravity.
  
The present chapter analyses some basic aspects about the 2D Classical and Quantum Liouville Theory.
We explicitly use some results taken from Chapter \ref{Chapter2}.

\section{The beginning of the Liouville Theory}\label{Section1.3}

\begin{figure}[htbp]
  \centering
    \includegraphics[scale=0.9]{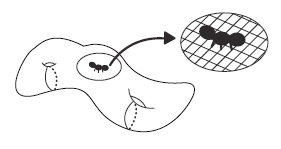}
    \rule{30em}{0.5pt}
  \caption[A manifold]{For an ant walking around a neighbourhood of a point on a manifold, it looks like a plane.}
  \label{fig:Manifold}
\end{figure}

Liouville theory initially arose at the end of nineteenth century into Poincare's works \cite{Poincare}. Liouville theory was extensively studied because of its connection with the uniformization problem of Riemann surfaces. In this section we make and outline this mathematical point of view.
First of all, let us define a Riemannian manifold \footnote{A manifold is a topological space that remind us an Euclidean space in the neighbourhood of each point as it is shown in Fig. \ref{fig:Manifold}.}\footnote{A Riemannian manifold or (smooth) Riemannian space $(M,g)$ is a real smooth manifold $M$ equipped with an inner product $g_p$ on the tangent space $T_{p} M$ at each point $p$ that varies smoothly from point to point in the sense that if $X$ and $Y$ are vector fields on $M$, then $p\mapsto g_p(X(p),Y(p))$ is a smooth function.} $X$ with a {\it local} chart $(U_{\alpha},z_{\alpha})$ \footnote{Combination of open subsets $U_{\alpha}$ on $X$ and the coordinates $z_{\alpha}$ associate to each patch is called a local chart.} and a local metric associated to $U_{\alpha}$,
\begin{equation}
ds^{2} = e^{{\varphi}_{\alpha}}dz_{\alpha}^{2}. 
\label{conformalmetricalpha}
\end{equation}

This metric is sometimes called a hyperbolic metric. It has a constant negative curvature $R=-1$, that is,
\begin{equation}
R=-2e^{-{\varphi}_{\alpha}}\partial\overline{\partial}\varphi=-1.
\label{escalardecurvatura1}
\end{equation}
From this last expression we can find the following differential equation by taking into account the latter equality,
\begin{equation}
\partial\overline{\partial}\varphi_{\alpha}= \frac{1}{2}e^{{\varphi}_{\alpha}}.
\label{liouvilleequation}
\end{equation}
This is the well known Liouville equation and the origin of the name ``Liouville'' for the present theory.
On the other hand, let us denote by $D$ either the Riemann sphere $\mathbb{P}^{1} = \mathbb{C} \cup \{\infty\}$, the complex plane $\mathbb{C}$, or the
upper half plane $H = \{w \in \mathbb{C}|\text{Im}\, w > 0\}$. The uniformization theorem states that every
Riemann surface $\Sigma$ is conformally equivalent to the quotient $D/\Gamma$ with $\Gamma$ a freely acting discontinuous group of fractional transformations preserving $D$.

Let us consider the case of Riemann surfaces with universal covering H and denote by
$\tau$ the complex analytic covering $ \tau: H \rightarrow \Sigma$. In this case $\Gamma$ (the automorphism group of $\tau$ ) is a finitely generated group \footnote{Called the Fuchsian group.} $\Gamma \subset PSL(2, \mathbb{R}) = SL(2, \mathbb{R})/\mathbb{Z}_{2}$\footnote{Remember that we will extend $SL(2, \mathbb{R})$ to $SL(2, \mathbb{C}) $. } acting on $H$ by linear fractional transformations as defined in Chapter \ref{Chapter2}.

Following \cite{Matone}, we can express $e^{\varphi_{\alpha}}$ as
\begin{eqnarray}
\frac{ \|(\tau^{-1})'\|^{2}}{(\text{Im} \tau^{-1})^{2}}.
\end{eqnarray}

Therefore, it is evident that from the inverse map $\tau^{-1}$ we can find $\varphi_{\alpha}$. Conversely to find  $\tau^{-1}$ in terms of $\varphi_{\alpha}$, we use the Schwarzian equation 
\begin{eqnarray}\label{nonlinearised}
  \left\lbrace \tau^{-1}, z_{\alpha} \right\rbrace = T_{z_{\alpha}} 
\end{eqnarray}
where $\left\lbrace \,\,\, , \,\,\, \right\rbrace$ is the Schwarzian derivative \footnote{$\left\lbrace f, z \right\rbrace = \frac{f'''}{f'}-\frac{3}{2}\left( \frac{f''}{f'}\right)^{2}. $} and $T_{z_{\alpha}}$ is identified as the energy-momentum tensor and explicitly it takes the form
\begin{equation}
T_{z_{\alpha}}=\frac{\partial^{2}\varphi_{\alpha}}{\partial z_{\alpha}^{2}}-\frac{1}{2}(\frac{\partial\varphi_{\alpha}}{\partial z_{\alpha}})^{2}.
\label{tensor1}
\end{equation}
The linearised version of eq.(\ref{nonlinearised}) is the so called {\it uniformization equation} \cite{Matone}
\begin{equation}
\frac{d^{2}u_{\alpha}}{dz_{\alpha}^{2}}+ \frac{1}{2}T_{z_{\alpha}}u_{\alpha}=0.
\label{secondorder1}
\end{equation}
with the property that its two independent solutions $u_{\alpha}^{1}$ and $u_{\alpha}^{2}$ define the inverse map $\tau^{-1}$ in the way
\begin{eqnarray}
\tau^{-1}=\frac{u_{\alpha}^{2}}{u_{\alpha}^{1}}.
\end{eqnarray}
As a result, solutions of eq.(\ref{secondorder1}) also define $\phi$. On the other hand, from eqs.(\ref{liouvilleequation}) and (\ref{tensor1}), it is easy to see that $\overline{\partial}\, T_{z_{\alpha}}=0$. Indeed,
\begin{eqnarray*}
\overline{\partial}\, T_{z_{\alpha}}&=&\partial(\partial\overline{\partial}\varphi_{\alpha})+\frac{1}{2}\overline{\partial}(\partial \varphi_{\alpha})^{2}\\
&=&\partial(\frac{1}{2}e^{{\varphi}_{\alpha}})-(\partial\varphi_{\alpha})(\frac{1}{2}e^{{\varphi}_{\alpha}})\\
&=& 0.
\end{eqnarray*}
In this way $T_{z_{\alpha}}$ does not depend on $\overline{z}_{\alpha}$. In other words, $T_{z_{\alpha}}$ is a holomorphic function.
  
Let us define now an $n$-{\it punctured} sphere. An $n$-punctured sphere is defined as the manifold $X=\mathbb{P}^{1} \diagdown \left\{z_{1}, z_{2},..., z_{n}\right\}$, where $\left\{z_{1}, z_{2},..., z_{n}\right\}$ are removed points. 

For the very special cases $n=1,2$, we have:

\begin{enumerate}
\item  $\mathbb{P}^{1} \diagdown \left\{z\right\}$\hspace{0.3cm} for $n=1$.
\item  $\mathbb{P}^{1} \diagdown \left\{z_{1}, z_{2}\right\}$\hspace{0.3cm} for $n=2$.
\end{enumerate}

The first case is homeomorphic to $\mathbb{C}$ and the other to $\mathbb{C}^{*}\equiv \mathbb{C}- \left\{0\right\}$. However, no case carries us to a hyperbolic metric. Thus, the first non-trivial case appears as $n=3$.

The Riemann sphere with three punctures $\mathbb{P}^{1} \diagdown \left\{z_{1}, z_{2}, z_{3}\right\}$ is homeomorphic to $\mathbb{C}-\left\{0,1\right\}$. Furthermore, it is a non-compact Riemann surface whose total area is finite. We are interested in this particular case.

For $n\geq 3$, we would have a Riemann sphere with $n$ - punctures $\mathbb{P}^{1}\diagdown {z_{1},\dots,z_{n}}$. Additionally, such a set of points is homeomorphic to $\mathbb{C}-\left\{\widetilde{z}_{1}, \widetilde{z}_{2}, \dots, \widetilde{z}_{n-1} \right\}$. Hence, we are now interested in studying how the hyperbolic metric eq.(\ref{conformalmetricalpha}) behaves over the Riemann sphere with $n$ - punctures.

\begin{figure}[htbp]
  \centering
    \includegraphics[scale=0.4]{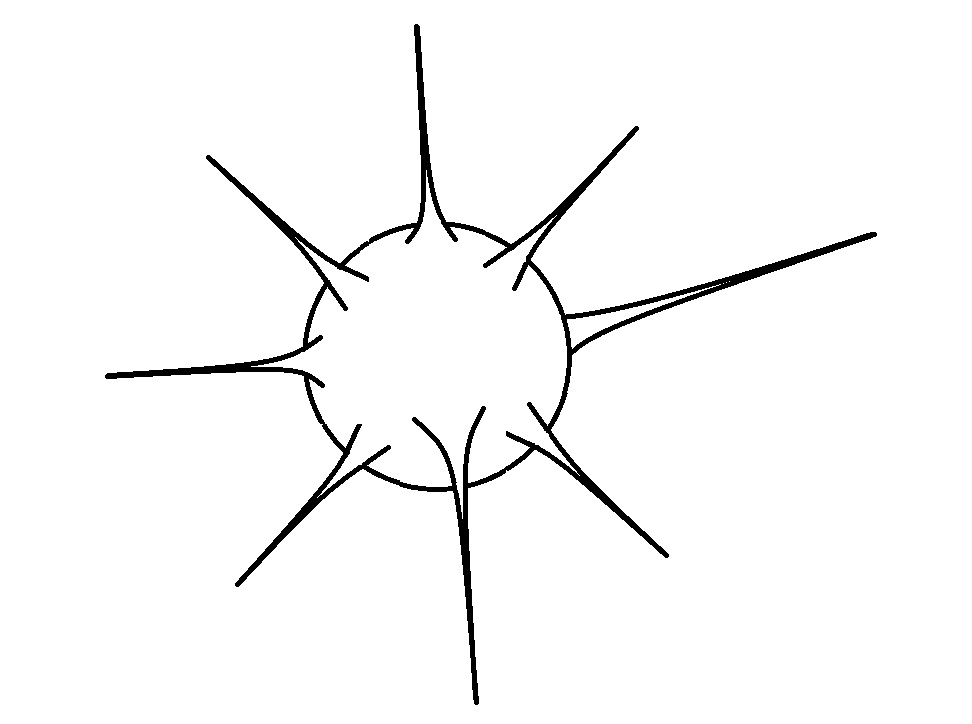}
    \rule{30em}{0.5pt}
  \caption[Punctured sphere]{A Riemann sphere with insertions of infinite positive curvature metrics, at its {\it punctured} points.}
  \label{fig:sphere}
\end{figure}

On the Riemann sphere, which has a constant positive curvature $R=+1$, the points removed are placed at the infinity. Hence, it is necessary to insert infinity positive curvature metrics at the punctured points, in order to balance the negative curvature of the hyperbolic geometry described by $ds^{2}=e^{\varphi_{\alpha}}dz_{\alpha}d\overline{z}_{\alpha}$. This process is illustrated in Fig.\ref{fig:sphere}.

Near these insertions, the metric density given by the conformal factor in eq.($\ref{conformalmetricalpha}$) has an asymptotic behaviour given by
\begin{equation}
e^{\varphi_{(z,\overline{z})}}\simeq \frac{1}{\left|z-z_{i}\right|^{2}\log^{2}\left|z-z_{i}\right|},\hspace{0.3cm}\text{as}\hspace{0.3cm} z\rightarrow z_{i}.
\label{metricainfinita}
\end{equation}
Moreover, as $z\rightarrow z_{\infty}$ \footnote{$z_{\infty}$ is representing the point at the infinity.} this metric density has a similar form,
\begin{equation}
e^{\varphi}\simeq \frac{1}{|z|^{2}\log^{2}|z|}.
\end{equation}
All these results will be reinterpreted in the framework of two dimensional Classical and Quantum Liouville theory \cite{2}\cite{bosonicstring}\cite{nakayama}\cite{ginsparg}\cite{Seiberg}\cite{zamo}\cite{dual}\cite{dual2}. There, $\varphi$ is reinterpreted as a scalar field \footnote{Properly $\varphi$ is not an scalar field, due to its anomalous transformation law, under a change of coordinates. For instance, if $z \rightarrow w(z)$, we have $\varphi(z)\rightarrow \varphi(w)+2\ln|\partial w/\partial z|$.} living on a Riemann surface, along with a local metric tensor $g^{\mu\nu}$.

\section{Classical Liouville Theory}

Let us start by equipping a two dimensional surface $\Sigma$ with a metric $g_{\mu\nu}$. As such, Liouville theory is defined as the theory of metrics $g_{\mu\nu}$ on $\Sigma$. The Liouville field $\varphi$, also called the Liouville mode, is defined as the exponent of the conformal factor. Another way to contemplate that conformal factor is as a link between the metrics $g_{\mu\nu}$ and $\widehat{g}_{\mu\nu}$ \footnote{Sometimes $g_{\mu\nu}$ is called physical metric and $\widehat{g}_{\mu\nu}$ as reference metric or fiducial metric.}. Thus, they are related by the familiar expression:
\begin{equation}\label{gegrelation}
\underbrace{g_{\mu\nu}}_{\text{Physical metric}} =  \overbrace{e^{\varphi}}^{\text{Conformal factor}} \,\times \underbrace{\widehat{g}_{\mu\nu}}_{\text{Reference metric}}
\end{equation}
Following the notation given in \cite{Seiberg} \cite{ginsparg}, the Classical Liouville action is given by
\begin{equation}\label{classicalLiouvilleaction}
S_{L}[\varphi,\widehat{g}]=\frac{1}{2\pi}\int_{\Sigma} d^{2}z\,\sqrt{\widehat{g}}\left(\frac{1}{2}\widehat{g}^{\mu\nu}\partial_{\mu}\varphi\partial_{\nu}\varphi+\widehat{R}\varphi+\Lambda\mathit{e}^{\varphi}\right)  
\end{equation}
The Euler-Lagrange equation leads to the next equation of motion
\begin{equation}\label{EulerLagrange}
\Delta_{\widehat{g}}\varphi+\widehat{R}+\Lambda\mathit{e}^{\varphi}=0,
\end{equation}
where $\Delta_{\widehat{g}}$ is the Beltrami-Laplace operator associated to $\widehat{g}$ \footnote{$\Delta_{\widehat{g}}=-\frac{1}{\sqrt{\widehat{g}}}\partial_{\mu}\widehat{g}^{\mu\nu}\sqrt{\widehat{g}}\partial_{\nu}$.}. From eq.(\ref{gegrelation}) we have the next useful and crucial result
\begin{equation}\label{RRrelation}
\sqrt{g} R = \sqrt{\widehat{g}}\left(\widehat{R}+\Delta_{\widehat{g}}\varphi\right), 
\end{equation}
where $R=R_{[g]}$ and $\widehat{R}=R_{[\widehat{g}]}$ are the scalars of curvature for the physical and the reference metric respectively. In addition, from eqs.(\ref{EulerLagrange}) and (\ref{RRrelation}), it is easy to see that the scalar of curvature of the physical metric is given by
\begin{equation}
R(\sigma)=-\Lambda.
\label{Rlambda}
\end{equation} 
Thus, this classical configuration describes a two dimensional surface $\Sigma$ with constant curvature equal to $-\Lambda$. If we fix $\Lambda=1$, we obtain the curvature described at the beginning of this section. 

As in the last section, we have taken the line element as $ds^{2}=e^{\varphi} dz d\overline{z}$ \footnote{Here we disappear the sub-index $\alpha$, because an $n\geq 3$ - punctured sphere is always homeomorphic to $\mathbb{C}-\{ z_{1},\dots,z_{n-1}\} $. Thus, it is possible to cover these manifolds by a single coordinate chart $\mathbb{C}$, except at the removed points.}. Hence, the reference metric is taken as an Euclidean two dimensional flat metric expressed in complex coordinates $(z,\overline{z})$. Following this reasoning we fix the reference metric $\widehat{g}$ as
\begin{eqnarray}\label{gALPHABETA}
\widehat{g}_{\mu\nu}=
\left( 
\begin{matrix}
0&&1/2\\
1/2&&0
\end{matrix}
\right) 
\end{eqnarray}
and the reference line element is given by
\begin{equation}
d\widehat{s}^{2}=dz\,d\overline{z}.
\end{equation}
Using these coordinates the Laplace-Beltrami operator is reduced to $\Delta_{\widehat{g}}=-4\partial\overline{\partial}$, and we have $\widehat{R}=0$. Replacing these results into eq.(\ref{EulerLagrange}), we obtain the second order differential equation 
\begin{equation}\label{LiouvilleEQ}
4\partial\,\overline{\partial}\varphi = \Lambda \mathit{e}^{\varphi}.
\end{equation}
By comparing eqs.(\ref{liouvilleequation}) and (\ref{LiouvilleEQ}), we recognize this last expression as the Liouville equation again. Its solutions can be written in terms of the solutions of the associated differential equation mentioned in the first section. Such an equation is rewritten as
\begin{equation}\label{DiffEQ}
4\partial^{2}_{z}U(z)+T(z)U(z)=0,
\end{equation}
where $T(z)$ is the holomorphic part of energy-momentum tensor associated to the Liouville action eq.(\ref{classicalLiouvilleaction}). If $U_{1}(z)$ and $U_{2}(z)$ are two linearly independent solutions of eq.(\ref{DiffEQ}), from the theory of differential equations their Wronskian, denoted by $\mathcal{W}_{1,2}$, is a constant
\begin{eqnarray}
\mathcal{W}_{1,2}=\left|
\begin{matrix}
U_{1}(z)&&U_{2}(z)\\
\partial_{z}U_{1}(z)&&\partial_{z}U_{2}(z)
\end{matrix}
\right| = \text{constant.}
\end{eqnarray}  
Moreover, by a suitable choice of the basis in the space of solutions we can choose this constant as unit. In addition, let us define $\mathbf{U}(z)$ and $\overline{\mathbf{U}}(\overline{z})$ as:
\begin{eqnarray}
\mathbf{U}(z)=\left( 
\begin{matrix}
U_{1}(z)\\
U_{2}(z)
\end{matrix}
\right)=\left( U_{1}(z),U_{2}(z)\right)^{T},\hspace{0.3cm}\text{and}\hspace{0.3cm}\overline{\mathbf{U}}(\overline{z})=\left(\overline{U}_{1}(\overline{z}),\overline{U}_{2}(\overline{z})\right)  
\end{eqnarray}
Using the Wrosnkians $\mathcal{W}_{1,2}=1$ and $\overline{\mathcal{W}}_{1,2}=1$ we can build a combination of $\mathbf{U}(z)$ and $\overline{\mathbf{U}}(\overline{z})$ which satisfies the Liouville differential eq.(\ref{LiouvilleEQ}). Such a combination also uses an Hermitian matrix $\mathbf{\Lambda}$ which has determinant $-\Lambda$. We have  
\begin{equation}\label{EquationAH}
\varphi(z,\overline{z}) = -2\log \left(\overline{\mathbf{U}}(\overline{z}) \mathbf{\Lambda} \mathbf{U}(z)\right)+\log 8 
\end{equation}
From the latter equation, we realize that it solves the Liouville equation (\ref{LiouvilleEQ}). Because if we replace eq.(\ref{EquationAH}) into eq.(\ref{LiouvilleEQ}), we obtain
\begin{eqnarray}
4\partial\overline{\partial}\varphi &=& 4\partial\overline{\partial}(-2\log \left(\overline{\mathbf{U}}(\overline{z}) \mathbf{\Lambda} \mathbf{U}(z)\right)+\log 8)\\
&=& -8 \partial \frac{\overline{\partial}(\overline{U}\mathbf{\Lambda}U)}{\overline{U}\mathbf{\Lambda}U}\\
&=& -8 \frac{(\overline{\partial}(\overline{U}\mathbf{\Lambda})\partial U)(\overline{U}\mathbf{\Lambda}U)-(\overline{\partial}(\overline{U}\mathbf{\Lambda})U)(\overline{U}\partial U)}{(\overline{U}\mathbf{\Lambda}U)^{2}}.
\end{eqnarray}
In the latter expression, the numerator is reduced to the product of the Wrosnkian $\mathcal{W}_{1,2}$ and the relation
\begin{eqnarray}
\overline{U}_{1}\overline{\partial}\,\overline{U}_{2}-\overline{U}_{2}\overline{\partial}\,\overline{U}_{1}=
\det(\mathbf{\Lambda})=-\Lambda.
\end{eqnarray}
Combining these results we obtain 
\begin{eqnarray}\label{feb7}
4\partial\overline{\partial}\varphi &=& 8\frac{\Lambda}{(\overline{U}\mathbf{\Lambda}U)^{2}}\\
&=& \Lambda e^{\varphi}.
\end{eqnarray}
The matrix $\mathbf{\Lambda}$ could assume a convenient form depending on if we are working with a positive or negative curvature. For instance, the most useful forms are \cite{zamo}
\begin{eqnarray}
\text{Positive curvature}\,(\Lambda < 0)&:& \mathbf{\Lambda}=\sqrt{-\Lambda}
\left( 
\begin{matrix}
1 & & 0\\
0 & & 1
\end{matrix}
\right)\label{PositiveCurvature}\\
\text{Negative curvature}\,(\Lambda > 0)&:& \mathbf{\Lambda}=\,\,\sqrt{\Lambda}\,\,
\left( 
\begin{matrix}
1 & & 0\\
0 & & -1
\end{matrix}
\right)\label{NegativeCurvature} 
\end{eqnarray}
Some elementary solutions of Liouville equation with positive curvature eq.(\ref{PositiveCurvature}) are described below.

\subsubsection{Sphere}

This is the simplest classical solution. In this case the functions $U_{1}(z)$ and $U_{2}(z)$ are
\begin{eqnarray}\label{Solutions}
U_{1}(z)&=&1\nonumber\\
U_{2}(z)&=&z
\end{eqnarray}
From eq.(\ref{EquationAH}), we have
\begin{eqnarray}
\varphi(z, \overline{z})=-2\log(1+z\overline{z})+\log\left(\frac{-8}{\Lambda}\right) 
\end{eqnarray}
where $\Lambda <0$, for positive curvature. This solution has a corresponding metric given by
\begin{eqnarray}\label{24}
e^{\varphi(z, \overline{z})}dz\,d\overline{z}=\frac{-8}{\Lambda}\frac{dz\,d\overline{z}}{(1+z\overline{z})^{2}}.
\end{eqnarray}
The latter metric describes a sphere of constant area $-8\pi/\Lambda$.

\subsubsection{Pseudosphere}

This solution is also called Poincaré disk. It is the elementary solution for negative curvature eq.(\ref{NegativeCurvature}) ($\Lambda >0$). In this case we also use $U_{1}(z)$ and $U_{2}(z)$ given by  eq.(\ref{Solutions}). However, instead of eq.(\ref{PositiveCurvature}) we will use eq.(\ref{NegativeCurvature}), and we obtain
\begin{eqnarray}
\varphi(z, \overline{z})=-2\log(1-z\overline{z})+\log\left(\frac{8}{\Lambda}\right). 
\end{eqnarray}
In contrast to eq.(\ref{24}), the metric for the pseudosphere is given by
\begin{eqnarray}
e^{\varphi(z, \overline{z})}dz\,d\overline{z}=\frac{8}{\Lambda}\frac{dz\,d\overline{z}}{(1-z\overline{z})^{2}}.
\end{eqnarray}
So far, we revised some basic features about classical Liouville theory. Starting from the next section, we will see how quantum Liouville theory possesses an anomalous term called the {\it central charge} and later on derive the Liouville action from string theory.

\subsubsection{Liouville theory and the Virasoro algebra}

The energy-momentum tensor associated to eq.(\ref{classicalLiouvilleaction}) can be split into three parts as 
\begin{eqnarray}
T_{z\,z} &=& - \frac{\partial^{2}\varphi}{\partial z^{2}}+ \frac{1}{2}(\frac{\partial\varphi}{\partial z})^{2}
\label{Stressenergy1},\\
T_{\overline{z}\,\overline{z}} &=& - \frac{\partial^{2}\varphi}{\partial \overline{z}^{2}}+ \frac{1}{2}(\frac{\partial\varphi}{\partial \overline{z}})^{2},
\\
T_{\overline{z}\,z} &=& T_{z \,\overline{z}} = 0.\label{OHH}
\end{eqnarray}
In order to get eq.(\ref{OHH}) we have used the equation of motion explicitly.

On the other hand, to maintain our conformal transformation $e^{\varphi(z)dz^{2}}=e^{\varphi(w)}dw^{2}$, we have to impose an unusual transformation law for the Liouville field $\varphi$, under an arbitrary analytic transformation $z \longrightarrow f(z)$,
\begin{equation}
\varphi(z) \longrightarrow \varphi(f) = \varphi(z) - 2\ln\left|\frac{\partial f}{\partial z}\right|.
\end{equation}  
As pointed out by Ginsparg and Moore \cite{ginsparg}, this linear shift in $\varphi$ under a conformal transformation allows its interpretation as a Goldstone boson for broken Weyl invariance.

Using eq.(\ref{Stressenergy1}) and  $\partial_{z}\varphi(z) \longrightarrow \partial_{z}\varphi(z) = \frac{d f}{d z}\partial_{f} \varphi(f) - \frac{d}{d z} \frac{1}{\gamma}\ln\left|\frac{\partial f}{\partial z}\right|$, we get that the energy-momentum tensor transforms as
\begin{equation}\label{anomalous}
T_{zz} \longrightarrow \left( \frac{df}{dz} \right)^{2}T_{ff} + \frac{c}{12}\mathcal{S}[f;z], 
\end{equation}
where $\mathcal{S}[f;z]$ is a new notation for the already known Schwarzian derivative, and $c$ is the central charge. At the beginning of this chapter we just gave an expression for this derivative, but here we can derive that expression for the Schwarzian derivative 
\begin{eqnarray}
\mathcal{S}[f;z]&=&-\frac{1}{2}\left(\partial_{z}\ln\left|\frac{\partial f}{\partial z}\right| \right)^{2} + \frac{d^{2}}{dz^{2}}\ln\left|\frac{\partial f}{\partial z}\right|\\
&=& \frac{f'''}{f'}-\frac{3}{2}\left(\frac{f''}{f'}\right)^{2}. 
\end{eqnarray}
The central charge $c$ arises because the anomalous transformation for the energy-momentum tensor eq.(\ref{anomalous}). Therefore, using the decomposition of the energy-momentum tensor in term of generators $\widehat{L}_{n}$,
\begin{equation}
\widehat{L}_{n} = \oint_{\mathcal{C}_{z_{0}}}\frac{dz}{2\pi i}(z-z_{0})^{n+1}T_{zz}
\end{equation}
and its anti-holomorphic counterpart, we find that they satisfy the Virasoro algebra,
\begin{eqnarray}
\left[\widehat{L}_{n},\widehat{L}_{m}\right] =(n-m)\widehat{L}_{n+m}+\frac{c}{12}n(n^{2}-1)\delta_{n+m,0},\\
\left[\widehat{\overline{L}}_{n},\widehat{\overline{L}}_{m}\right] =(n-m)\widehat{\overline{L}}_{n+m}+\frac{c}{12}n(n^{2}-1)\delta_{n+m,0}.
\end{eqnarray}
We already saw this algebra at the end of the Chapter \ref{Chapter2}. The anomalous term called central charge which disappears in the global conformal group (for $n=-1,0,1$) arose from the anomalous tranformation of the energy-momentum tensor. As also pointed out in Chapter \ref{Chapter2}, the central charge $c$ of a conformal theory depends on the particular lagrangian used. As we progress, we will discover what its value in Liouville theory. 

In the next section, we will investigate how Liouville theory appears in the context of string theory from the Polyakov action. 

\section{Non-critical string theory: From Polyakov to Liouville action}

The well known principle of general covariance allows us to establish that any quantum field theory may be coupled to gravity. The result is an action of the form $S[g,X]$. The fields $X^{\mu}$ are called ``matter fields'', and as usual $g_{\alpha\beta}$ denotes the world-sheet metric.

At classical level, $S[g,X]$ coupled to gravity in two dimensions is always a conformal field theory, because the Liouville equation of motion implies that $T_{\mu}^{\mu}=0$. In complex coordinates this condition becomes $T_{z\overline{z}}=0$. However, at quantum level we study correlation functions computed by using the path integral formalism. Thus, we are interested in expressions of the kind
\begin{eqnarray}
\langle\mathcal{O}_{1}\dots\mathcal{O}_{n}\rangle = \frac{1}{Z}\int\frac{\mathcal{D}g\,\mathcal{D}X}{V_{\text{Diff}}}e^{-k\int R-\mu A-S[g,X]}\mathcal{O}_{1}\dots\mathcal{O}_{n}.
\end{eqnarray}
In this expression we have labelled by $\mathcal{O}_{i}$ the covariant operators and $V_{\text{Diff}}$ represents the volume of the orbit of the diffeomorphism  group. By using the usual conformal gauge, defined as $g_{\alpha\beta}=e^{\phi}\delta_{\alpha\beta}$, Polyakov \cite{bosonicstring} showed that the initial matter-gravity theory could be expressed as a tensor product of the Liouville theory and the matter theory $S_{\text{Matter}}$. Note that in order to use the conformal gauge, it is assumed that the original matter theory is also conformally invariant. Otherwise the resulting theory could not be a simple tensor product \cite{ginsparg}.

Recalling that the Polyakov action is
\begin{equation}
S_{B}=\frac{1}{4\pi}\int d^{2}\xi\sqrt{g}g^{\alpha\beta}\partial_{\alpha}X^{\mu}\partial_{\beta}X_{\mu},
\end{equation}
where we fix the Regge slope $\alpha'$ to be one. Additionally, we adopt the convention that the first Greek letters $\alpha, \beta, \dots$ denote world-sheet coordinates and the middle ones $\mu, \nu,\dots $ space-time coordinates. Therefore, we will consider the Euclidean partition function \footnote{We arrive to this partition function after a Wick rotation.}
\begin{eqnarray}
Z=\int\frac{\mathcal{D}g\,\mathcal{D}X}{V_{\text{Diff}}}e^{-S_{B}[g,X]-\frac{\mu_{0}}{4\pi}\int d^{2}\xi\sqrt{g}}.
\end{eqnarray}
Before we continue with the treatment of the latter partition function, it is important to point out some things about the measures $\mathcal{D}X$ and $\mathcal{D}g$. First, the measure $\mathcal{D}X$ of matter fields is determined by
\begin{equation}
\int \mathcal{D}g\,\delta X\,e^{-||\delta X||^{2}_{g}}=1.
\end{equation}
In this equation we are denoting by $||\delta X||_{g}$ to the norm of $X$ fields. This norm is defined as $||\delta X||^{2}_{g}=\int d^{2}\xi \sqrt{g}(\delta X)^{2}$. 

Such a measure remains invariant under the action of diffeomorphic transformations. However, the same measure is not more invariant under conformal transformations because the exponential conformal factor spoils the linear behaviour of $\mathcal{D}_{g}X$. Fortunately, as pointed out by Y. Nakayama \cite{nakayama}, the non-trivial measure transformation under conformal mappings is obtained by using some algebraic techniques developed by T. Fujikawa et. al \cite{Fujikawa}. Following their results, we use such a measure transformation
\begin{equation}\label{DX1}
\mathcal{D}_{g}X=e^{\frac{c_{M}}{48\pi}\widetilde{S}_{L}[\phi]}\mathcal{D}_{\widehat{g}}X
\end{equation}   
where $\widetilde{S}_{L}$ is the Liouville action \footnote{Note that we change $\sigma \rightarrow \phi$.} and $c_{M}$ is the central charge of the matter sector.
On the other hand, the metric measure $\mathcal{D}g$ is defined by:
\begin{eqnarray}
1&=&\int \mathcal{D}_{g}\delta g \,e^{-\frac{1}{2}||\delta g||_{g}^{2}}\label{Dgmeasure}\\
||\delta g||_{g}^{2}&=&\int d^{2}\xi \sqrt{g}(g^{\alpha\gamma}g^{\beta\lambda}+\mathcal{C} g^{\alpha\beta}g^{\gamma\lambda})\delta g_{\alpha\beta}\delta g_{\gamma\lambda}\label{Dgnorm}
\end{eqnarray}
A variation on the metric, say $\delta g(x)$, has two contributions. The first is due to diffeomorphisms, and the second to Weyl invariance. Under an infinitesimal changes of coordinates $x^{\alpha}\rightarrow x^{\alpha}+\varepsilon^{\alpha}$, reparametrization invariance gives rise to the metric variation $\delta g_{\alpha\beta}(x)=\nabla_{\alpha}\varepsilon_{\beta}(x)+\nabla_{\beta}\varepsilon_{\alpha}(x)$. In addition, local Weyl invariance contributes with a variation of the metric tensor which is proportional to the local metric $g_{\alpha\beta}$ and the variation of Liouville mode, this is $\delta g_{\alpha\beta}=g_{\alpha\beta} \delta \phi$. If both symmetries are present, the total metric variation is
\begin{equation}\label{TotalGG}
\delta g_{\alpha\beta}=g_{\alpha\beta} \delta \phi+\nabla_{\alpha}\varepsilon_{\beta}(x)+\nabla_{\beta}\varepsilon_{\alpha}(x).
\end{equation}
We need to point out that in the definition of $\mathcal{D}g$, it was divided by the volume of the space of diffeomorphisms $V_{\text{Diff}}$. Because the infinitesimal change of coordinates can be contemplated as a gauge transformation. Sometimes it is useful to contemplate this space as the space of the vector fields $\varepsilon^{\alpha}(x)$ \cite{zamo}. 

It is also important to point out that the value $\mathcal{C}$ in eq.(\ref{Dgnorm}) does not affect our results \cite{1} \cite{Tong} \cite{Polchinski1} \cite{zamo} \cite{hatfield}. 

Eq.(\ref{TotalGG}) can be rewritten by adding and subtracting the term $g_{\alpha\beta}\nabla\cdot\varepsilon$, thus we have
\begin{equation}
\delta g_{\alpha\beta}=g_{\alpha\beta}(\delta \phi + \nabla\cdot\varepsilon) + (\nabla_{\alpha}\varepsilon_{\beta}(x)+\nabla_{\beta}\varepsilon_{\alpha}(x)-g_{\alpha\beta}\nabla\cdot\varepsilon).
\end{equation} 
Here, we can redefine the Liouville mode to be $\delta \phi'=\delta \phi+\nabla\cdot\varepsilon$, thus the first term is only related to conformal transformations. On the other hand, if we call by $\mathcal{E}_{\alpha\beta}$ to  $\nabla_{\alpha}\varepsilon_{\beta}(x)+\nabla_{\beta}\varepsilon_{\alpha}(x)-g_{\alpha\beta}\nabla\cdot\varepsilon$, we realize that $g^{\alpha\beta}\mathcal{E}_{\alpha\beta}=0$. Hence, it is describing a space orthogonal to the metric. This fact allows us to rewrite eq.(\ref{Dgnorm}) as
\begin{equation}\label{NORMGFII}
||\delta g||_{g}^{2} = \int d^{2}x \sqrt{g}\left[\mathcal{P}(\delta \phi(x))^{2} + \mathcal{E}_{\alpha\beta}\mathcal{E}^{\alpha\beta} \right], 
\end{equation}
where $\mathcal{P}=2(1+2\mathcal{C})$. Because of the Gaussian form of $\mathcal{E}_{\alpha\beta}\mathcal{E}^{\alpha\beta}$ in the metric norm, it gives rise to a functional determinant, as is the case for QFT. The only remaining dependence is on the Liouville mode $\phi$. Therefore, it defines a measure $\mathcal{D}\phi$. In this way, we obtain the next contribution from both symmetries,
\begin{equation}
\det\left[ \mathcal{E}\right] \mathcal{D}\phi.
\end{equation}
It is customary in QFT to write the functional determinant as a Gaussian functional integral over fermionic variables. We follow this procedure and as usual call this Gaussian integral as the ghost partition function. It is given by
\begin{equation}
\det\left[ \mathcal{E}\right]  = Z_{\text{ghost}}=\int \mathcal{D}b\mathcal{D}c\,\, e^{-S_{\text{ghost}}[b,c]}
\end{equation}  
The ghost action is 
\begin{equation}
S_{\text{ghost}}[b,c]=\frac{1}{2\pi}\int d^{2}x\,\sqrt{g}\,b_{\alpha\beta}\nabla^{\alpha}c^{\beta}
\end{equation}
where $b_{\alpha\beta}$ is symmetric and traceless.

As we mentioned in the first part of the present chapter, we deal with a theory defined on a sphere. For others topologies we would need to deal with a measure of the moduli space $\int_{\text{moduli}} d\mu(g)$. Hence, in the present case we do not deal with moduli.

In the same way as the matter fields, there is a transformation law of $\mathcal{D}_{g}b\mathcal{D}_{g}c$ under the conformal transformations $g\rightarrow e^{\phi}\widehat{g}$. Such a transformation is given by
\begin{equation}\label{DbDc1}
\mathcal{D}_{g}b\mathcal{D}_{g}c = e^{-\frac{26}{48\pi}\widetilde{S}_{L}[\phi]}\mathcal{D}_{\widehat{g}}b\mathcal{D}_{\widehat{g}}c.
\end{equation}
From now on we write $\mathcal{D}(gh)=\mathcal{D}_{g}b\mathcal{D}_{g}c $ for the ghost measure. 

A non-trivial task is about the measure $\mathcal{D}\phi$. From eq.(\ref{NORMGFII}), and after integrating the Gaussian part, we can define the norm for the Liouville mode,
\begin{equation}
||\delta \phi||^{2}_{g}=\mathcal{P}\int d^{2}x \,\sqrt{g}\, (\delta \phi)^{2}.
\end{equation}   

Due to Weyl invariance in the total partition function, we observe that this norm gives rise to a non-linear measure. Explicitly, we have, under a conformal transformation $g\rightarrow e^{\phi}\widehat{g}$,
\begin{equation}\label{NORM2}
||\delta \phi||^{2}_{g}=\mathcal{P}\int d^{2}x \,\sqrt{\widehat{g}}\,e^{\phi} (\delta \phi)^{2}.
\end{equation}   

Following the Distler-Kawai hypothesis \cite{DistlerKawai}, we change this non-linear measure to a linear one by adding some additional terms which are proportional to those already present in the action $\widetilde{S}_{L}$. We will use these additional terms to build our renormalized Liouville action in the next section. Hence, this hypothesis allows us to change eq.(\ref{NORM2}) to a linear norm,
\begin{equation}
||\delta \phi||^{2}_{g}=\mathcal{P}\int d^{2}x \,\sqrt{\widehat{g}}\,(\delta \phi)^{2},
\end{equation}
which is obviously invariant under $\phi(x)\rightarrow \phi(x) + f(x)$, where $f(x)$ is an arbitrary but fixed function.

Now the total partition function has contributions coming from matter fields $X$, ghosts $b_{\alpha\beta},c_{\alpha}$ and the Liouville mode $\phi$. Therefore, such a partition function is described as 

\begin{equation}
Z=\int \mathcal{D}_{g}X\mathcal{D}_{g}(gh)\mathcal{D}_{g}\phi\,e^{-S_{B}[X,g]-S_{ghost}[b,c,g]}.
\end{equation} 

Under the conformal transformation $g=e^{\phi}\widehat{g}$, and using eqs.(\ref{DX1}), (\ref{DbDc1}) we get the next transformation for the total measure:

\begin{eqnarray}
\mathcal{D}_{g}X\mathcal{D}_{g}(gh)\mathcal{D}_{g}\phi = e^{\frac{c_{M}-26}{48\pi}\int d^{2}x \sqrt{\widehat{g}}\left[ \frac{1}{2}\widehat{g}^{\alpha\beta}\partial_{\alpha}\phi\partial_{\beta}\phi +\widehat{R}\phi\right] } \mathcal{D}_{\widehat{g}}X\mathcal{D}_{\widehat{g}}(gh)\mathcal{D}_{\widehat{g}}\phi
\end{eqnarray}

Thinking of a future renormalized theory, we introduce two local terms with dimensional coupling constants $\Lambda$ and $\Gamma$. They make up the action $S_{\text{grav}[g]}$ which depends on the metric $g$ only. In fact, this action could have many other local terms, which could be powers of the scalar of curvature $R$. However, these terms have some problems in defining the theory \cite{zamo}. Therefore, we only take into account the first two terms
\begin{equation}
S_{\text{grav}}[g]=\Lambda \int d^{2}x\,\sqrt{g}+\Gamma\int d^{2}x\,\sqrt{g}R(x)
\end{equation}
where the first term is identified as the cosmological term. By appealing to the Gauss-Bonnet theorem we find that the second term is a topological invariant term, thus for the sphere this term does not contribute. 

After this analysis the final form of the total partition function is
\begin{equation}\label{TotalZ}
Z=\int \mathcal{D}_{\widehat{g}}X\mathcal{D}_{\widehat{g}}(gh)\,e^{-S_{B}[X,\widehat{g}]-S_{ghost}[b,c,\widehat{g}]}\underbrace{\int \mathcal{D}_{\widehat{g}}\phi\,e^{\frac{c_{M}-26}{48\pi}\int d^{2}x \sqrt{\widehat{g}}\left[ \frac{1}{2}\widehat{g}^{\alpha\beta}\partial_{\alpha}\phi\partial_{\beta}\phi +\widehat{R}\phi + \Lambda e^{\phi}\right]}}_{\text{Liouville Partition function}}
\end{equation} 
This is the way Liouville theory arises in the context of non-critical string theory $c_{M}\neq 26$. Liouville partition function is given by the last term in eq.(\ref{TotalZ}).

\section{Quantum Liouville Theory}

In this section we analyse the quantum aspects of Liouville theory. Quantum Liouville Theory was extensively studied because its connection to non-critical String Theories \cite{1} \cite{bosonicstring} \cite{Seiberg} \cite{witten2012} \cite{zamo} \cite{russo}. As we stated in the previous section, Quantum Liouville Theory is also viewed as the quantum theory of the world-sheet. Here we follow the notation given in \cite{witten2012} \cite{zamo2}.

To start, let us recall our Liouville action obtained from eq.(\ref{TotalZ}),
\begin{equation}\label{QNRLiouville}
S_{L}=\frac{26-c_{M}}{48\pi}\int d^{2}x \sqrt{\widehat{g}}\left[ \frac{1}{2}\widehat{g}^{\alpha\beta}\partial_{\alpha}\phi\partial_{\beta}\phi +\widehat{R}\phi + \Lambda e^{\phi}\right].
\end{equation}
As already mentioned, the Distler-Kawai hypothesis states that we can get a linear measure $\mathcal{D}\phi$ by adding additional terms of the type presented in eq.(\ref{QNRLiouville}). Thus, the renormalized Liouville action is given by
\begin{equation}\label{RNLiouvilleAction}
S_{L}[\phi]=\frac{1}{8\pi b^{2}}\int d^{2}x\,\left[\frac{1}{2}\widehat{g}^{\alpha\beta}
\partial_{\alpha}\phi\partial_{\beta}\phi+q\widehat{R}\phi+\lambda\,e^{\phi}\right]. 
\end{equation}
We change our notation to the one commonly used in general literature. First, we rewrite the field $\phi$ as $2b\phi$ and define the constants $Q=q/b$ and $\mu=\lambda/(8\pi b^{2})$. With these definitions the Liouville action becomes
\begin{equation}
S_{L}=\frac{1}{4\pi}\int d^{2}x \sqrt{\widehat{g}} \left[ \widehat{g}^{\alpha\beta} \partial_{\alpha} \phi \partial_{\beta} \phi+Q \widehat{R} \phi + 4\pi \mu e^{2b\phi}\right]. 
\label{liouvilleaction}
\end{equation}
Here, the conformal factor relating the reference metric and the physical metric is  $e^{2b\phi}$. Therefore, the usual expression relating to conformal metrics is given by
\begin{equation}
g_{\alpha\beta}= e^{2b\phi} \widehat{g}_{\alpha\beta}. 
\label{Weylsymmetry}
\end{equation} 
The reference metric  $\widehat{g}$ is just an auxiliary one. In other words, it can be chosen at will. As pointed out by A. and Al. Zamolodchikov \cite{zamo}, the physical results should not depend on what choice we make. Such a statement is called {\it background} independence.

Using the background independence we derive a relation between the parameters $Q$ and $b$ defined above. To get that relation we start redefining the reference metric $\widehat{g}$. This is made by using a conformal transformation. In this case our conformal factor has the same form as in eq.(\ref{Weylsymmetry}), but the function $\sigma=2b\phi$ changes. We label the new exponent of the conformal factor as $\widetilde{\sigma}$. Thus, we define
\begin{equation}
\widehat{g}_{\alpha \beta}= e^{\widetilde{\sigma}} \widetilde{g}_{\alpha\beta}.
\label{arbitrarymetric}
\end{equation} 
It is useful to analyse this problem in two parts. First, we treat the case $\mu=0$. Additionally, we take into account the well known relation between the scalars of curvature $\sqrt{\widehat{g}}\,\widehat{R}=\sqrt{\widetilde{g}}\,(\widetilde{R}-\Delta_{[\widetilde{g}]}\widetilde{\sigma})$. By replacing this curvature relation and eq.(\ref{arbitrarymetric}) into eq.(\ref{liouvilleaction}), we obtain
\begin{equation}\label{2feb}
S_{L}[e^{\widetilde{\sigma}} \widetilde{g},\widehat{\phi}]|_{\mu=0}=\frac{1}{4\pi}\int
d^{2}\xi\sqrt{\widetilde{g}}\left[ \widetilde{g}^{\alpha\beta}\partial_{\alpha}\widehat{\phi}\partial_{\beta}\widehat{\phi}+Q\widetilde{R}\widehat{\phi}-\frac{Q^{2}}{2}\left( \widetilde{R}\widetilde{\sigma}+\frac{1}{2}\widetilde{g}^{\alpha\beta}\partial_{\alpha}\widetilde{\sigma}\partial_{\beta}\widetilde{\sigma}\right)\right]  
\end{equation}
where $\widehat{\phi}=\phi+(Q/2)\,\widehat{\sigma}$. Due to the Distler-Kawai hypothesis, the measure associated to $\phi$ is invariant under this shift, $\mathcal{D}\phi=\mathcal{D}\widehat{\phi}$. In the derivation of eq.(\ref{TotalZ}) we saw that under a conformal transformation in the integrand of the partition function there appeared an exponential term whose power is proportional to Liouville action. We apply the same argument here, but now with the conformal transformation $\widehat{g}_{\alpha\beta}=e^{\widetilde{\sigma}}\widetilde{g}_{\alpha\beta}$. Thus, under conformal transformations this measure transforms as
\begin{equation}\label{3feb}
\mathcal{D}_{\widehat{g}}\phi=\,e^{\frac{1}{48\pi}\int d^{2}x\sqrt{\widetilde{g}}\left[ \frac{1}{2}\widetilde{g}^{\alpha\beta}\partial_{\alpha}\widetilde{\sigma}\partial_{\beta}\widetilde{\sigma}+
\widetilde{R}\widetilde{\sigma}\right] }\,\mathcal{D}_{\widetilde{g}}\phi.
\end{equation}
Combining eqs.(\ref{TotalZ}), (\ref{2feb}) and (\ref{3feb}) we get, 
\begin{equation}
\exp\left\lbrace \frac{c_{M}-26+1+6Q^{2}}{48\pi}\int d^{2}x\sqrt{\widetilde{g}}\left[ \frac{1}{2}\widetilde{g}^{\alpha\beta}\partial_{\alpha}\widetilde{\sigma}\partial_{\beta}\widetilde{\sigma}+\widetilde{R}\widetilde{\sigma}
\right] \right\rbrace . 
\end{equation}
The requirement of background independence is satisfied if $c_{M}-26+1+6Q^{2}=0$. This leads us to the relation
\begin{equation}\label{feb4}
Q^{2}=\frac{25-c_{M}}{6}.
\end{equation} 
Now, we return to  $\mu\neq 0$. In this case, we need to take into account the term $\mu\int \sqrt{\widehat{g}}e^{2b\phi}d^{2}x$. Under a conformal transformation $\widehat{g}(x)\rightarrow \widetilde{g}(x)$ there are three different sources which contribute to this term. The first comes from 
\begin{equation}\label{EQ1} 
\sqrt{\widehat{g}}=e^{\widetilde{\sigma}}\sqrt{\widetilde{g}}.
\end{equation}
Our second source is due to the shift $\widehat{\phi}\rightarrow \phi+(Q/2)\widetilde{\sigma}$. We have:
\begin{equation}\label{EQ2}
e^{2b\phi}=e^{-bQ\widetilde{\sigma}}e^{2b\widehat{\phi}}
\end{equation}
Finally, due to the Distler-Kawai hypothesis, we know that all terms have to be renormalized, as in eq.(\ref{RNLiouvilleAction}), by adding a proportional term to the ones already present there. In this way, the exponential part of the action has an additional dependence on the background metric
\begin{equation}\label{EQ3}
\left[e^{2b\phi} \right]_{\widehat{g}} =e^{b^{2}\widetilde{\sigma}}\left[e^{2b\phi} \right]_{\widetilde{g}}. 
\end{equation}
From eqs.(\ref{EQ1}), (\ref{EQ2}) and (\ref{EQ3}), we obtain for the exponential term
\begin{equation}
\mu\int d^{2}x\, e^{(1-bQ+b^{2})\widetilde{\sigma}}e^{2b\phi}.
\end{equation}
In order to maintain this term invariant, we obtain the famous relation between $Q$ and $b$, 
\begin{equation}
Q= b+\frac{1}{b}. 
\label{Qb}
\end{equation} 
We may also say that $Q$ and $b$ are related in this way to ensure background independence. 

\subsection{Liouville energy-momentum tensor}

The energy momentum tensor in Liouville theory is defined  as usual by
\begin{equation}\label{DEFT}
T_{\mu\nu}=-\frac{4\pi}{\sqrt{g}}\frac{\delta S_{L}}{\delta g^{\mu\nu}}.
\end{equation}
In order to find the expression for the  energy-momentum tensor in Liouville theory, we separate this action in three parts and use eq.(\ref{DEFT}) to calculate this tensor.

From eq.(\ref{liouvilleaction}), we have 
\begin{equation}\label{Stotal}
S_{L}= S_{1}+S_{2}+S_{3}
\end{equation}
where
\begin{eqnarray}
S_{1}&=&\frac{1}{4\pi}\int d^{2}x \sqrt{g}\,g^{\alpha\beta} \partial_{\alpha} \phi \partial_{\beta} \phi \label{S1}\\
S_{2}&=&\frac{Q}{4\pi}\int d^{2}x \sqrt{g} R \phi\label{S2}\\
S_{3}&=&\mu\int d^{2}x \sqrt{g} e^{2b\phi}\label{S3}.
\end{eqnarray}
We will consider each contribution to the energy-momentum tensor separately.

\subsubsection{The free Field $S_{1}$}

It is easy to see that this is a free field action for $\phi$. We use the fact that the path integral of a functional derivative vanishes, leading to the standard result:
\begin{equation}
\langle\phi(x)\phi(x')\rangle=-\frac{1}{2}\log|x-x'|^{2}.
\end{equation} 
It is possible also to express this relation in complex coordinates, 
\begin{equation}\label{FreeLiouvillePropagator}
\langle\phi(z)\phi(w)\rangle=-\frac{1}{2}\log|z-w|.
\end{equation}
where $z=x_{1}+ix_{2}$, $\overline{z}=\overline{x}_{1}+i\overline{x}_{2}$ and  $w=x'_{1}+ix'_{2}$, $\overline{w}=\overline{x}'_{1}+i\overline{x}'_{2}$. We will use this result later.

We recall the well known variation of the metric tensor,
\begin{equation}\label{DELTAg}
\delta g=g \,g^{\alpha\beta}\delta g_{\alpha\beta},
\end{equation}
or, by using $g^{\alpha\gamma}g_{\gamma\beta}=\delta^{\alpha}_{\beta}$, we obtain $\delta g=-g \,g_{\alpha\beta}\delta g^{\alpha\beta}$.

Combining eqs.(\ref{DEFT}), (\ref{S1}) and (\ref{DELTAg}), it is easy to obtain
\begin{equation}\label{FREEfieldT}
T_{\alpha\beta}=\frac{1}{2}\delta_{\alpha\beta}(\partial\phi)^{2}-\partial_{\alpha}\phi\partial_{\beta}\phi.
\end{equation} 

\subsubsection{The curvature term $S_{2}$}

In a naive derivation of Liouville energy-momentum tensor, we could chose the flat conformal gauge first, and later to compute $T_{\alpha\beta}$. However, if we proceed in this way, we will lose a contribution from the curvature term $\phi\delta R$. To see this, we need some identities from Differential Geometry. 

The Christoffel symbols and the Riemann curvature tensor, using the Levi-Civita connexion, are given by
\begin{eqnarray}
\Gamma^{\mu}_{\nu\lambda}&=&\frac{1}{2}g^{\mu\gamma}\left(\partial_{\nu}g_{\gamma\lambda}+\partial_{\lambda}g_{\gamma\nu}-\partial_{\nu\lambda} \right),\\
R^{\alpha}_{\mu\beta\nu}&=&\partial_{\beta}\Gamma^{\alpha}_{\mu\nu}-\partial_{\mu}\Gamma^{\alpha}_{\beta
\nu}+\Gamma^{\alpha}_{\tau\beta}\Gamma^{\tau}_{\mu\nu}-\Gamma^{\alpha}_{\tau\nu}\Gamma^{\tau}_{\mu
\beta}.\label{Riemannexpr}
\end{eqnarray}   
In addition, the Ricci tensor is defined as $R^{\alpha}_{\mu\alpha\nu}$. Other useful identities involving $\Gamma^{\alpha}_{\mu\nu}$ are
\begin{eqnarray}
\Gamma^{\alpha}_{\alpha\lambda}&=&\partial_{\lambda}(\log\sqrt{g})\label{GAMMA1},\\
g^{\alpha\beta}\Gamma^{\mu}_{\alpha\beta}&=&-\frac{1}{\sqrt{g}}\partial_{\nu}(g^{\mu\nu}\sqrt{g}).\label{GAMMA2}
\end{eqnarray}
The next step is to analyse the variation of $R$. Due to $R=g^{\alpha\beta}R_{\alpha\beta}$, we have $\delta R=\delta g^{\alpha\beta}\,R+g^{\alpha\beta}\delta R_{\alpha\beta}$. Nonetheless, because the first term of $\delta R$ involves derivative of $g$, they will vanish after taking the flat limit case. Hence, we only consider the second term, $\delta R=g^{\alpha\beta}\delta R_{\alpha\beta}$. Now, we recall that $R_{\alpha\beta}=R_{\alpha\mu\beta}^{\mu}$, thus we have from eq.(\ref{Riemannexpr}) for the Ricci tensor,
\begin{equation}\label{EQahhh}
g^{\alpha\beta}R_{\alpha\beta}=g^{\alpha\beta}\partial_{\mu}\Gamma^{\mu}_{\alpha\beta}-g^{\alpha\beta}\partial_{\alpha}\Gamma^{\mu}_{\mu\beta}+
\Gamma^{\mu}_{\lambda\mu}g^{\alpha\beta}\Gamma^{\lambda}_{\alpha\beta}-g^{\alpha\beta}\Gamma^{\mu}_{\lambda\beta}
\Gamma^{\lambda}_{\alpha\mu}.
\end{equation}
We need to keep in mind that at the end of this calculation we use the conformal flat gauge. We have to be aware that we are using a normal coordinate system. Because of this normal coordinate system, the last two terms in eq.(\ref{EQahhh}) disappear. Taking this reasoning into account and using eqs.(\ref{GAMMA1}), (\ref{GAMMA2}), the last equation is reduced to 
\begin{equation}
g^{\alpha\beta}\delta R_{\alpha\beta}=\partial_{\lambda}(g^{\alpha\beta}\delta \Gamma^{\lambda}_{\alpha\beta}-g^{\alpha\lambda}\Gamma^{\gamma}_{\beta\gamma}).
\end{equation}
Hence, we are able to compute this curvature contribution to the energy-momentum tensor,
\begin{eqnarray}
\delta S_{1}&=&\frac{Q}{4\pi}\int d^{2}x\sqrt{g}\,\phi\, \delta R\\
&=&\frac{Q}{4\pi}\int d^{2}x\sqrt{g}\,\phi\, \partial_{\lambda}(g^{\alpha\beta}\delta \Gamma^{\lambda}_{\alpha\beta}-g^{\alpha\lambda}\Gamma^{\gamma}_{\alpha\gamma})\\
&=&\frac{Q}{4\pi}\int d^{2}x\sqrt{g}\,(\partial_{\lambda}\phi)\,\delta\left(\frac{1}{\sqrt{g}}\partial_{\gamma}(g^{\lambda\gamma}\sqrt{g})+ g^{\lambda\alpha}\partial_{\alpha}(\ln\sqrt{g}) \right)\\
&=&\frac{Q}{4\pi}\int d^{2}x\sqrt{g}\,(\partial_{\lambda}\partial_{\gamma}\phi)\delta g^{\lambda\gamma}+\mathcal{O}(\partial g).
\end{eqnarray}
Finally, the flat conformal gauge causes $\mathcal{O}(\partial g)\rightarrow 0$, we get
\begin{equation}\label{CurvatureT}
T_{\alpha\beta}=Q\,\partial_{\alpha}\phi\,\partial_{\beta}\phi.
\end{equation} 

\subsubsection{The exponential term $S_{3}$}

In order for Liouville action to remain invariant under a conformal transformation, we found that the parameters $Q$ and $b$ are related by the expression $Q=b+1/b$, as in eq.(\ref{Qb}). This duality between $Q$ and $b$ ensures that the exponential term in $S_{3}$ should not depend on what choice of field $\phi$ we make. In particular, if we choose $\phi \ll 0$ the exponential contribution $e^{2b\phi}$ turns off. It means the exponential term $S_{3}$ does not contribute to the energy-momentum tensor. 

Finally, we are able to combine the contributions from $S_{1}, S_{2}$ and $S_{3}$ to build the energy-momentum tensor. From eqs.(\ref{FREEfieldT}) and (\ref{CurvatureT}), we have in a general background metric $g_{\alpha\beta}$
\begin{equation}\label{GeneralT}
T_{\alpha\beta}=\frac{1}{2}\delta_{\alpha\beta}(\partial\phi)^{2}-\partial_{\alpha}\phi\partial_{\beta}\phi+Q\,\partial_{\alpha}\phi\,\partial_{\beta}\phi.
\end{equation}
Considering the flat metric expressed in complex coordinates which is described in eq.(\ref{gALPHABETA}), we have for $T_{zz}, T_{\overline{z}\,\overline{z}}, T_{\overline{z}\,z}$ and $T_{z\,\overline{z}}$,
\begin{eqnarray}
T_{z\,z} &=& - \frac{\partial^{2}\phi}{\partial z^{2}}+ Q(\frac{\partial\phi}{\partial z})^{2}
,\label{tensor2z}\\
T_{\overline{z}\,\overline{z}} &=& - \frac{\partial^{2}\phi}{\partial \overline{z}^{2}}+ Q(\frac{\partial\phi}{\partial \overline{z}})^{2}
,\label{tensor3zbarra}\\
T_{\overline{z}\,z} &=& T_{z \,\overline{z}} = 0.
\end{eqnarray}

\section{The central charge $c_{L}$}

As we saw in Chapter \ref{Chapter2}, the central charge is one of the most important objects in CFT due to its direct relation with the Virasoro algebra. Here, we are interested in computing the Liouville central charge. This central charge is usually denoted by $c_{L}$.

We thus compute the OPE $T_{zz}T_{ww}$. In this order, we use the free field correlator given in eq.(\ref{FreeLiouvillePropagator}) and replace it into eq.(\ref{tensor2z}). Thus, we have
\begin{eqnarray}\label{TTLiouvilleOPE}
T_{zz}T_{ww}&=&\left[- \frac{\partial^{2}\phi(z)}{\partial z^{2}}+ Q(\frac{\partial\phi(z)}{\partial z})^{2} \right] \left[- \frac{\partial^{2}\phi(w)}{\partial w^{2}}+ Q(\frac{\partial\phi(w)}{\partial w})^{2} \right]\\
&=&\left( \frac{1}{2}+3Q^{2}\right)\frac{1}{(z-w)^{4}}+\dots , 
\end{eqnarray} 
where dots mean non-singular terms as usual. By comparing with eq.(\ref{TTOPE}), we find 
\begin{eqnarray}\label{feb5}
c_{L}=1+6Q^{2}.
\end{eqnarray}
Based on Polyakov partition function we get a total central charge placed in front of  Liouville's contribution equal to $c_{M}-26$. Hence, this anomalous contribution from the central charge could disappear if we take $c_{M}=26$. It gives rise to the so called critical String Theory. On the other hand, if $c_{M}\neq 26$ we need to be careful because of the contribution of the Liouville action in the Polyakov partition function. However, are we doomed to have a non-vanishing central charge? To see this better let us to take into account all central charge contributions that come from the matter, ghosts and Liouville sector, thus we observe that 
\begin{eqnarray}\label{feb6}
c_{\text{Tot}}=c_{M}-26+c_{L}.
\end{eqnarray}
By replacing eqs.(\ref{feb4}) and (\ref{feb5}) into (\ref{feb6}) we obtain that the total central charge vanishes. Therefore, we do not have this anomalous term anymore.

\section{Liouville primary fields  $e^{2\alpha_{i}\phi}$}

In addition to eq.(\ref{TTLiouvilleOPE}), we usually analyse another important OPE, the operator product between the energy momentum tensor and the primary fields of the concerned theory. 

Are the fields $\phi$ in Liouville theory primaries? To answer this question we need to remember that a primary field has a well defined transformation law under an arbitrary coordinate transformation, say $z\rightarrow w$. This transformation law was studied in Chapter \ref{Chapter2}. Nonetheless, in order to maintain conformal invariance in Liouville action we find that the field $\phi$ transforms as
\begin{eqnarray}\label{PHItransf}
\phi(w,\overline{w})=\phi(z,\overline{z})-\frac{Q}{2}\ln\left|\frac{\partial w}{\partial z}\right|^{2}.
\end{eqnarray}
It is easy to see that such a field does not have a suitable behaviour to be a primary field. However, due to the logarithmic term in the right hand side of eq.(\ref{PHItransf}), we consider an exponential expression of $\phi$ which could have a more suitable behaviour. In fact, we define Liouville primary fields also called vertex operator as $e^{2\alpha_{i}\phi(z,\overline{z})}$, where $i=1,2,\dots$. From eq.(\ref{PHItransf}) we can arrive at the transformation law
\begin{eqnarray}\label{LiouvillePO}
e^{2\alpha_{i}\phi(w,\overline{w})}=\left(\frac{\partial w}{\partial z} \right)^{-\alpha_{i}Q} \left(\frac{\partial\overline{w}}{\partial\overline{z}} \right)^{-\alpha_{i}Q}\,e^{2\alpha_{i}\phi(z,\overline{z})} 
\end{eqnarray}
and thus define our primary fields as such exponential functions. In addition, from eq.(\ref{LiouvillePO}) we can infer that the classical holomorphic and anti-holomorphic conformal dimensions are given by $\alpha Q$. From now on, instead $h$ and $\overline{h}$, we use the usual notation for conformal dimensions $\Delta$ and $\overline{\Delta}$ \cite{zamo} \cite{witten2012}. Hence, the last statement means $\Delta_{\text{class}}=\overline{\Delta}_{\text{class}}=\alpha\,Q$.

It is interesting to compare the classical conformal dimensions with its quantum counterpart. We analyse the OPE between $T_{zz}$ with $e^{2\alpha\phi(w)}$ (despite the sub-index $i$ from $\alpha_{i}$), in the same way as we did in Chapter \ref{Chapter2}. We have,
\begin{eqnarray}\label{ConformalLiouvilledimension}
T_{zz}e^{2\alpha\phi(w)}&=&\left[ -(\frac{\partial \phi}{\partial z})^{2}+Q\frac{\partial^{2}\phi }{\partial z^{2}} \right]\left[\sum_{k=0}^{\infty}\frac{(2\alpha)^{k}}{k!}\phi^{k}(w)\right]\nonumber\\
&=&-\left[-(2\alpha)^{2}\left(-\frac{1}{2(z-w)}\right)^{2}+\frac{Q\alpha}{(z-w)^{2}}   \right]\sum_{k=0}^{\infty}\frac{(2\alpha)^{k}}{k!}\phi^{k}(w)+\dots\nonumber\\
&=&\frac{\alpha(Q-\alpha)}{(z-w)^{2}}e^{2\alpha\phi(w)}+\dots
\end{eqnarray}
To arrive from the first to the second line, we have used eq.(\ref{FreeLiouvillePropagator}). Therefore, in contrast to $\Delta_{\text{class}}=\alpha\,Q$ analysing eq.(\ref{ConformalLiouvilledimension}) we get that the quantum conformal dimension is given by 
\begin{equation}\label{Liouvilledimension}
\Delta_{i}=\alpha_{i}(Q-\alpha_{i}).
\end{equation}
In this last expression we come back to label Liouville operators by $i=1, 2, \dots$.

On the other hand, if we consider the exponential term in Liouville action and apply the latter result on the conformal dimension, it is easy to see that we get $\Delta_{b}=b(Q-b)$. Now, considering that this term is placed into the Liouville action, to be integrated correctly with the surface term it must have a conformal dimension $\Delta_{b}=1$, because of the definition eq.(\ref{primarytransformation}). Therefore, we have $b(Q-b)=1$ and thus it is straightforward to obtain the well known relation:  
\begin{equation}
Q=b+\frac{1}{b}.
\end{equation}
       
\chapter{Liouville three point function} 

\label{Chapter4} 

\lhead{Chapter 4. \emph{Liouville three point function}} 

\section{Liouville correlation functions}

In Chaper \ref{Chapter3} we already pointed out that the structure of Liouville primary operators are given by exponential functions. Classically, these operators have the following transformation law
\begin{equation}
e^{2\alpha \phi^{'}_{(z^{'},\overline{z}^{'})}}= \left(\frac{\partial w}{\partial z}\right)^{-\alpha Q}\left(\frac{\partial \overline{w}}{\partial \overline{z}}\right)^{-\alpha Q} e^{2\alpha \phi_{(z,\overline{z})}}.
\label{fieldtransformation}
\end{equation}
We denote these operators as $V_{\alpha}\equiv e^{2\alpha \phi_{(z,\overline{z})}} $. We find that $V_{\alpha}$ is a conformal operator with classical conformal weights  $\Delta_{class}=\overline{\Delta}_{class}= \alpha Q$. The constant $\alpha$ is called the Liouville momentum. Surprisingly, such weights change in the quantum case. In other words, conformal weights are modified by the introduction of radial quantization. This fact can be seen clearly from the OPE between $e^{2\alpha \phi_{(z,\overline{z})}}$ and the energy-momentum tensor $T_{z z}$, as done in Chapter \ref{Chapter2}. Quantum conformal weights are modified by a factor of $-\alpha^{2}$, that is,
\begin{eqnarray}
\Delta &=& \Delta_{class}-\alpha^{2},\\
\overline{\Delta}&=& \overline{\Delta}_{class} - \alpha^{2}. 
\end{eqnarray}
Thereby, Liouville quantum conformal weights are given by the expressions
\begin{eqnarray}\label{quantumCW}
\Delta &=& \alpha(Q-\alpha),\\
\overline{\Delta} &=& \alpha(Q-\alpha).
\end{eqnarray}
We now consider correlation functions in Liouville theory by using the path integral formulation of QFT. Such correlation functions are defined as \cite{zamo} \cite{1}
\begin{equation}
\left\langle e^{2\alpha_{1} \phi_{(z_{1},\overline{z}_{1})}} ... e^{2\alpha_{n} \phi_{(z_{n},\overline{z}_{n})}}\right\rangle  \equiv \int \mathcal{D} \phi_{c} \; e^{-S_{L}} \prod^{n}_{i=1} \exp \left(2\alpha_{i}\phi(z_{i},\overline{z}_{i})\right).
\label{green1}
\end{equation}
We can also write $V_{\alpha}(z,\overline{z}) = e^{2\alpha \phi_{(z,\overline{z})}}$ \cite{zamo} and thus obtain
\begin{equation}
\left\langle V_{\alpha_{1}}(z_{1},\overline{z}_{1})... V_{\alpha_{n}}(z_{n},\overline{z}_{n})\right\rangle \equiv \int \mathcal{D} \phi\; e^{-S_{L}} \prod^{n}_{i=1} \exp \left(2\alpha_{i}\phi(z_{i},\overline{z}_{i})\right).
\label{green2}
\end{equation}
Following \cite{dual}, for purposes of calculation it is useful to label eq.(\ref{green2}) as $\mathcal{H}_{N}(x_{I},\alpha_{I})$. Additionally, we will rewrite $\phi$ by $\widetilde{\phi}$. Hence, eq.(\ref{green1}) or eq.(\ref{green2}) are rewritten as
\begin{equation}
\mathcal{H}_{N}(x_{I},\alpha_{I})=\left\langle \prod_{I=1}^{N} e^{2\alpha_{I}\widetilde{\phi}_{(x_{I})}}\right\rangle = \int \mathcal{D} \widetilde{\phi}\; e^{-S_{L}+2\sum_{I=1}^{N}\alpha_{I}\widetilde{\phi}_{(x_{I})}},
\label{DOZZ1}
\end{equation}
where $S_{L}$ is given by eq.(\ref{liouvilleaction}). Later on, we will compute this function explicitly for $N=3$, which means computing the Liouville three-point function. Such a correlator was derived  independently for the first time by H. Dorn \& H. Otto \cite{dotto1} and A. Zamolodchikov \& Al. Zamolodchikov \cite{zamo2}, for this reason that expression is also called the DOZZ formula.

\section{Computing Liouville N - point functions}

\subsection{Coulomb gas method}\label{Coulomb}

The Coulomb gas method is the usual way to compute correlation functions in Liouville theory. Shortly we will explain this method and how it is applied.

From eq.(\ref{liouvilleaction}), we realize that the basic problem for computing correlation functions will be to find a way to deal with the exponential function arising from Weyl invariance $\mu e^{2b\phi}$. Such a term could be contemplated as a kind of exponential potential. The Coulomb gas method consists of expanding such an exponential term, which is placed within the path integral, as an infinite sum,
\begin{eqnarray}
\mathcal{H}_{N}=\int \mathcal{D}\phi e^{-\int d^{2}z \sqrt{g} \mu e^{2b\phi}}...=\sum_{s}\frac{(-\mu)^{s}}{s!}\int \mathcal{D}\phi \,\left( \int d^{2}z\,\sqrt{g} \mu e^{2b\phi}\right)^{s}...=\sum_{s}\mathcal{H}_{N}^{s}.\nonumber\\   
\end{eqnarray}
It is therefore possible to extract the familiar expression $(-1)^{s}/s!$, which is nothing but the pole of the Gamma function $\Gamma(z)$ at $z=-s$. Poles are important for two reasons. First, the remnant part of the correlator is raised to powers of $s$, thus $s \in \mathbb{N}$. Hence we must deal with the poles of the Gamma function. The second reason lies in the fact that we are dealing with a two dimensional theories which is isomorphic to the complex plane, where the residue theorem is a beautiful cornerstone. In fact, it was proposed by several authors \cite{zamo} \cite{zamo2} \cite{dotto1} \cite{dual} \cite{nakayama} \cite{3} \cite{Rashkov} (we will also adopt this statement) that at the end of the computation of the Liouville $N$-point function, the following relation holds:
\begin{eqnarray}\label{RESRES}
\text{Res}(\sum_{s=(Q-\sum_{I}\alpha_{I})/b}\mathcal{H}_{N})= \mathcal{H}_{N}^{s}.
\end{eqnarray}
The left hand side is the residue of the entire correlator $\mathcal{H}_{N}$ evaluated at the poles $s=(Q-\sum_{I}\alpha_{I})/b$, while the right hand side is the $s^{\text{th}}$ of the expansion of $\mathcal{H}_{N}$ as an infinite sum, such a expansion arises because the exponential potential is Taylor expanded. 

In contrast to the Coulomb gas method, L. O'Raifeartaigh, M. Pawlowski, and V. Sreedhar (OPS) \cite{dual} proposed a novel way to perform this computation. Instead of expanding the exponential potential as an infinite sum, they used the Sommerfeld-Watson transform \cite{WATSON}. We follow this path to compute the Liouville three point function and call it the OPS method. 
 
\subsection{The OPS method}
 
In Chapter \ref{Chapter3} we concluded that to maintain conformal invariance on the quantum side of Liouville theory, it is required that the holomorphic conformal dimension eq.(\ref{quantumCW}) of the exponential term in the Liouville action $S_{\text{SL}}$ must be equal to $1$. In this way a relation between $Q$ and $b$ could be derived. Let us recall that expression:
\begin{equation}\label{Qb1}
Q=b+\frac{1}{b}.
\end{equation}
From this relation it is easy to see that $Q$ is invariant under $b\longrightarrow 1/b$. Hence, this invariance makes reference to a type of quantum duality within Liouville theory, which is not present in the classical regime. Such a duality makes its second appearance in the fact that conformal weights are invariant with respect to the transformation
\begin{equation}
\alpha \longrightarrow (Q-\alpha).
\end{equation}
An immediate consequence of this duality is that there are two primaries of the form $e^{2 \alpha\phi}$, for each conformal weight $\Delta$. That is, using eq.(\ref{quantumCW}), we obtain 
\begin{equation}\label{Deltaduality}
\alpha_{1,2}=\frac{Q \pm \sqrt{Q^{2}-4\Delta}}{2},
\end{equation}
as well as another analogous relation for its anti-holomorphic counterpart.

O'Raifeartaigh et. al \cite{dual} used this fact to re-derive the DOZZ formula in a slightly different way as compared to the traditional one, which usually uses the Coulomb gas method. They argued that the lattice of poles, which appears in the DOZZ formula, has a quantum origin in the duality given by eq.(\ref{Deltaduality}). Such a duality is introduced within Liouville theory through the insertion of another exponential potential of the form $e^{2\frac{1}{b}\phi}$. However, this is not the first time where such an idea is used, in fact there are previous works which deal with a double potential theory, we cite \cite{DotsenkoDual} for example.

Following \cite{dual}, we re-derive the DOZZ formula. In this process we will find some subtleties to be highlighted.

To begin with, we want to get a compact expression for the  $N$ - point function. As stated above, we  consider not only contributions from the potential $\mu e^{2 b \widetilde{\phi}}$, but also contributions from $\mu e^{2 \frac{1}{b} \widetilde{\phi}}$ \cite{dual} \cite{dual2}. Therefore, for purposes of calculation, we write these potentials as $W_{b}$ and $W_{\frac{1}{b}}$:
\begin{eqnarray}
W_{b}&=& \mu_{b} e^{2 b \widetilde{\phi}}, 
\label{potencial1}\\
W_{1/b}&=& \mu_{\frac{1}{b}} e^{2 \frac{1}{b}\widetilde{\phi}}.
\label{potential2}
\end{eqnarray}
Moreover, in a two dimensional compact \footnote{Recall that this surface was ``compacted" with the insertion of infinity positive metrics on its punctured points.} Riemann surface it is possible to separate the field $\widetilde{\phi}_{(x)}$ as the sum of two pieces $\phi_{0}$ and $\phi(x)$ \cite{Goulian}. In this way for the field $\widetilde{\phi}(x)$ we have:
\begin{equation}
\widetilde{\phi}_{(x)}= \phi_{0} + \phi_{(x)}. 
\label{separated}
\end{equation}
In eq.(\ref{separated}), $ \phi_{0}$  is called the zero mode and the other field $\phi_{(x)}$ belongs to the Kernel of the scalar Laplacian operator $\nabla^{2}$. Hence, $\phi_{0}$ is understood as a constant field, while  $\phi_{(x)}$ generates the space of functions orthogonal to the Kernel. This means,
\begin{equation}
\nabla^{2}\phi_{0}=0 \;\;\;\; ;\;\;\;\;\int d^{2}x \;\phi_{(x)}=0.
\label{conditions1}
\end{equation} 
On the other hand, from eq.(\ref{separated}) we have the following relation between the measures of $\widetilde{\phi}(x), \phi_{0}$ and $\phi(x)$ \cite{dual2} \cite{nakayama} \cite{Goulian},
\begin{equation}
\mathcal{D}\widetilde{\phi} = \mathcal{D} \phi_{0} \, \mathcal{D} \phi_{(x)},
\label{measureAA}
\end{equation}
where we are also using the fact that all measures in Liouville theory are linear (see Chapter \ref{Chapter3}). 

By taking into account the insertion of the dual potential $W_{1/b}$, we have for the Liouville action:
\begin{equation}
S_{L}=\frac{1}{4\pi}\int d^{2}x \sqrt{\widehat{g}} \left[ \widehat{g}^{\alpha\beta} \partial_{\alpha} \phi \partial_{\beta} \phi+Q \widehat{R} \phi + 4\pi \mu_{b} e^{2b\phi}+4\pi \mu_{1/b} e^{2(\frac{1}{b})\phi}\right]. 
\label{liouvilleaction3}
\end{equation}
Now, we insert eq.(\ref{separated}) into eq.(\ref{liouvilleaction3}) and obtain
\begin{eqnarray}
S_{L} &=&\frac{1}{4\pi}\int d^{2} x \sqrt{\widehat{g}} \; \phi \Delta \phi_{(x)}+Q\widehat{R} (\phi_{0} + \phi_{(x)})\nonumber\\
 &+&4\pi \mu_{b} e^{2b(\phi_{0}+\phi_{(x)})}+4\pi\mu_{1/b} e^{2(\frac{1}{b})(\phi_{0}+\phi_{(x)})},
\label{liouvilleaction2}
\end{eqnarray}  
where we dropped the surface term coming from the kinetic part and rewrote the Laplacian operator $\nabla^{2}$ as
\begin{equation}
\Delta=-\frac{1}{\sqrt{\widehat{g}}}\partial_{\alpha}(\sqrt{\widehat{g}} \; \widehat{g}^{\alpha\beta}\; \partial_{\beta}). 
\label{delta}
\end{equation}
For simplicity we relabel $\mu_{1/b}=\mu_{\widetilde{b}}$ in eq.(\ref{liouvilleaction2}), from now on. Thus, we write for the exponential potentials 
\begin{equation}
U_{b}= \int d^{2} x \sqrt{\widehat{g}}\; \mu_{b} e^{2b\phi_{(x)}},\;\;\; \text{and}\;\;\; U_{\widetilde{b}}= \int d^{2}x \sqrt{\widehat{g}} \;\mu_{{\widetilde{b}}} e^{2({\frac{1}{b}})\phi_{(x)}}.
\label{r1}
\end{equation}
After replacing eq.(\ref{r1}) into eq.(\ref{liouvilleaction2}) we realize that the following potentials $U_{b}\,e^{2b\phi_{0}}$, $U_{\widetilde{b}}\,e^{2(\frac{1}{b})\phi_{0}}$ are now part of the Liouville action. In this way, eq.(\ref{liouvilleaction2}) becomes:
\begin{eqnarray}
S_{L}&=&\frac{1}{4\pi}\int d^{2} x \sqrt{\widehat{g}}\phi_{(x)} \Delta \phi_{(x)}+\underbrace{\frac{1}{4\pi}\int d^{2} x\sqrt{\widehat{g}}\,Q\widehat{R}\phi_{0}}_{\text{Gauss-Bonnet term}}+\underbrace{\frac{1}{4\pi}\int d^{2} x\sqrt{\widehat{g}}\,Q\widehat{R}\,\phi_{(x)}}_{\text{Coupling to the curvature}}\nonumber\\
&+& U_{b}\,e^{2b \phi_{0}}+U_{\widetilde{b}}\,e^{2(\frac{1}{b})\phi_{0}}.
\label{liouvilleaction4}
\end{eqnarray}  
The Gauss-Bonnet term has this name because when we separate $Q$ and $\phi_{0}$ out of the integrand, we still have a remnant contribution $\int d^{2}x\,\sqrt{\widehat{g}}\widehat{R}$. Such expression is part of the well known mathematical statement called the Gauss-Bonnet Theorem. Such a theorem states that the total curvature over a surface $\Sigma$ is characterized completely by a topological invariant quantity called the Euler characteristic $\chi_{\Sigma}$ . This theorem is explicitly written as:
\begin{equation}\label{gaussbonnet}
\int_{\Sigma} d^{2}x\sqrt{\widehat{g}}\widehat{R} = 4\pi \chi_{\Sigma}.
\end{equation} 
For a space topologically equivalent to a $2$-sphere, $\chi_{sphere} = 2$. In consequence, $\int d^{2}x\sqrt{\widehat{g}}\widehat{R}$ only contributes with a constant factor.

Taking into account eqs.(\ref{measureAA}),(\ref{r1}) and (\ref{gaussbonnet}), $\mathcal{H}_{N}(x_{I},\alpha_{I})$ becomes
\begin{eqnarray}\label{r2}
\hspace{-1cm}\mathcal{H}_{N}(x_{I},\alpha_{I})&=&\int \mathcal{D} \phi_{0} e^{-2Q\phi_{0}+2\sum_{I}\alpha_{I}\phi_{0}}\int\mathcal{D} \phi_{(x)}\;e^{-U_{b}e^{2b\phi_{0}}}e^{-U_{\widetilde{b}}e^{2(\frac{1}{b})\phi_{0}}}\times\\
&&\hspace{-1.9cm}\exp\left\lbrace -\int d^{2} x \sqrt{\widehat{g}}\left[\frac{1}{4\pi} \phi_{(x)} \Delta \phi_{(x)}+\frac{Q}{4\pi}\widehat{R} \phi_{(x)}\right] + 2\sum_{I}\alpha_{I}\phi_{(x_{I})}\right\rbrace \nonumber. 
\end{eqnarray}
As mentioned in subsection \ref{Coulomb}, in the context of conformal theories, the so called Coulomb gas expansion is usually used to solve this type of path integrals. 


Instead, the OPS method makes use of the Sommerfeld-Watson transform \cite{dual} of the exponential function, which is expressed in the parameter space $s$ as:
\begin{equation}
e^{t}=\frac{1}{2\pi i}\int^{\infty}_{-\infty}ds \; \frac{(-t)^{i s}}{\Gamma (1+ i s)}\frac{\pi}{\sinh(\pi s)}.
\label{SW}
\end{equation}
It is worth commenting that in contrast to the Coulomb gas expansion, where singularities are not obvious because they are placed within $U_{b}$ and $U_{\widetilde{b}}$, the Sommerfeld-Watson transform has manifest singularities in the denominator.

Additionally, we use the following notation
\begin{equation}\label{theta}
\Theta = Q - \sum_{I} \alpha_{I},
\end{equation}
in eq.(\ref{r2}). Such a term forms part of the exponential part of $\mathcal{H}_{N}(x_{I},\alpha_{I})$ associated to the zero mode $\phi_{0}$. That exponential factor is
\begin{equation}\label{THETA}
e^{-2\Theta\phi_{0}}.
\end{equation}
The field $\phi$ has a smooth behaviour over the sphere or a topologically equivalent surface, except at its punctured points.  The field $\phi$ near such points have the following expression \cite{zamo} \cite{zamo2} \cite{dotto1} \cite{russo} \cite{nakayama}
\begin{equation}\label{PuncturedPoints}
\phi_{i}(z,\overline{z}) = - 2 Q \log|z-z_{i}|,
\end{equation}
where $z_{i}$ are the punctured points. These points produce an important restriction relating $\Theta$ and $\alpha_{I}$. In order to see it let us replace eq.(\ref{PuncturedPoints}) into eq.(\ref{THETA}), we have
\begin{equation}\label{restriction}
\mathit{e}^{-2\Theta\phi_{i}}= |z-z_{i}|^{4Q\Theta}.
\end{equation}
The expression above shows how the field $\phi_{i}$ behaves near the punctures. In particular, we compute eq.(\ref{restriction}) for the zero mode $\phi_{0}$. It is easy to see from eq.(\ref{r2}) that the constraint of $\Theta$ is a consequence of the integral over the zero mode. From eq.(\ref{restriction}) and the requirement that the zero mode integration has to be finite, we obtain
\begin{equation}\label{thetacondition}
\Theta > 0.
\end{equation}
By using eq.(\ref{theta}), we must have
\begin{equation}\label{Constraint11}
Q-\sum_{I}\alpha_{I} > 0.
\end{equation}
The equation (\ref{Constraint11}) is obtained in a different manner in \cite{zamo} \cite{witten2012}. Following \cite{dual} we replace eq.(\ref{SW}) into eq.(\ref{r2}) and perform the analytic continuation $\Theta \longrightarrow i \Theta$. Therefore, we get for $\mathcal{H}_{N}(z_{I},\alpha_{I})$ the functional integral:
\begin{eqnarray}\label{W1}
\mathcal{H}_{N}(z_{I},\alpha_{I}) &=& - \frac{1}{4}\int\ \mathcal{D}\phi_{0}\mathcal{D}\phi\,  d\,s \,\, d\,l \frac{\mathit{e}^{-2i(\Theta - b \,s-\frac{l}{b})\phi_{0}}}{\sinh(\pi s)\sinh(\pi l)}\frac{(U_{b})^{is}}{\Gamma(1+is)}\times\\
&&\hspace{-2.5cm}\frac{(U_{\widetilde{b}})^{il}}{\Gamma(1+il)} \times \exp\left\lbrace -\int d^{2}z\sqrt{g_{(z)}}\left[ \frac{1}{4\pi}\phi\Delta\phi + \frac{Q}{4\pi}\widehat{R}\phi\right] + 2\alpha_{I}\phi(z_{I}) \right\rbrace. \nonumber 
\end{eqnarray}
In order to evaluate eq.(\ref{W1}), we define $\mathcal{F}_{N}(b,\widetilde{b};\alpha_{I}; i s, i l)$ as
\begin{eqnarray}\label{nse}
\mathcal{F}_{N}(b,\widetilde{b};\alpha_{I}; i s, i l)&=& \int \mathcal{D}\phi  \frac{(U_{b})^{is}}{\Gamma(1+is)}
\frac{(U_{\widetilde{b}})^{il}}{\Gamma(1+il)}\times\nonumber\\
&&\exp\left\lbrace -\int d^{2}z\sqrt{g_{(z)}}\left[ \frac{1}{4\pi}\phi\Delta\phi + \frac{Q}{4\pi}\widehat{R}\phi\right] + 2\alpha_{I}\phi(z_{I}) \right\rbrace.\nonumber\\
\end{eqnarray}
Therefore, eq.(\ref{W1}) becomes
\begin{equation}\label{F37}
\mathcal{H}_{N}(z_{I},\alpha_{I}) = - \frac{1}{4}\int\ \mathcal{D}\phi_{0}\,d\,s \,d\,l\, \frac{\mathit{e}^{-2i(\Theta-bs-\frac{l}{b})\phi_{0}}}{\sinh(\pi s)\sinh(\pi l)}\mathcal{F}_{N}(b,\widetilde{b};\alpha_{I}; i s, i l).
\end{equation}
Our next step will be computing the integral over the zero mode $\phi_{0}$. This integration gives rise to a delta function $\delta^{2}(\Theta-bs-(\frac{1}{b})l)$. Hence, eq.(\ref{F37}) becomes 
\begin{equation}\label{important1}
\mathcal{H}_{N}(z_{I},\alpha_{I}) = - \frac{1}{8}\int \,d\,s \,d\,l\,\frac{\delta^{2}(\Theta-bs-(\frac{1}{b})l)}{\sinh(\pi s)\sinh(\pi l)}\mathcal{F}_{N}(b,\widetilde{b};\alpha_{I}; i s, i l).
\end{equation}
We must point out that to compute eq.(\ref{F37}), the values of $is$ and $il$ are restricted to be integers, $m$ and $n$ respectively, otherwise it is not possible to compute that integral for the general case \cite{dual}. 

Substituting eq.(\ref{r1}) into eq.(\ref{nse}), it is easy to obtain
\begin{equation}\label{F38}
\mathcal{F}_{N}(b,\widetilde{b};\alpha_{I}; i s, i l) =  \frac{\mu_{b}^{m}\mu_{\widetilde{b}}^{n}}{m!n!}\int \prod_{i=1}^{m}d^{2}x_{i}\sqrt{g(x_{i})}\prod_{r=1}^{n}d^{2}y_{r}\sqrt{g(y_{r})}\mathcal{J}([\phi]),
\end{equation}
where
\begin{equation}\label{J}
\mathcal{J}([\phi])= \int \mathcal{D}\phi\mathit{e}^{-\int d^{2}z\sqrt{g}(\phi\frac{\Delta}{4\pi}\phi + \frac{Q}{4\pi}\widehat{R}(z)\phi) - 2\sum_{I=1}^{N}\alpha_{I}\phi(z_{I}) - 2\sum_{i=1}^{m}b\phi(x_{i}) - 2\sum_{r=1}^{N}\frac{1}{b}\phi(y_{r}) }.
\end{equation}
Some terms of eq.(\ref{J}) are transformed by applying the basic property of delta functions $\int \phi(y)\delta(x-y)=\phi(x)$. Thus, we have
\begin{eqnarray}
2\sum_{I=1}^{N}\alpha_{I}\phi(z_{I}) &=& \int d^{2}z\sqrt{g}(\sum_{I=1}^{N}\frac{2\alpha_{I}}{\sqrt{g}}\delta^{2}(z-z_{I}))\phi(z),\label{j1}\\
2\sum_{i=1}^{m} b \,\, \phi(x_{i}) &=& \int d^{2}z \sqrt{g} (\sum_{i=1}^{m}\frac{2 b}{\sqrt{g}}\delta^{2}(z-x_{i}))\phi(z),\label{j2}\\
2\sum_{r=1}^{n}\frac{1}{b}\phi(y_{r}) &=& \int d^{2}z\sqrt{g}(\sum_{r=1}^{n}\frac{2}{b \sqrt{g}}\delta^{2}(z-y_{r}))\phi(z).\label{j3}
\end{eqnarray}
If we replace eqs.(\ref{j1}), (\ref{j2}) and (\ref{j3}) into eq.(\ref{J}), it is not difficult to obtain 
\begin{eqnarray}\label{J1}
\mathcal{J}([\phi]) &=& \int \mathcal{D}\phi\exp\left\lbrace -\int d^{2}z\sqrt{g}\left[ \phi\frac{\Delta}{4\pi}\phi + \mathit{j}\phi\right]\right\rbrace, 
\end{eqnarray}
where we have introduced a source $j(z)$. This function is given by
\begin{equation}\label{current}
j(z) = \frac{Q}{4\pi}\widehat{R}(z) - \sum_{I=1}^{N}\frac{2\alpha_{I}}{\sqrt{g}}\delta^{2}(z-z_{I}) - \sum_{i=1}^{m}\frac{2 b}{\sqrt{g}}\delta^{2}(z-x_{i}) - \sum_{r=1}^{n}\frac{2}{b \sqrt{g}}\delta^{2}(z-y_{r}).
\end{equation}
Hence, eq.(\ref{J1}) can be seen as a Gaussian path integral. In order to compute it, we define the functional inner product $\langle\hspace{0.2cm},\hspace{0.2cm}\rangle$ as
\begin{equation}
\langle\psi,\mathbb{A} \xi\rangle = \int d^{2}z \sqrt{g_{(z)}} \psi(z) \mathbb{A} \xi(z), 
\end{equation} 
where $\mathbb{A}$ is an arbitrary operator. Thus, we can rewrite eq.(\ref{J1}) as
\begin{eqnarray}
\mathcal{J}([\phi]) &=& \int \mathcal{D}\phi\, e^{\left\lbrace -(\frac{\Delta}{4\pi})\langle\phi + \frac{1}{2}(\frac{\Delta}{4\pi})^{-1}j,\phi + \frac{1}{2}(\frac{\Delta}{4\pi})^{-1}j\rangle + \langle\frac{1}{2}(\frac{\Delta}{4\pi})^{-1}j, \frac{1}{2}(\frac{\Delta}{4\pi})^{-1}j\rangle\right\rbrace}, \label{J2}\\
 &=& \int \mathcal{D}\phi\,e^{-(\frac{\Delta}{4\pi})\left\lbrace\langle\phi + \frac{1}{2}(\frac{\Delta}{4\pi})^{-1}j,\phi + \frac{1}{2}(\frac{\Delta}{4\pi})^{-1}j\rangle\right\rbrace} e^{(\frac{\Delta}{4\pi})\left\lbrace  \langle\frac{1}{2}(\frac{\Delta}{4\pi})^{-1}j, \frac{1}{2}(\frac{\Delta}{4\pi})^{-1}j\rangle\right\rbrace}.\nonumber
\end{eqnarray}
It is easy to see from the latter equation that only the first exponential term is a functional of $\phi$. Therefore, the second exponential factor could be put out of the functional integral. We have
\begin{equation}\label{EXPJ}
\mathcal{J}([\phi]) = \frac{1}{\sqrt{\det(\frac{\Delta}{4\pi})}}\times\exp\left\lbrace (\frac{\Delta}{4\pi}) \left\langle \frac{1}{2}(\frac{\Delta}{4\pi})^{-1}j, \frac{1}{2}(\frac{\Delta}{4\pi})^{-1}j\right\rangle  \right\rbrace. 
\end{equation}
As in QFT we identified the inverse of $\Delta$, as the Green function $G(x,y)$. This fact arises from the Green equation,
\begin{equation}\label{Deltadefinition}
\Delta G(x,y)=\frac{\pi}{\sqrt{g}}\delta^{2}(x,y).
\end{equation}
In operator form, we can build from eq.(\ref{Deltadefinition}) the Green function $G(x,y)$ as follows
\begin{eqnarray}
G(x,y) &=& \frac{\pi}{\sqrt{g}}\frac{1}{\Delta},\nonumber\\
\sqrt{g}\, G(x,y) &=& \frac{1}{4}(\frac{\Delta}{4\pi})^{-1}.\label{Gdefinition}
\end{eqnarray}
From eq.(\ref{Gdefinition}). we observe in eq.(\ref{EXPJ}) that the exponent could be written as
\begin{eqnarray}
 \hspace{-0.5cm}\left\langle \frac{1}{2}j, \frac{1}{2}(\frac{\Delta}{4\pi})^{-1}j\right\rangle &=& \int d^{2}z\sqrt{g(z)}\int d^{2}z'\sqrt{g(z')} j(z)G(z,z')j(z'), 
\end{eqnarray}
and finally eq.(\ref{EXPJ}) becomes,
\begin{equation}\label{Jz}
\mathcal{J}([\phi]) = \frac{1}{\sqrt{\det(\frac{\Delta}{4\pi})}}\times\mathit{e}^{\int d^{2}z\sqrt{g(z)}\int d^{2}z'\sqrt{g(z')} j(z)G(z,z')j(z')}.
\end{equation}
Replacing eq.(\ref{Jz}) into eq.(\ref{F38}), 
\begin{eqnarray}\label{FF}
\mathcal{F}_{N}(b,\widetilde{b};\alpha_{I}; m, n) &=&  \frac{\mu_{b}^{m}\mu_{\widetilde{b}}^{n}}{m!n!} \frac{1}{\sqrt{\det(\frac{\Delta}{4\pi})}} \int \prod_{i=1}^{m}d^{2}x_{i}\sqrt{g(x_{i})}\prod_{r=1}^{n}d^{2}y_{r}\sqrt{g(y_{r})}\nonumber\\
&&\times\mathit{e}^{\int d^{2}z\sqrt{g(z)}\int d^{2}z'\sqrt{g(z')} j(z)G(z,z')j(z')},
\end{eqnarray}
and by substituting the currents $j(z)$, $j(z')$ into eq.(\ref{FF}), we get:
\begin{eqnarray}\label{FNpolyakovterm}
 \mathcal{F}_{N}(b,\widetilde{b};\alpha_{I}; m, n) &=& \frac{\exp\left\lbrace \frac{Q^{2}}{16\pi^{2}} \int d^{2}z\sqrt{g(z)}\int d^{2}z'\sqrt{g(z')}\widehat{R}(z)G(z,z')\widehat{R}(z')\right\rbrace}{\sqrt{\det(\frac{\Delta}{4\pi})}}\nonumber\\
 && \times \frac{\mu^{m}_{b}\mu^{n}_{\widetilde{b}}}{m!n!}\exp\left\lbrace \sum_{I\neq J}^{N} 4\alpha_{I}\alpha_{J} G(z_{I}, z_{J}) \right\rbrace\nonumber \\
 && \exp \left\lbrace -2Q\sum_{I=1}^{N}\alpha_{I} P(z_{I}) \right\rbrace \times \mathcal{I}_{N}.
\end{eqnarray}  
Functions $P(z)$ and $\mathcal{I}_{N}$ are defined in eq.(\ref{FNpolyakovterm}) and are given as:
\begin{eqnarray}\label{Pfunction}
P(z)&=&\frac{1}{\pi}\int d^{2}z'\sqrt{g(z')} G(z, z')\widehat{R}(z'),\label{Pfunction}\\
\mathcal{I}_{N}(b,\widetilde{b};\alpha_{I};m,n) &=& \int \prod_{i=1}^{m}d^{2}x_{i}\sqrt{g(x_{i})}e^{-2bQP(x_{i})} \int \prod_{i=1}^{n}d^{2}y_{r}\sqrt{g(y_{r})}\nonumber\\
&&e^{-2\frac{1}{b}QP(y_{r})}\times e^{\left[  \mathcal{G}_{b}(x_{i},x) + \mathcal{G}_{\widetilde{b}}(y_{r},y) + 8 G(x_{i},y_{r})\right] }.\label{INfunction1}
\end{eqnarray}
Additionally, $\mathcal{G}_{b}$, $\mathcal{G}_{\widetilde{b}}$ are defined as: 
\begin{eqnarray}
\mathcal{G}_{b}(x_{i},x) &=& 8 b \sum_{I=1}^{N}\alpha_{I} G(x_{i}, z_{I}) + 4 b^{2} \sum_{j=1}^{m}G(x_{i},x_{j}),\label{Bus1}\\
\mathcal{G}_{\widetilde{b}}(y_{r},y) &=& 8 \left( \frac{1}{b}\right)  \sum_{I=1}^{N}\alpha_{I} G(y_{r}, z_{I}) + 4 \left( \frac{1}{b} \right)^{2} \sum_{s=1}^{n}G(y_{r},y_{s}).\label{Bus2}
\end{eqnarray}
O’Raifeartaigh et. al \cite{dual} points out that the factor proportional to $Q^{2}$ placed in front of the integral eq.(\ref{FNpolyakovterm}) gives rise to the contribution $1 + 6 Q^{2}$ which is the central charge of the Virasoro algebra in Liouville theory, as mentioned before. Moreover, the denominator is a functional determinant which comes from the integration over the zero mode. All these factors do not play any other role in our calculations. Thus, to simplify our computation we drop them.

Let $P(z)$ be expressed in conformal coordinates, thus it could be reduced to (see Appendix \ref{AppendixA} ),   
\begin{equation}\label{Pz}
P(z) \longrightarrow  \frac{1}{2} \ln\sqrt{g(z)}.
\end{equation}
On the other hand, the Green function $G(x,y)$ defined in eq.(\ref{Deltadefinition}), has a singular ultraviolet behaviour. It means that this function becomes divergent when its points are infinitesimally close. This fact can be seen from the next expression for the Green function \cite{dual}
\begin{eqnarray}\label{Bus3}
G(x,y)=-\frac{1}{2}\ln\frac{|x-y|}{\Lambda},
\end{eqnarray}
where the parameter $\Lambda$ is a dimensional cut-off (do not confuse this with the cosmological constant introduced in the classical Liouville action). When we want to evaluate what happens to $G(x,y)$ as $x\rightarrow y$, the introduction of $\Lambda$ as a renormalization scale is useful. Therefore, $\lim_{x\rightarrow y}G(x,y)$ becomes $G(x,x)_{0}^{R}$. The renormalized Green function as $x\rightarrow y$ is given by (see Appendix \ref{AppendixB})
\begin{equation}\label{Gz}
G(x_{i},x_{i})_{0}^{R}=\frac{1}{4}\ln\sqrt{g(x_{i})}.
\end{equation}
where $i= 1, 2$. By inserting eqs.(\ref{Pz}), (\ref{Gz}) into eq.(\ref{INfunction1}), it is not difficult to show that this integral could be written in conformal coordinates as
\begin{eqnarray}\label{INfunction2}
\mathcal{I}_{N}(b,\widetilde{b};\alpha_{I};m,n) &=& \int \prod_{i=1}^{m}d^{2}x_{i}\sqrt{g(x_{i})}^{\beta_{b}} \int \prod_{i=1}^{n}d^{2}y_{r}\sqrt{g(y_{r})}^{\beta_{\widetilde{b}}}\nonumber\\
&&\times\mathit{e}^{\left[  \mathcal{G}^{R}_{b}(x_{i},x) + \mathcal{G}^{R}_{\widetilde{b}}(y_{r},y) + 8 G(x_{i},y_{r})\right] },
\end{eqnarray}
where $\beta_{b}= 1 - Qb + b^{2}$ and $\beta_{\widetilde{b}}= 1 - Q(\frac{1}{b})+ (\frac{1}{b})^{2}$.  Therefore, since $Q=b+1/b$, $\beta_{b}$ and $\beta_{\widetilde{b}}$ must be equal to zero. This result could also be obtained by appealing to the constraints imposed by Weyl invariance. 

\subsubsection*{Constraints imposed by Weyl Invariance}

Weyl symmetry offers a good way to impose constraints in eq.(\ref{INfunction2}). To see this, we perform a Weyl transformation
\begin{equation}
\sqrt{g} \longrightarrow \mathit{e}^{2\alpha\varphi} \sqrt{g},
\end{equation}
where $\mathit{e}^{2\alpha\varphi}$ is our conformal factor. In order to simplify our calculation, we represent the conformal factor by $\Omega = \mathit{e}^{2\alpha\varphi}$. By using this convention, we have to figure out how the function $\mathcal{I}_{N}$ transforms under Weyl transformations; it is easy to realize that the contributions to the integral come from $\sqrt{g(x_{i})}^{\beta_{b}}$ and $\sqrt{g(y_{r})}^{\beta_{\widetilde{b}}}$. Hence, $\mathcal{I}_{N}$ transforms as:
\begin{equation}\label{In1}
\mathcal{I}_{N} \longrightarrow \Omega^{m \beta_{b} + n \beta_{\widetilde{b}}}\mathcal{I}_{N}.
\end{equation} 
However, this factor can be reabsorbed into the dimensional parameters $\mu_{b}$ and $\mu_{\widetilde{b}}$ by imposing the suitable scaling
\begin{equation}\label{MU1}
\mu_{b} \longrightarrow \Omega^{-\beta_{b}}\mu_{b}\,\,\,\, \text{and}\,\,\,\,\mu_{\widetilde{b}} \longrightarrow \Omega^{-\beta_{\widetilde{b}}}\mu_{\widetilde{b}}.
\end{equation}
Analysing eqs.(\ref{In1}) and (\ref{MU1}), we find that the expression
\begin{equation}
\mu^{m}_{b}\mu^{n}_{\widetilde{b}}\,\mathcal{I}_{N},
\end{equation}
is Weyl invariant. However, we still have the factors $W_{b}$ and $W_{\widetilde{b}}$ into $\mathcal{F}_{N}$. These factors transform to:
\begin{eqnarray}
W_{b} &\longrightarrow & \Omega^{-\beta_{b}}W_{b},\\
W_{\widetilde{b}} &\longrightarrow & \Omega^{-\beta_{\widetilde{b}}}W_{\widetilde{b}}.
\end{eqnarray}
In order to ensure Weyl invariance of the N-point function, we impose the constraints:
\begin{eqnarray}
\beta_{b} &=& 0,\label{BETAcondition1}\\
\beta_{\widetilde{b}} &=& 0.\label{BETAcondition2}
\end{eqnarray}
We realize from these relations that it is another way to re-derive the well known duality equation between $Q$ and $b$,
\begin{equation}\label{Q=b+1b}
Q = b + \frac{1}{b}.
\end{equation}
It is worthwhile to recall that this requirement is a necessary condition for background independence (see Chapter \ref{Chapter3}).

After performing the renormalization program and using eqs.(\ref{Bus3}), (\ref{BETAcondition1}), (\ref{BETAcondition2}) we are able to rewrite eq.(\ref{INfunction2}) as,
\begin{eqnarray}\label{INfunction3}
\mathcal{I}_{N}(b,\widetilde{b};\alpha_{I};m,n) &=& \int \prod_{i=1}^{m}d^{2}x_{i} \int \prod_{i=1}^{n}d^{2}y_{r}
\mathit{e}^{\left[  \mathcal{G}^{R}_{b}(x_{i},x) + \mathcal{G}^{R}_{\widetilde{b}}(y_{r},y) - 4 \ln\frac{|z-z'|}{\Lambda}\right] }.
\end{eqnarray}
Renormalized functions $ \mathcal{G}^{R}_{b}(x_{i},x)$ and $\mathcal{G}^{R}_{\widetilde{b}}(y_{r},y)$ are defined by inserting the Green function given by eq.(\ref{Bus3}), into eqs.(\ref{Bus1}) and (\ref{Bus2}). Therefore, these equations become:
\begin{eqnarray}
\mathcal{G}_{b}^{R}(x_{i},x) &=& -4 b \sum_{I=1}^{N}\alpha_{I} \ln\frac{|x_{i}-z_{I}|}{\Lambda} - 2 b^{2} \sum_{i\neq j}^{m}\ln\frac{|x_{i}-x_{j}|}{\Lambda},\nonumber\\
\mathcal{G}_{\widetilde{b}}^{R}(y_{r},y) &=& -4(\frac{1}{b}) \sum_{I=1}^{N}\alpha_{I} \ln\frac{|y_{r}-z_{I}|}{\Lambda} - 2 (\frac{1}{b})^{2} \sum_{r \neq s}^{n}G\ln\frac{|y_{r}-y_{s}|}{\Lambda}.\nonumber
\end{eqnarray}
Replacing the latter equations and also eqs.(\ref{BETAcondition1}), (\ref{BETAcondition2}) into (\ref{INfunction3}), we get for $\mathcal{I}_{N}$:
\begin{eqnarray}\label{INOMEGA1}
\mathcal{I}_{N}(b,\widetilde{b};\alpha_{I};m,n) &=& \int \prod_{i<j}^{m}\prod_{r<s}^{n}d^{2}x_{i}d^{2}y_{r}|x_{i}-x_{j}|^{-4b^{2}}|y_{r}-y_{s}|^{-4(\frac{1}{b})^{2}}|x_{i}-y_{r}|^{-4}\nonumber\\
&&\times\prod_{i=1}^{m}\prod_{r=1}^{n}\prod_{I=1}^{N}|x_{i}-z_{I}|^{-4b\alpha_{I}}|y_{r}-z_{I}|^{-4(\frac{1}{b})\alpha_{I}}.
\end{eqnarray} 
Let us define the following measure $\mathcal{D}\mho(x_{i},y_{r})$ as,
\begin{equation}\label{MEASUREOOO}
\mathcal{D}\mho(x_{i},y_{r}) = \prod_{i<j}^{m}\prod_{r<s}^{n}d^{2}x_{i}d^{2}y_{r}|x_{i}-x_{j}|^{-4b^{2}}|y_{r}-y_{s}|^{-4(\frac{1}{b})^{2}}|x_{i}-y_{r}|^{-4}.
\end{equation}
Therefore, we rewrite eq.(\ref{INOMEGA1}) by using the measure $\mathcal{D}\mho(x_{i},y_{r})$, 
\begin{eqnarray}\label{INOMEGA2}
\mathcal{I}_{N}(b,\widetilde{b};\alpha_{I};m,n) = \int \mathcal{D}\mho(x_{i},y_{r})
\prod_{i=1}^{m}\prod_{r=1}^{n}\prod_{I=1}^{N}|x_{i}-z_{I}|^{-4b\alpha_{I}}|y_{r}-z_{I}|^{-4(\frac{1}{b})\alpha_{I}}.\nonumber\\
\end{eqnarray} 
The Dirac delta function in eq.(\ref{important1}) imposes a restriction over $m$ and $n$,
\begin{equation}\label{Constraint}
mb + n(\frac{1}{b}) = \Theta = Q - \sum_{I}\alpha_{I}.
\end{equation}
From eqs.(\ref{INOMEGA1}) and (\ref{INOMEGA2}), we realize that the cut-off $\Lambda$ does not appear. This fact occurs because if we put all $\Lambda$ contributions together, the next expression is obtained,
\begin{equation}\label{MUrenorm}
\left(\frac{\Lambda}{ds} \right)^{2Q^{2}}\times\left(\mu_{b}(ds)^{2} \right)^{m}\times \left(\mu_{\widetilde{b}}(ds)^{2} \right)^{n}\times\left(\frac{|\Lambda_{0}|}{ds} \right)^{-2\left[ \sum_{I}\alpha_{I}^{2}-(\sum_{I}\alpha_{I})^{2}\right] },     
\end{equation}
where $ds$ is the usual ``space-time'' interval and $\Lambda_{0}$ is the unit length in which the external points $z_{I}$ are measured. Consequently, we can absorb the first term in the normalization of the partition function because it does not depend on which points the correlator is evaluated. On the other hand, the other three terms could be reabsorbed in the normalization of $\mu_{b}$, $\mu_{\widetilde{b}}$ and unit length $\Lambda_{0}$. O'Raifeartaigh et al. \cite{dual} after normalizing theses cosmological constants write eq.(\ref{MUrenorm}) as
\begin{equation}
\left( \frac{\mu^{R}_{b}}{\mu}\right)^{m} \times \left( \frac{\mu^{R}_{\widetilde{b}}}{\mu} \right)^{n}. 
\end{equation}
In contrast, to make contact with the usual notation given in the literature of the DOZZ formula we just write $\mu^{m}_{b}\times\mu_{\widetilde{b}}^{n}$ for this parametric dependence. However, we must always keep in mind that they are renormalized quantities. 
Translational invariance of eq.(\ref{INOMEGA2}) is explicit because it only involves terms which are functions of the difference between two points. Nevertheless, the scaling $z_{I} \rightarrow \lambda z_{I}$ is not a trivial issue.

\subsubsection*{Scale covariance of $\mathcal{I}_{N}$}

The scaling of $z_{I}$ coordinates is written as,
\begin{equation}\label{zIscaling}
z_{I} \longrightarrow \lambda z_{I}, 
\end{equation}   
where $\lambda$ is chosen as a scaling parameter. Now, we analyse how the measure defined in eq.(\ref{MEASUREOOO}) transforms under that scaling.

As pointed out by O'Raifeartaigh et al. \cite{dual}, substituting eqs.(\ref{Constraint}), (\ref{zIscaling}) into eq.(\ref{INOMEGA2}) we get the next transformation for $\mathcal{I}_{N}$ under the scaling of $z_{I}$:
\begin{equation}\label{ScalingIN}
\mathcal{I}_{N}\rightarrow |\lambda|^{-2(Q-\sum_{I}\alpha_{I})\sum_{I}\alpha_{I}} \mathcal{I}_{N}.
\end{equation}
Eq.(\ref{ScalingIN}) only involves a scaling factor which could be seen as the Jacobian of the scaling transformation given above, thus it is considered that $\mathcal{I}_{N}$ is a covariant expression. 

\subsubsection*{SL(2,C) covariance}

It is well known that transformations of the full conformal group could be expressed as a $SL(2, \mathbb{C})$ transformation \cite{pginsparg}. Such transformation is defined as the map, 
\begin{equation}\label{SL2C}
z \rightarrow z'=\frac{az+b}{cz+d},
\end{equation} 
where $ad-bc=1$. From eq.(\ref{SL2C}), it is easy to see that the differential $dz$ and the difference between two points are given by:
\begin{eqnarray}
dz&\rightarrow &\frac{dz}{(cz+d)^{2}},\label{dzTransf}\\
z_{i}-z_{j}&\rightarrow &\frac{z_{i}-z_{j}}{(cz_{i}+d)(cz_{j}+d)}.\label{zIzJ}
\end{eqnarray}
Substituting the latter equations into eqs.(\ref{INOMEGA1}) and (\ref{MEASUREOOO}), it is not difficult to show that the integral $\mathcal{I}_{N}$ transforms according to:
\begin{equation}\label{INSL2C}
\mathcal{I}_{N}\rightarrow\prod_{I=1}^{N} \left( \frac{1}{|c z_{I}+d|}\right)^{-4(Q-\sum_{I}\alpha_{I})\alpha_{I}} \mathcal{I}_{N}.
\end{equation} 
We must realize that due to eq.(\ref{dzTransf}), we can rewrite eq.(\ref{INSL2C}) in a more suitable way. Hence, $\mathcal{I}_{N}$ is given by
\begin{equation}
\mathcal{I}_{N}\rightarrow\prod_{I=1}^{N} \left| \frac{\partial z'_{I}}{\partial z_{I}}\right|^{-(Q-\sum_{I}\alpha_{I})\alpha_{I}} \mathcal{I}_{N},
\end{equation}
where $\mathcal{I}_{N}$ has the expected covariant expression for a point function (see Chapter \ref{Chapter2}). Moreover, we observe that it has a conformal dimension $\Delta_{\mathcal{I}_{N}}= \sum_{I}\alpha_{I}(Q-\sum_{I}\alpha_{I})$.

From this fact, we conclude that $\mathcal{I}_{N}$ is $SL(2, \mathbb{C})$ covariant. If we only consider the real part of those transformations that group has $3$ free parameters. Therefore, in the complex case we have a total of $6$ real free parameters, or $3$ pair of points over the Riemann sphere. By using this fact, we can fix three points of $\mathbb{C}$ freely. We use this freedom to choose three suitable points $z'_{1}=R$, $z'_{2}=0$ and $z'_{3}=1$, where at the end of the calculation we will take $R \rightarrow \infty$.

By using $SL(2, \mathbb{C})$ covariance, we can change to arbitrary coordinates $z_{1}, z_{2}$ and $z_{3}$. Hence, we have 
\begin{eqnarray}
z'_{1}=\frac{ a z_{1} + b }{ c z_{1} + d } &=& R,\label{Eq1} \\
z'_{2}=\frac{ a z_{2} + b }{ c z_{2} + d } &=& 0, \label{Eq2}\\
z'_{3}=\frac{ a z_{3} + b }{ c z_{3} + d } &=& 1,\label{Eq3}
\end{eqnarray}   
where $a, b, c$ and $d$ are related by the condition $ad-bc=1$ as in eq.(\ref{SL2C}). Such a system of equations, along with the condition cited before between $a, b, c$ and $d$, becomes solvable. After solving eqs.(\ref{Eq1}), (\ref{Eq2}) and (\ref{Eq3}), we obtain for $a, b, c $ and $d$ \cite{dual}:
\begin{eqnarray}
a&=&\sqrt{\frac{R z_{31}}{(R -1)z_{23}z_{12}}},\\
b&=&-z_{2}\sqrt{\frac{R z_{31}}{(R -1)z_{23}z_{12}}},\\
c&=&\frac{R z_{23}+z_{12}}{\sqrt{R(R -1)z_{12}z_{23}z_{31}}},\\
d&=&\frac{R z_{1}z_{23}+z_{3}z_{12}}{\sqrt{R(R -1)z_{12}z_{23}z_{31}}}.
\end{eqnarray}
In order to use eq.(\ref{INSL2C}), we explicitly build its transformation factor under $SL(2,\mathbb{C})$ transformations. This factor is given by
\begin{equation}\label{FactorIN}
\prod_{I=1}^{N} \left( \frac{1}{|c z_{I}+d|}\right)^{-4(Q-\sum_{I}\alpha_{I})\alpha_{I}}=\prod_{I=1}^{N} \left|\frac{\sqrt{R(R-1)z_{12}z_{23}z_{31}}}{z_{12}z_{3I}+R z_{23}z_{1I}}\right|^{-4(Q-\sum_{I}\alpha_{I})\alpha_{I}},
\end{equation}
and the integral $\mathcal{I}_{N}$ evaluated at the points $z'_{1},z'_{2}$ and $z'_{3}$:
\begin{eqnarray}\label{INOOOOO}
\mathcal{I}_{N}(z'_{1},z'_{2},z'_{3})&=&\int \mathcal{D}\mho(x_{i},y_{r})\prod_{i,r}\prod_{j=4}^{N}|x_{i}-r_{j}|^{-4b\alpha_{j}}|y_{r}-r_{j}|^{-4(\frac{1}{b})\alpha_{j}}|x_{i}-R|^{-4b\alpha_{1}}\nonumber\\
&&|x_{i}|^{-4b\alpha_{2}}|x_{i}-1|^{-4b\alpha_{3}}|y_{r}-R|^{-4(\frac{1}{b})\alpha_{1}}|y_{r}|^{-4(\frac{1}{b})\alpha_{2}}|y_{r}-1|^{-4(\frac{1}{b})\alpha_{3}}.\nonumber\\
\end{eqnarray}
For the latter equation we have defined $r_{j}$ as,
\begin{equation}
r_{j}=\frac{R z_{j2}z_{31}}{z_{12}z_{3j}+R z_{23}z_{1j}},
\end{equation} 
where $j\geq 4$, because the first three points were already fixed.
 
By inserting eqs.(\ref{FactorIN}), (\ref{INOOOOO}) into eq.(\ref{INSL2C}) and taking the limit as $R \rightarrow \infty$, it is possible to obtain:
\begin{eqnarray}\label{INHome}
\mathcal{I}_{N}&=&\left|\frac{z_{12}z_{31}}{z_{23}}\right|^{-2(Q-\sum_{I}\alpha_{I})\alpha_{1}}\prod_{I=2}^{N}\left|\frac{z_{12}z_{31}}{z_{23}}\right|^{2(Q-\sum_{I}\alpha_{I})\alpha_{I}}|z_{1I}|^{4(Q-\sum_{I}\alpha_{I})\alpha_{I}}\times\nonumber\\
&&\int \mathcal{D}\mho(x_{i},y_{r})\prod_{i,r}\prod_{j=4}^{N}|x_{i}-r_{j}|^{-4b\alpha_{j}}|y_{r}-r_{j}|^{-4(\frac{1}{b})\alpha_{j}}\times\nonumber\\
&&|x_{i}|^{-4b\alpha_{2}}|x_{i}-1|^{-4b\alpha_{3}}|y_{r}|^{-4(\frac{1}{b})\alpha_{2}}|y_{r}-1|^{-4(\frac{1}{b})\alpha_{3}}.\nonumber\\
\end{eqnarray}
It is worthwhile to point out that the $R$ dependence is cancelled by putting together all the contributions coming from eqs.(\ref{FactorIN}) and (\ref{INOOOOO}). Hence, there is no divergent terms after taking $R \rightarrow \infty$.

\section{Computing N=3 by the OPS method}

After replacing eqs.(\ref{FNpolyakovterm}) and (\ref{INHome}) into eq.(\ref{important1}) and fixing the number of Liouville \hspace{2cm} primaries to be $3$, we find that the function $\mathcal{H}_{N}(x_{I},\alpha_{I})$ can be rewritten as:
\begin{equation}
\mathcal{H}_{3}(z_{1}, z_{2}, z_{3};\alpha_{1}, \alpha_{2}, \alpha_{3})= \mathcal{N}_{3} \left|z_{12}\right|^{2(\Delta_{3}- \Delta_{1}-\Delta_{2})}\left|z_{23}\right|^{2(\Delta_{1}- \Delta_{3}-\Delta_{2})}\left|z_{31}\right|^{2(\Delta_{2}- \Delta_{3}-\Delta_{1})}.
\label{G3}
\end{equation}
In this equation, $\mathcal{N}_{3}$ is defined by the following relation,
\begin{eqnarray}
\mathcal{N}_{3}(\alpha_{1}, \alpha_{2}, \alpha_{3})= -\frac{1}{8}\int ds dl \frac{\mathcal{I}_{3}(is, il)}{\Gamma (1+ i s)\Gamma (1+ i l)}\frac{\delta(bs+(\frac{1}{b})l-\Theta)}{\sinh(\pi s)\sinh(\pi l)}
\mu_{b}^{is}\mu_{\widetilde{b}}^{il}.
\label{H3}
\end{eqnarray}
We have to recall that $\Theta = Q - \sum_{I}\alpha_{I}$. The next task is to compute eq.(\ref{H3}). In order to do this, we take integers values of $is$ and $il$ again, this means $is\rightarrow m$ and $il\rightarrow n$, where $m$ and $n$ are integers. After this replacement, from eq.(\ref{INHome}) as $N=3$ we find that $\mathcal{I}_{3}$ becomes, 
\begin{equation}
\mathcal{I}_{3}(m, n)=\int \mathcal{D}\mho_{(x_{i},y_{r})}\prod_{i=1}^{m}\prod_{r=1}^{n}\left|x_{i}\right|^{-4b\alpha_{2}}\left|x_{i}-1\right|^{-4b\alpha_{3}}\left|y_{r}\right|^{-4(1/b)\alpha_{2}}\left|y_{r}-1\right|^{-4(1/b)\alpha_{3}}.
\label{I3}
\end{equation}
To clarify eq.(\ref{I3}), we have to recall that $\mathcal{D}\mho_{(x_{i},y_{r})}$ is defined as an integral measure. Such a measure was defined by the following expression,
\begin{equation}
\mathcal{D}\mho_{(x_{i},y_{r})}=\prod_{i < j}^{m}\prod_{r < s}^{n} d^{2}x_{i}  d^{2}y_{r} \left|x_{i}-x_{j}\right|^{-4b^{2}}\left|y_{r}-y_{s}\right|^{-4(1/b)^{2}}\left|x_{i}-y_{r}\right|^{-4}.
\label{K3}
\end{equation}
We are now interested in solving the integral defined by eqs.(\ref{I3}) and (\ref{K3}). To achieve this we will use the result given by Dotsenko and Fateev \cite{dofa} (see Appendix \ref{AppendixC}) and use the Zamolodchikovs function $\Upsilon(x)$  \cite{zamo2}:
\begin{equation}
 \mathcal{I}_{3}(m, n)= - m! n! \phi^{m}_{b}\phi^{n}_{1/b}\frac{\Upsilon'_{(0)}}{\Upsilon'_{(-bm-(\frac{1}{b})n)}}\prod_{I=1}^{N=3}\frac{\Upsilon_{(2\alpha_{I})}}{\Upsilon_{(\sum_{J}\alpha_{J}-2\alpha_{I})}}.
\label{I31}
\end{equation}
The Zamolodchikovs function is defined in the real strip $0<Re(x)<Q$ by the following integral: 
\begin{equation}
\log \Upsilon_{(x)}= \int^{\infty}_{0}\frac{dt}{t}\left[\left(\frac{Q}{2}-x\right)^{2}e^{-t}-\frac{\sinh^{2}\left(\frac{Q}{2}-x\right)\frac{t}{2}}{\sinh \frac{bt}{2}\sinh\frac{t}{2b}}\right].
\label{upsilon}
\end{equation} 
Functions $\phi_{b}$ and $\phi_{1/b}$, given in eq.(\ref{I31}), are defined as:
\begin{eqnarray}
\phi_{b}&=&\pi b \frac{\Upsilon_{(2b)}}{\Upsilon_{(b)}},
\label{phib}\\
\phi_{1/b}&=&\frac{\pi }{b} \frac{\Upsilon_{(2(\frac{1}{b}))}}{\Upsilon_{(\frac{1}{b})}}.\label{phi1b}
\end{eqnarray}
Eq.(\ref{upsilon}) along with its simple zeros is defined as quasi-periodic \cite{zamo2}\cite{dual}, in the sense that under a displacement $x\rightarrow x+b$, it is possible to obtain \cite{witten2012}
\begin{equation}
 \Upsilon_{(x+b)} = \Upsilon_{(x)} \gamma_{(bx)}b^{1-2bx}, \;\;\; \text{with} \;\;\; \gamma_{(x)}= \frac{\Gamma_{(x)}}{\Gamma_{(1-x)}}.
\label{upsilonAhhh}
\end{equation}
In addition to the definition of Zamolodchikovs function, we mention that $\Upsilon_{(x)}$ is an entire function with simple zeros at $x=-(m b+ n(1/b))$ and $x= (m+1) b+ (n+1)(1/b))$ for all $m,n \geq 0$ \cite{zamo2}. This fact can be seen from eq.(\ref{upsilonAhhh}). Moreover, it is easy to see from eq.(\ref{upsilon}) that $\Upsilon_{(x)}$ has the so called reflection symmetry property $\Upsilon_{(x)}=\Upsilon_{(Q-x)}$. We must also realize that eq.(\ref{upsilon}) remains invariant under the inversion $b\rightarrow 1/b$.

As a result, the analytic continuation of $\,m! n! \phi^{m}_{b}\phi^{n}_{1/b}$ to $\Gamma_{(1+is)}\Gamma_{(1+il)}\phi^{is}_{b}\phi^{il}_{1/b}$ is performed in a straightforward manner. Therefore, we replace the latter result into eq.(\ref{I31}) to get:
\begin{eqnarray}
 \mathcal{I}_{3}(is, il)= -\Gamma_{(1+is)}\Gamma_{(1+il)}\phi^{is}_{b}\phi^{il}_{1/b}\,\frac{\Upsilon'_{(0)}}{\Upsilon'_{(-bs-(\frac{1}{b})l)}}\prod_{I=1}^{N=3}\frac{\Upsilon_{(2\alpha_{I})}}{\Upsilon_{(\sum_{J}\alpha_{J}-2\alpha_{I})}}.
\label{I32}
\end{eqnarray}
By substituting eq.(\ref{I32}) into eq.(\ref{H3}), we have
\begin{eqnarray}\label{CertainN3}
\mathcal{N}_{3}=\frac{1}{8}\int dsdl\frac{\delta(bs+(1/b)l-\Theta)}{\sinh(\pi s)\sinh(\pi l)}\left(\mu \phi_{b} \right)^{is}\left(\mu \phi_{1/b} \right)^{il}\frac{\Upsilon'_{(0)}}{\Upsilon'_{(-bs-(\frac{1}{b})l)}}\prod_{I=1}^{N=3}\frac{\Upsilon_{(2\alpha_{I})}}{\Upsilon_{(\sum_{J}\alpha_{J}-2\alpha_{I})}}.\nonumber\\ 
\end{eqnarray}
The integral $\mathcal{P}(\Theta)$, which has all contributions from $s$ and $l$, is defined by:
\begin{eqnarray}\label{Pint1}
\mathcal{P}(\Theta)=\int ds dl \frac{\delta(bs+(1/b)l-\Theta)}{\sinh(\pi s)\sinh(\pi l)}\left(\mu \phi_{b} \right)^{is}\left(\mu \phi_{1/b} \right)^{il}.
\end{eqnarray}
From this integral expression, we can derive a transformation under the displacement $\Theta \rightarrow \Theta + ib$, 
\begin{eqnarray}\label{PFunctionFalse}
\mathcal{P}(\Theta + i b)=-\frac{1}{\mu \phi_{b}}\left\lbrace \mathcal{P}(\Theta)-2\frac{(\mu\phi_{1/b})^{i b\Theta}}{\sinh(\pi b\Theta)}\right\rbrace,  
\end{eqnarray}
for more details about this procedure see \cite{dual}. We must realize that eq.(\ref{PFunctionFalse}) has poles at $\Theta=\pm (mb+ n/b)$. The latter equation has a similar form that $\Upsilon(\Theta)$ under the displacement of $\Theta \rightarrow \Theta + b$. From eq.(\ref{upsilonAhhh}) we have \cite{dual} \cite{zamo} \cite{zamo2}
\begin{equation}\label{UPSilonFunctionVerdadera}
\frac{\Upsilon'(\Theta + b)}{\Upsilon(\Theta + b)}= \frac{\Upsilon'(\Theta)}{\Upsilon(\Theta)}+\frac{\gamma'(b\Theta)}{\gamma(b\Theta)}-2\ln b. 
\end{equation}
Therefore, following the assumption of L.O'Raifeartaigh et. al \cite{dual} we identified that under a  suitable limit
both functions behave similarly. The constant part placed at the end in eq.(\ref{UPSilonFunctionVerdadera}) is identified as an integration constant in eq.(\ref{PFunctionFalse}). Such a suitable limit mentioned above is defined by using eqs.(\ref{upsilon}) and (\ref{Pint1}), because while $\Upsilon'(\Theta)/\Upsilon$ evaluated at its poles has a residue equal to $1$, $\mathcal{P}(\Theta)$ has a residue equal to $(\mu\, \phi_{b})^{m}(\mu\,\phi_{1/b})^{n}$. As such, we have
\begin{eqnarray}
\lim_{\Theta\rightarrow mb+n/b}\frac{\mathcal{P}(\Theta)}{\Upsilon'(\Theta)/\Upsilon(\Theta)}=(\mu\, \phi_{b})^{m}(\mu\,\phi_{1/b})^{n}.
\end{eqnarray}  
By using the OPS {\it gauge}, which is defined as $(\mu\, \phi_{b})^{1/b}=(\mu\, \phi_{1/b})^{b}$ \cite{dual} \cite{zamo2}, eq.(\ref{Pint1}) becomes, 
\begin{eqnarray}\label{Pint2}
\mathcal{P}(\Theta)=\left(\mu \phi_{b} \right)^{\frac{\Theta}{b}}\int ds dl \frac{\delta(bs+(1/b)l-\Theta)}{\sinh(\pi s)\sinh(\pi l)}.
\end{eqnarray}
It is useful to define from eq.(\ref{Pint2}) the following integral,
\begin{eqnarray}\label{P0}
\mathcal{P}_{0}(\Theta)=\int ds dl \frac{\delta(bs+(1/b)l-\Theta)}{\sinh(\pi s)\sinh(\pi l)},
\end{eqnarray}
where we only take into account the remnant contributions of $s$ and $l$.
Insertion of eqs.(\ref{Pint2}) and (\ref{P0}) into eq.(\ref{CertainN3}) lead to
\begin{eqnarray}\label{N3.1}
\mathcal{N}_{3}=\frac{1}{8}\left(\mu \phi_{b} \right)^{\frac{\Theta}{b}}\Upsilon'_{(0)}\frac{\mathcal{P}_{0}(\Theta)}{\Upsilon'_{(-\Theta)}}\prod_{I=1}^{N=3}\frac{\Upsilon_{(2\alpha_{I})}}{\Upsilon_{(\sum_{J}\alpha_{J}-2\alpha_{I})}}.
\end{eqnarray}
After taking the OPS gauge, $\mathcal{P}_{0}(\Theta)$ has a residue equal to $\pm 1$ depending on if we are taking either positive or negative values for $m$ and $n$. In contrast, $\Upsilon'(\Theta)/\Upsilon(\Theta)$ has a residue equal to $1$ valid for positive values of $m$ and $n$. Hence, we have the following limit  
\begin{equation}
\lim_{\Theta\rightarrow m b+n/b}\frac{\mathcal{P}_{0}(\Theta)}{\Upsilon'(\Theta)/\Upsilon(\Theta)}=1,
\end{equation}
for positive $m$ and $n$. We extrapolate this limit to positive and negative values of the poles $\Theta=\pm (mb+n/b)$. Hence, we can write
\begin{eqnarray}\label{CH1A}
\lim_{\Theta\rightarrow m b+n/b}\frac{\mathcal{P}_{0}(\pm\Theta)}{\Upsilon'(\pm\Theta)/\Upsilon(\pm\Theta)}=1.
\end{eqnarray}
In addition, we consider that because the poles of $\mathcal{P}_{0}(\Theta)$ are $\Theta= \pm(mb+n/b)$, the integral $\mathcal{P}_{0}(\Theta)$ evaluated close to them satisfies 
\begin{eqnarray}\label{CH1B}
\lim_{\Theta \rightarrow mb+n/b} \frac{\mathcal{P}_{0}(\Theta)}{\mathcal{P}_{0}(-\Theta)}= 1.
\end{eqnarray} 
Using eqs.(\ref{CH1A}) and (\ref{CH1B}) into eq.(\ref{N3.1}) evaluated at its poles, we have
\begin{eqnarray}
\mathcal{N}_{3}&=&\lim_{\Theta\rightarrow mb+n/b}\frac{1}{8}\left(\mu \phi_{b} \right)^{\frac{\Theta}{b}}\Upsilon'(0)\frac{\mathcal{P}_{0}(\Theta)}{\Upsilon'(-\Theta)}\frac{\mathcal{P}_{0}(-\Theta)}{\mathcal{P}_{0}(\Theta)}\,\prod_{I=1}^{N=3}\frac{\Upsilon(2\alpha_{I})}{\Upsilon(\sum_{J}\alpha_{J}-2\alpha_{I})}\nonumber\\
&=&\lim_{\Theta\rightarrow mb+n/b}\frac{1}{8}\left(\mu \phi_{b} \right)^{\frac{\Theta}{b}}\Upsilon'(0)\frac{\mathcal{P}_{0}(-\Theta)}{\Upsilon'(-\Theta)}\,\prod_{I=1}^{N=3}\frac{\Upsilon(2\alpha_{I})}{\Upsilon(\sum_{J}\alpha_{J}-2\alpha_{I})}\nonumber\\
&=&\lim_{\Theta\rightarrow mb+n/b =Q-\sum_{I}\alpha_{I}}\frac{1}{8}\left(\mu \phi_{b} \right)^{\frac{\Theta}{b}}\frac{\Upsilon'(0)}{\Upsilon(-\Theta)}\,\prod_{I=1}^{N=3}\frac{\Upsilon(2\alpha_{I})}{\Upsilon(\sum_{J}\alpha_{J}-2\alpha_{I})}\label{AAAAA}.
\end{eqnarray}
In this way, from eq.(\ref{AAAAA}) we obtain for $\mathcal{N}_{3}$ 
\begin{eqnarray}\label{N3Final}
\mathcal{N}_{3}=\frac{1}{8}\left(\mu \phi_{b} \right)^{\frac{Q-\sum_{I}\alpha_{I}}{b}}\frac{\Upsilon'(0)}{\Upsilon(\sum_{I}\alpha_{I}-Q)}\prod_{I=1}^{N=3}\frac{\Upsilon(2\alpha_{I})}{\Upsilon(\sum_{J}\alpha_{J}-2\alpha_{I})}.
\end{eqnarray}
Additionally, from eq.(\ref{phib}) we have
\begin{eqnarray}\label{phibNew}
\phi_{b}&=&\pi b \frac{\Upsilon(2b)}{\Upsilon(b)}\nonumber\\
&=&\pi b \gamma(b^{2})\,b^{1-2b^{2}}.
\end{eqnarray} 
Finally, substituting eqs.(\ref{phibNew}), (\ref{N3Final}) into eq.(\ref{G3}) we achieve the DOZZ formula
\begin{eqnarray}
\mathcal{H}_{3}(z_{1}, z_{2}, z_{3}; \alpha_{1}, \alpha_{2}, \alpha_{3})&=&\left[\pi\mu\gamma_{(b^{2})}b^{(2-2b^{2})}\right]^{(Q-\sum_{I}\alpha_{I})/b}
\frac{\Upsilon'(0)}{\Upsilon(\sum_{I}\alpha_{I}-Q)}\prod_{I=1}^{3}\frac{\Upsilon(2\alpha_{I})}{\Upsilon(\sum_{J}\alpha_{J}-2\alpha_{I})}\times\nonumber\\
&&\left|z_{12}\right|^{2(\Delta_{3}- \Delta_{1}-\Delta_{2})}\left|z_{23}\right|^{2(\Delta_{1}- \Delta_{3}-\Delta_{2})}\left|z_{31}\right|^{2(\Delta_{2}- \Delta_{3}-\Delta_{1})}.\nonumber\\
\label{DOZZ}
\end{eqnarray}
In order for eq.(\ref{DOZZ}) to make contact with the usual form of the DOZZ formula, we reabsorbed the $1/8$ factor by redefining $\mu$. 
\chapter{$\mathcal{N}=1$ Supersymmetric Liouville Theory} 

\label{Chapter5} 

\lhead{Chapter 5. \emph{$\mathcal{N}=1$ Supersymmetric Liouville Theory}} 

\section{Introduction}

So far we have studied the bosonic Liouville theory, our next task is to extend it to the fermionic case.

It is well known that particles can be classified as {\it bosons} and {\it fermions}. It is also known that bosons have an integer spin value $s \in \mathbb{Z}^{+}_{0}$. In contrast, fermions posses a half-integer spin value, this is $s \in \mathbb{Z}^{+}_{0}+\frac{1}{2}$. This fact gives rise to very different physical properties between bosons and fermions.

Additionally, particles such as electrons, neutrinos (leptons) and quarks, which form the usual matter, are fermions. Thus, in order to have a more realistic toy model about two dimensional quantum gravity, we need to incorporate fermions to Liouville theory. To achieve this, we will need to introduce a new kind of symmetry which relates bosons and fermions. Such a new symmetry is named {\it supersymmetry}. 

Supersymmetry gives rise to a conserved charge by appealing to the Noether theorem. We label such a conserved charge by $Q$. As usual it is promoted to the status of the generator of this symmetry. At this point, we find the basic feature of supersymmetry. The generator $\widehat{Q}$ acting on a bosonic state turns it into a fermionic one, and vice versa. Mathematically it means,
\begin{eqnarray}
\widehat{Q}|\text{Boson}\rangle &=& |\text{Fermion}\rangle,\\
\widehat{Q}|\text{Fermion}\rangle &=& |\text{Boson}\rangle.
\end{eqnarray}    

\section{Supersymmetry}

\subsection{Poincaré group}

In Chapter \ref{Chapter1}, we studied the group arising from conformal invariance, it was called Conformal group. It was also noted that such a group includes the so-called Poincaré group as a sub-group. 

In addition, it is well known that the Poincaré group is built up by the Lorentz group and translations. For instance, under a transformation $x^{\mu}\rightarrow x'^{\mu}$, we have   
\begin{equation}\label{Poincaresymmetry1}
x'^{\mu}=\underbrace{\Lambda^{\mu}_{\nu}\,x^{\nu}}_{\text{Lorentz transf.}}+\underbrace{\xi^{\mu}}_{\text{Translations}},
\end{equation} 
where $\Lambda^{\mu}_{\nu}$ is a boost and $\xi^{\nu}$ is a translation in space-time. For a four dimensional flat space-time, where $\eta_{\mu\nu}$ is the Minkowski metric, we can express the Lorentz and translation generators as:
\begin{eqnarray}
(M^{\rho\sigma})^{\mu}_{\nu}&=&i(\eta^{\mu\sigma}\delta^{\rho}_{\nu}-\eta^{\mu\rho}\delta^{\sigma}_{\nu}),
\label{Mtensorrepr}\\
P_{\mu}&=&-i \partial_{\mu}.\label{Prepres}
\end{eqnarray}
These generators give rise to the following commutation relations
\begin{eqnarray}
\left[ P^{\mu}, P^{\nu}\right]   &=& 0\label{PPcommutator}\\
\left[ M^{\mu\nu}, P^{\sigma}\right]  &=& i(P^{\mu}\eta^{\nu\sigma}-P^{\nu}\eta^{\mu\sigma})\label{MPcommutator}\\
\left[ M^{\mu\nu}, M^{\rho\sigma}\right]  &=& i(M^{\mu\sigma}\eta^{\nu\rho}+M^{\nu\rho}\eta^{\mu\sigma}-M^{\mu\rho}\eta^{\nu\sigma}-M^{\nu\sigma}\eta^{\mu\rho})\label{MMcommutator},
\end{eqnarray}
which form the Poincar\'e algebra. From the latter commutators, we realize that $ M^{\mu\nu}$ or $P^{\nu}$ by themselves do not close the algebra. To do this eq.(\ref{MPcommutator}) is necessary. Such a mixture commutator is associated to the Thomas precession \cite{ryder}.

\subsection{Lorentz group and Spinors representation}

In four space-time dimensions, the Lorentz algebra has six generators. They are: three spatial rotations (combinations of $x y, yz$ and $zx$) and three boosts ({\it rotations} around $tx, ty$ and $tz$). Following the usual language given in the literature \cite{bilal}\cite{lambert}\cite{bertolini}\cite{quevedo} \cite{wess}, we denote these generators by $J_{i}$ and $K_{i}$, $i=1,2,3$ , for rotations and boosts, respectively. 

In addition, $J_{i}$ and $K_{i}$ satisfy the following commutation relations:
\begin{eqnarray}
\left[ J_{i}, J_{j}\right]  &=& i\epsilon_{ijk}J_{k},\label{JK1}\\
\left[ K_{i}, K_{j}\right]  &=& -i\epsilon_{ijk}J_{k},\label{JK2}\\
\left[ J_{i}, K_{j}\right]  &=& i\epsilon_{ijk}K_{j}.\label{JK3}
\end{eqnarray}
At this point it is usual to introduce the complex linear combinations $J^{\pm}_{i}$. These operators are defined by the following expressions,
\begin{equation}\label{Jmasmenos}
J^{\pm}_{i}=\frac{1}{2}\left(J_{i}\pm i K_{i} \right). 
\end{equation}
By replacing eq.(\ref{Jmasmenos}) into eqs.(\ref{JK1}), (\ref{JK2}) and (\ref{JK3}), it is not hard to see that the commutation relations which describe the Lorentz group are described now by two commuting $SU(2)$ algebras. We have
\begin{eqnarray}
\left[ J^{\pm}_{i}, J^{\pm}_{j}\right] &=& i \epsilon_{ijk}J^{\pm}_{k},\\
\left[ J^{\pm}_{j}, J^{\mp}_{j} \right] &=& 0. 
\end{eqnarray} 
As a result, the Lorentz group $SO(3,1)$ can be expressed as the direct sum $SU(2) \otimes SU(2)$.

We also point out that eqs.(\ref{JK1}), (\ref{JK2}) and (\ref{JK3}), together with the following equations
\begin{eqnarray}
\left[ J_{i}, P_{j} \right] &=& i \epsilon_{ijk}P_{k}\hspace{0.2cm},\hspace{1.3cm} \left[ J_{i}, P_{0}\right] = 0,\\
\left[ K_{i}, P_{j} \right] &=& -i P_{0}\hspace{0.44cm},\hspace{1.3cm} \left[ K_{i}, P_{0}\right] = -i P_{j}.
\end{eqnarray}
give rise to the more compact expressions eqs.(\ref{MPcommutator}) and (\ref{MMcommutator}), by identifying $M_{0i}=K_{i}$ and $M_{ij}=\epsilon_{ijk}J_{k}$.

Moreover, a homomorphism exists between the Lorentz group $SO(3,1)$ and $SL(2,\mathbb{C})$. In this particular case, for each element of $SO(3,1)$ there are two elements of $SL(2, \mathbb{C})$.  In mathematical terms, $SL(2, \mathbb{C})$ is the universal cover of the Lorentz group. To explicitly see the homomorphism between $SO(3,1)$ and $SL(2,\mathbb{C})$, we make use of the Pauli matrices which are given by
\begin{eqnarray}
\sigma^{0}&=&
\left( \begin{matrix}
1&&0\\
0&&1
\end{matrix}\right),
\hspace{1cm}
\sigma^{1}=
\left( \begin{matrix}
0&&1\\
1&&0
\end{matrix}\right),\label{Pauli1}\\
\vspace{2cm}
\sigma^{2}&=&
\left( \begin{matrix}
0&&-i\\
i&&0
\end{matrix}\right),
\hspace{0.8cm}
\sigma^{3}=
\left( \begin{matrix}
1&&0\\
0&&-1
\end{matrix}\right).\label{Pauli2}
\end{eqnarray}
In addition, the function $X=x_{\mu}\sigma^{\mu}$ is defined together with the usual expression for the four-position $x^{\mu}=(x^{0}, x^{1}, x^{2}, x^{3})$. Such a function $X$ is explicitly written by using Pauli matrices as,
\begin{eqnarray}\label{DeterminantX}
X=\left( 
\begin{matrix}
x_{0}+x_{3} && x_{1}-ix_{2}\\
x_{1}+ix_{2}&& x_{0}-x_{3}
\end{matrix}\right). 
\end{eqnarray}
We also know that the four position norm $|x|$ under a Lorentz transformation $x \rightarrow \Lambda x$ is an invariant. Thus, we have:
\begin{equation}\label{invariantinterval}
|x'|^{2}=|\Lambda x|^{2}=x_{0}^{2}-x_{1}^{2}-x_{2}^{2}-x_{3}^{2} 
\end{equation}
Furthermore, by calling $G$ to an element belonging to $SL(2, \mathbb{C})$, we find that the mapping $X \rightarrow G X G^{\dagger}$ preserves the determinant of eq.(\ref{DeterminantX}), which is exactly the invariant interval given in eq.(\ref{invariantinterval}):
\begin{equation}\label{Gdeterminant}
\det X'=\det (G X G^{\dagger}) = x_{0}^{2}-x_{1}^{2}-x_{2}^{2}-x_{3}^{2}.  
\end{equation}
As already mentioned, in order to obtain eq.(\ref{Gdeterminant}) we require the homomorphism between $SO(3,1)$ and $SL(2, \mathbb{C})$ must be $1-2$. For example, in case of the trivial identity transformation of the Lorentz group $\Lambda=\mathbf{1} \in SO(3,1) $, the corresponding elements for $SL(2, \mathbb{C})$ are $G=\pm \mathbf{1}$ \cite{quevedo}. This is basically the origin of the two inequivalent representations when we write the Lorentz group by using $G\in SL(2, \mathbb{C})$. In the literature that representation is called spinor representation. 

In this way, we conclude that in spinor representation there are two inequivalent representations of the Lorentz group. 

Now, let us introduce the mathematical objects called {\it spinors} by defining them as the objects $\psi$ transforming under a matrix transformation $G\in SL(2, \mathbb{C})$. On the other hand, because elements of $SL(2, \mathbb{C})$ are $2\times 2$ matrices, the spinor $\psi$ has to be of the form:
\begin{eqnarray}
\psi=
\left( \begin{matrix}
\psi_{1}\\
\psi_{2}
\end{matrix}\right). 
\end{eqnarray}
Therefore, it is usual in QFT to express the transformation of $\psi$, in spinor representation, by labelling its components. For instance, if $G\in SL(2, \mathbb{C})$ we have
\begin{eqnarray}\label{SpinortransfG}
\psi_{\alpha}\rightarrow \psi'_{\alpha}=G_{\alpha}^{\beta}\psi_{\beta},
\end{eqnarray}
where $\alpha, \beta=1, 2$. 

Recalling that we have found two inequivalent representations in $SL(2, \mathbb{C})$. We will label the matrices belonging to each representation by $G$ and $G^{*}$. These representations are called \cite{quevedo}:
\begin{enumerate}
\item Fundamental representation 
\begin{equation}
\psi'_{\alpha}=G_{\alpha}^{\beta}\psi_{\beta}.
\end{equation}
\item Conjugate representation\footnote{It is usual to denote indices in the conjugate representation by dotted indices; however we will not care about this point from next section on.}
\begin{equation}
\overline{\chi}'_{\dot{\alpha}} = G_{\dot{\alpha}}^{*\,\dot{\beta}}\overline{\chi}_{\dot{\beta}}.
\end{equation}
\end{enumerate}
In addition, we have another non-independent representation:
\begin{enumerate}
\item[3.] Contravariant representation
\begin{eqnarray}
\psi'^{\alpha}=\psi^{\beta}(G^{-1})^{\alpha}_{\beta}\hspace{0.5cm},\hspace{1,2cm}\overline{\chi}'^{\alpha} = \overline{\chi}^{\beta}(G^{*\,-1})^{\alpha}_{\beta}.
\end{eqnarray}
\end{enumerate}
Sometimes, the representation carried out by the matrices $G_{\alpha}^{\beta}$ is denoted as $\left(\frac{1}{2},0 \right) $. In contrast, for the matrices $G_{\alpha}^{*\beta}$ we have $\left(0, \frac{1}{2} \right)$.

Analogously to the metric tensor $\eta^{\mu\nu}$, which is an invariant under $SO(3,1)$, to raise or lower indices within $SL(2,\mathbb{C})$, we use 
\begin{eqnarray}\label{fermionicmetric}
g^{\alpha\beta}=
\left( 
\begin{matrix}
0&&1\\
-1&&0
\end{matrix}\right), 
\end{eqnarray}
where $g^{\alpha\beta}=-g_{\alpha\beta}$. In the context of particle physics, eq.(\ref{fermionicmetric}) is written as $\varepsilon$ instead of $g$. Once in a while eq.(\ref{fermionicmetric}) is called the flat {\it fermionic} metric. Nonetheless, to build a generalization of the bosonic metric in the supersymmetric case it is convenient to contemplate eq.(\ref{fermionicmetric}) as a metric which raises or lower indices. Hence, we have the next transformations for the fundamental and conjugate representations
\begin{eqnarray}
\psi^{\alpha}=g^{\alpha\beta}\psi_{\beta},\label{Psi1}\\
\overline{\chi}^{\alpha}=g^{\alpha\beta}\overline{\chi}_{\beta}.\label{Psi2}
\end{eqnarray}
Therefore, it is easy to see that contravariant representation is not independent.

We could also introduce the above results in a four-component formalism. For instance, if we define the Dirac matrices as
\begin{eqnarray}\label{DiracMatrix}
\gamma^{\mu}=\left( 
\begin{matrix}
0&&\sigma^{\alpha}\\
\overline{\sigma}^{\mu}&&0
\end{matrix}\right), 
\end{eqnarray}  
and $\gamma_{5}=i\gamma^{0}\gamma^{1}\gamma^{2}\gamma^{3}$. The four-component Dirac spinor is made from the two component spinors $\psi_{\alpha}$ and $\overline{\chi}^{\alpha}$:
\begin{eqnarray}
\left( 
\begin{matrix}
\psi_{\alpha}\\
\overline{\chi}^{\alpha}
\end{matrix}\right). 
\end{eqnarray}
These spinors, which has a defined chirality, are called Weyl spinors. Another kind of spinors are Majorana spinors, they are reached by imposing $\chi=\psi$ in the definition of Dirac spinors, thus the four component spinor is rewritten as
\begin{eqnarray}
\left( 
\begin{matrix}
\psi_{\alpha}\\
\overline{\psi}^{\alpha}
\end{matrix}\right). 
\end{eqnarray}
It is well known that only in a certain number of dimensions it can be achieved to have both type of spinors. For instance, it is possible in two and ten dimensions. This is an important point because we will deal with spinors in two dimensions, then they could be Majorana-Weyl spinors and this fact will reduce some further calculations.

The Lorentz generator given by eq.(\ref{Mtensorrepr}), in a tensor representation, is going to be expressed in spinor representation. It means that we rewrite $M^{\rho\sigma}$ by using the Dirac matrices. Hence, we have:
\begin{eqnarray}\label{Mspinorrep}
(M^{\rho\sigma})^{\mu}_{\nu}=\frac{i}{2}(\gamma^{\rho\sigma})^{\mu}_{\nu}=\frac{i}{4}\left(\gamma^{\rho}\gamma^{\sigma}-\gamma^{\sigma}\gamma^{\rho} \right)^{\mu}_{\nu}. 
\end{eqnarray}  
Of course, eq.(\ref{Mspinorrep}) satisfies the commutators given by eqs.(\ref{MMcommutator}) and (\ref{MPcommutator}). Replacing eq.(\ref{DiracMatrix}) into eq.(\ref{Mspinorrep}), it is no hard to obtain
\begin{equation}
(M^{\rho\sigma})^{\mu}_{\nu}=
\left( 
\begin{matrix}
(\sigma^{\rho\lambda})^{\beta}_{\alpha}&&0\\
0&&(\overline{\sigma}^{\rho\lambda})^{\dot{\beta}}_{\dot{\alpha}}
\end{matrix}\right),
\end{equation}
where we have defined its components as: 
\begin{equation} 
\begin{aligned}
(\sigma^{\rho\lambda})^{\beta}_{\alpha}& = \frac{i}{4}(\sigma^{\rho}\overline{\sigma}^{\lambda}-\sigma^{\lambda}\overline{\sigma}^{\rho})^{\beta}_{\alpha},\\
(\overline{\sigma}^{\rho\lambda})^{\dot{\beta}}_{\dot{\alpha}} & = \frac{i}{4}(\overline{\sigma}^{\rho}\sigma^{\lambda}-\overline{\sigma}^{\lambda}\sigma^{\rho})^{\dot{\beta}}
_{\dot{\alpha}},
\end{aligned}
\end{equation}
In the latter expressions for $(\sigma^{\rho\lambda})^{\beta}_{\alpha}$ and $(\overline{\sigma}^{\rho\lambda})^{\dot{\beta}}_{\dot{\alpha}}$, each one satisfies the same commutation relations as eq.(\ref{Mtensorrepr}) or eq.(\ref{Mspinorrep}). As we already said, it is because we can express the covering group $SL(2, \mathbb{C})$ as $SU(2)\otimes SU(2)$.

From the next section on, we will focus on spinors in two dimensions. Furthermore, they will be chosen to be Majora-Weyl spinors.   



\section{Spinors in flat and curved space-time}
 
As in the bosonic case, the supersymmetric extension of Liouville theory also lies in the field of string theory. Therefore, it is quite natural to think that we will need to incorporate fermionic degrees of freedom into the string. In order to perform this, we must couple fermions to the world-sheet; thus, we will deal with two dimensional spinors, as mentioned earlier.

Following the usual references in this field \cite{hatfield} \cite{wess} \cite{hoker1} \cite{Martinec} \cite{DistlerKawai} \cite{hoker2} \cite{3}, we adopt the next convention for index notation: 

\begin{itemize}
\item letters from the middle of the Greek alphabet, such as $\mu, \nu, \dots$, will be used to label space-time indices,

\item letters at the beginning of the Greek alphabet, such as $\alpha, \beta, \dots$ for spinorial indices,

\item lower-case letters at the beginning of the Latin alphabet, such as $a, b, \dots$, until $l$ for world-sheet indices,

\item lower-case letters at the middle of the Latin alphabet, such as $m, n, \dots$, until $y$ for vectors belong to $T_{x}(M)$.
\end{itemize}

To begin with, we consider two general mathematical statements about general coordinate transformations in a $d$-dimensional space-time \cite{hatfield} \cite{hoker2}:

\begin{enumerate}
\item Under a general space-time coordinate transformations, the bosonic fields will transform under a representation of $GL(d,\mathbb{R})$ \footnote{The general linear group $GL(d,\mathbb{R})$ is built up of $d\times d$-matrices of real numbers, whose determinant is non-zero.}.

\item There are no spinorial representation of $GL(d,\mathbb{R})$.

\end{enumerate}

Therefore, under a change of coordinates $x^{\mu}\rightarrow x'^{\mu}$ we have
\begin{equation}
v^{\nu'}=\frac{\partial x^{\nu'}}{\partial x^{\mu}} v^{\mu}= T^{\nu'}_{\mu} v^{\mu}.
\end{equation}
In the case of spinorial representations, the bilinear covariant expression $\overline{\psi}(x)\gamma^{\mu}\psi(x)$ needs a similar transformation:
\begin{equation}\label{SpinorsCurvedTransf}
\overline{\psi}(x')\gamma^{\nu'}\psi(x')= T^{\nu'}_{\mu}  \overline{\psi}(x)\gamma^{\mu}\psi(x).
\end{equation} 
From eq.(\ref{SpinorsCurvedTransf}), we realize that for the flat $d=4$ case the only requirement is Lorentz invariance. We saw that there is a spinorial representation of the Lorentz group $SO(3,1)$, this is given by
\begin{equation}
\overline{\psi}(x')\gamma^{n'}\psi(x') = \Lambda^{n'}_{m} \overline{\psi}(x)\gamma^{m}\psi(x),
\end{equation}
where $\Lambda^{m'}_{m}$ are Lorentz transformations expressed in the form of a boost or its Euclidean equivalent an $SO(n)$ element. However, without spinorial representations in the general case, we cannot accomplish it in a straightforward manner.

How can we avoid this problem?, we will see in the next subsection that the solution relays in the fact that any space-time is locally flat and introduce, at the tangent space, the so-called moving frame field.

\subsubsection{Moving frame field $e^{\mu}_{m}$}

As in general relativity, a moving frame field is a choice of an orthonormal basis of tangent vectors $e^{\mu}_{m}$ living on the tangent space $T_{x}(M)$ of a manifold $M$. That orthonormal basis field $e^{\mu}_{m}$ is called:

\begin{enumerate}
\item Zweibein in 2 dimensions
\item Vierbein in 4 dimensions
\item Vielbein in $n$ dimensions.
\end{enumerate}

We already mentioned above that the Greek indices from the middle of the alphabet indicate that $e^{\mu}_{m}$ transforms as an space-time vector embedded in a curved background. On the other hand, the lower-case latin letter indicates that $e^{\mu}_{m}$ transforms as a vector living on a flat tangent space. Hence, the well known orthonormality implies,
\begin{equation}
g_{\mu\nu}e^{\mu}_{m}e^{\nu}_{n}=\eta_{mn}.
\end{equation}
In addition, associated to each moving vector frame there is an orthonormal basis $e_{\nu}^{n}$ which belongs to the dual space or also called cotangent space $T^{*}_{x}(M)$. We have the next relations between $e^{\mu}_{m}$ and $e_{\nu}^{n}$:
\begin{eqnarray}
e^{\mu}_{m}e_{\nu}^{m}&=&\delta^{\mu}_{\nu},\label{munu}\\
e_{\mu}^{m}e^{\mu}_{n}&=&\delta^{m}_{n}.\label{mn}
\end{eqnarray}
Since the lower index $m$ on $e^{\mu}_{m}$ transforms under a Lorentz transformation as
\begin{equation}\label{Trans1}
e^{\mu}_{n'}=\Lambda^{m}_{n'} e^{\mu}_{m},
\end{equation}
the moving frame field is certainly not unique. Note that this transformation occurs in the tangent space of one point $x\in M$. 

Under a change of coordinates, the $\mu$ index still transforms as a space-time vector,
\begin{equation}\label{Transf2}
e^{\nu'}_{m}=T^{\nu'}_{\mu} e^{\mu}_{m}.
\end{equation}
Transformations given by eqs.(\ref{Trans1}) and (\ref{Transf2}) allow us to satisfy eq.(\ref{SpinorsCurvedTransf}). Therefore, we define the desired curved spinors as
\begin{equation}\label{gammaTranf}
\gamma^{\mu}(x)=e^{\mu}_{m}(x)\gamma^{m},
\end{equation} 
where $\gamma^{m}$ are the usual flat space Dirac matrices satisfying $\{\gamma^{m},\gamma^{n}\}=2\eta^{m\,n}$. By using these transformations, we can get
\begin{eqnarray}
\{\gamma^{\mu}(x),\gamma^{\nu}(x)\}&=&2g^{\mu\,\nu}(x),\\
\overline{\psi}(x')\gamma^{\nu'}\psi(x')&=& T^{\nu'}_{\mu}  \overline{\psi}(x)\gamma^{\mu}\psi(x).
\end{eqnarray} 
Therefore, to couple fermions in an invariant manner we must use $\gamma^{\mu}(x)$ as defined by eq.(\ref{gammaTranf}). 

Clearly, we are free to perform a different Lorentz transformation on each tangent space and still get the same result. To avoid this, we must gauge this local Lorentz transformation by introducing a gauge field which is called the spin connection.

\subsection{Covariant derivative $D_{\mu}$ and spin connection $s_{\mu}$.}

As stated above, in order to avoid the ambiguity arising from local Lorentz transformation in $T_{x}(M)$ we introduce a gauge field. That gauge field is a connection called the spin connection and written as $s_{\mu}$. In the same way as in general relativity, such a connection gives rise a covariant derivative. That covariant derivative is defined by,
\begin{equation}
D_{\mu}\psi=(\partial_{\mu}+s_{\mu})\psi.
\end{equation}
It is worthwhile to note that $s_{\mu}$ must transform as follows,
\begin{equation}
s_{\mu}\rightarrow G s_{\mu} G^{-1}-(\partial_{\mu}G)G^{-1},
\end{equation}
to maintain covariant $D_{\mu}$ under $\psi \rightarrow G(\Lambda)\psi$. Here, $G(\Lambda)$ is a spinorial representation of the Lorentz group as in eq.(\ref{SpinortransfG}).

As in the well known case of Yang-Mills theory, the connection $s_{\mu}$ is expanded in terms of the gauge group generators. For us, the spinorial representation of the Lorentz group
\begin{equation}\label{Spinconnection}
s_{\mu}=\frac{1}{4}s_{\mu}^{mn}\sigma_{mn},
\end{equation}
where $\sigma_{mn}=[\gamma^{m},\gamma^{n}]/2$. In eq.(\ref{Spinconnection}) the gamma matrices are the Dirac ones in an arbitrary number of dimensions. In particular, we will use $\gamma^{0}=\sigma^{2}$ and $\gamma^{1}=i\sigma^{1}$ for the two dimensional case.

Our next task will be to study all these features in the so-called Ramond-Neveu-Schwarz (RNS) action. That action possess the two dimensional supersymmetry we require and describes the superstring. Nonetheless, the RNS action is still a two dimensional supersymmetrical theory which does not explicitly have the supersymmetric behaviour yet. 

\subsection{The Ramond-Neveu-Schwarz action $S_{RNS}$}

We begin by writing the RNS action in Minkowski signature space-time \cite{hatfield}, then we change to Euclidean space by performing the well known Wick rotation. Therefore, the RNS action is given as follows
\begin{eqnarray}\label{RNS1}
S_{RNS}&=&-\frac{1}{2}\int d\sigma d\tau \sqrt{-g}\left(g^{ab}\partial_{a}x^{\mu}\partial_{b}x_{\mu}-i\overline{\psi}^{\mu}\gamma^{a}\partial_{a}\psi_{\mu}\right)\nonumber\\
&&-\frac{1}{2}\int d\sigma d\tau \sqrt{-g}\overline{\chi}_{a}\gamma^{b}\gamma^{a}\psi^{\mu}\partial_{b}x_{\mu}\nonumber\\
&&-\frac{1}{8}\int d\sigma d\tau \sqrt{g}\overline{\psi}_{\mu}
\psi^{\mu}\overline{\chi}_{a}\gamma^{b}\gamma^{a}\chi_{b}. 
\end{eqnarray}
In order to explicitly get the spin connection given in eq.(\ref{Spinconnection}), for the flat 2 dimensional Minkowski space. We use the two $\gamma$ matrices, 
\begin{equation}
\gamma^{0}=
\begin{pmatrix}
0 & -i\\
i & 0\\
\end{pmatrix}
,\hspace{1cm}\gamma^{1}=
\begin{pmatrix}
0 & i\\
i & 0\\
\end{pmatrix}, 
\end{equation}
these matrices satisfy $\{\gamma^{m},\gamma^{n}\}=2\eta^{mn}$. We should realize that they are related to the sigma matrices given in eqs.(\ref{Pauli1}) and (\ref{Pauli2}) by $\gamma^{0}=\sigma^{2}$ and $\gamma^{1}=i\sigma^{1}$. It is also useful to define $\gamma^{3} = \gamma^{0}\gamma^{1} = \sigma^{3}$ and $\sigma_{01} = \gamma_{3}$.

Hence, the covariant derivative in two dimensions is expressed as
\begin{equation}\label{covariantderivative}
D_{a}\psi=\partial_{a}\psi+\frac{1}{2}\omega_{a}\gamma_{5}\psi.
\end{equation} 
From the relation for curved $\gamma^{a}$ matrices,
\begin{equation}\label{curvedgamma}
\gamma^{a}=e^{a}_{m}\gamma^{m}
\end{equation}
where $e^{a}_{m}$ is the world-sheet zweibein (two dimensions). The moving frame is related to the world sheet metric by the relation:
\begin{equation}
g^{ab}=e^{a}_{m}e^{b}_{n}\eta^{mn}.
\end{equation}
From the latter expression, we are able to get
\begin{equation}\label{geeRelation}
\sqrt{-g}=\sqrt{|\det g_{ab}|}=e=\det(e^{m}_{a}).
\end{equation}
Eq.(\ref{geeRelation}) ensures that $S_{RNS}$ is a reparametrization invariant action.

It was already mentioned that the RNS action has the desired supersymmetry but in a non-explicit way. To see it, we will study the symmetries present into eq.(\ref{RNS1}). 

\subsection{Symmetries of the RNS action}

By defining an infinitesimal arbitrary Majorana spinor $\alpha(\sigma,\tau)$, the symmetries of the RNS action are given by the next relations: 
\begin{eqnarray}
\hspace{-1cm}\delta x^{\mu}&=&\overline{\alpha}\psi^{\mu},\hspace{0.5cm} \delta\psi^{\mu}=-i\gamma^{a}\alpha(\partial_{a}x^{\mu}-\frac{1}{2}\overline{\psi}^{\mu}\chi_{a}),\label{SS1}\\
\hspace{-1cm}\delta\chi_{a}&=&2D_{a}\alpha,\hspace{0.5cm}  \delta e^{m}_{a}=-i\overline{\alpha}\gamma^{m}\chi_{a}\label{SS2}.
\end{eqnarray}
It is easy to realize that these transformations are relating bosonic components, such as $x^{\mu}$, to fermionic ones, such as $\psi^{\mu}$. In order to get the invariance of $S_{RNS}$ under the latter equations, we make use of some gamma matrices properties in two dimensions, for instance we have
\begin{eqnarray}
\overline{\psi}\gamma^{m}\gamma_{n}\gamma_{p}\psi = 0,\label{2DrelationGamma1}\\
\omega^{np}_{m}\overline{\psi}\gamma_{m}\sigma_{np}\psi = 0\label{2DrelationGamma2}.
\end{eqnarray}
Eqs.(\ref{2DrelationGamma1}) and (\ref{2DrelationGamma2}) explains why the covariant derivative does not appear in the RNS action eq.(\ref{RNS1}).

As stated above, the symmetry arising from the invariance of the RNS action under eqs.(\ref{SS1}) and (\ref{SS2}) has a special name, this is called supersymmetry. In summary, from those equations we can realize that local world-sheet supersymmetry is parametrized by an infinitesimal arbitrary Majorana spinor and the RNS action is invariant under the local supersymmetry transformations given above.

In addition to local supersymmetry, the action is invariant under the Weyl transformations
\begin{eqnarray}
x^{\mu} &\rightarrow & x^{\mu},\hspace{0.8cm} e^{m}_{a} \rightarrow \Omega e^{m}_{a},\label{Weyl1}\\
\psi^{\mu} &\rightarrow & \Omega^{-1/2}\psi^{\mu},\hspace{0.4cm} \chi_{a} \rightarrow \Omega^{1/2}\chi_{a},\label{Weyl2}
\end{eqnarray}
where $\Omega=\Omega(\sigma, \tau)$ is an arbitrary function. 

We point out that in the context of supersymmetry, the $\chi_{a}$ field is called: the gravitino field. It is a field with spin $3/2$ and also the supersymmetric partner of $e_{a}^{m}$.

Finally, the RNS action is also invariant under 
\begin{equation}\label{Chitransf}
\chi_{a}\rightarrow \chi_{a} + \gamma_{a}\eta,
\end{equation}
where $\eta(\sigma,\tau)$ is also an arbitrary Majorana spinor. To derive the invariance eq.(\ref{Chitransf}), we use another property of two dimensional gamma matrices:
\begin{equation}
\gamma^{m}\gamma_{n}\gamma_{m}=0.
\end{equation}
The symmetry arising from eqs.(\ref{Weyl1}), (\ref{Weyl2}) and (\ref{Chitransf}) together means that the action describes a superconformal theory.

In summary, the RNS action possesses:
\begin{enumerate}
\item World-sheet reparametrization invariance 
\item Local Lorentz invariance at the world-sheet
\item Local world-sheet supersymmetry
\item Super Weyl invariance.
\end{enumerate}
In addition, $e^{m}_{a}$ and $\chi_{a}$ act like Lagrange multipliers. Therefore, their equations of motion generate the conservation of stress-energy tensor and supercurrent at the classical level, thus we have 
\begin{eqnarray}
T^{a}_{m}&=&-\frac{2}{e}\frac{\delta S_{RNS}}{\delta e^{m}_{a}} = 0,\\
J_{a}&=&\frac{1}{2}\gamma^{b}\gamma_{a}\psi^{\mu}\partial_{b}x_{\mu}=0.
\end{eqnarray}  
As our next point, we will deal with the Euclidean version of the RNS action. In order to achieve this, we will perform a Wick rotation over one of the two world-sheet coordinates. 

\subsection{The Euclidean RNS action}

After performing Wick rotation, we will identify the world-sheet coordinates $\sigma$ and $\tau$ as $\sigma=\xi^{1}$ and $\tau=-i\xi^{2}$ and to maintain the factor $\gamma^{m}\partial_{m}$ invariant, we must change $\gamma^{0}$ to be $\gamma^{0}=-i\gamma^{2}$, due to $\gamma^{2}\partial_{2}=\gamma^{0}\partial_{0}$. As a result, the anticommutation relation $\{\gamma^{p}, \gamma^{q}\}=2\eta^{pq}$ is now given by
\begin{equation}
\left\lbrace \gamma^{m}, \gamma^{n} \right\rbrace = -2\delta^{mn}.
\end{equation}
Moreover  $\gamma^{3}=i\gamma^{1}\gamma^{2}=\sigma^{3}$. Finally, the euclidean RNS action is as follows: 
\begin{eqnarray}
S^{E}_{RNS}&=&\frac{1}{2}\int d^{2}\xi \sqrt{g}\left( g^{ab}\partial_{a}x^{\mu}\partial_{b}x_{\mu} - i\psi_{\mu}\gamma^{a}\partial_{a}\psi^{\mu}\right)\nonumber\\
&&-\frac{1}{2}\int d^{2}\xi \sqrt{g}\left( \chi_{a}\gamma^{b}\gamma^{a}\psi_{\mu}\right)\left( \partial_{b}x^{\mu}-\frac{1}{4}(\chi_{b}\psi^{\mu})\right) .  
\end{eqnarray} 
Additionally, the superconformal gauge is defined as \cite{fermistring} \cite{hatfield}
\begin{equation}
e^{m}_{a}=\exp(\phi(\xi)/2)\delta^{m}_{a},\hspace{0.4cm} \chi^{a}=\gamma^{a}\chi,
\end{equation}
where $\chi$ is a constant spinor. In this gauge, $\chi$ disappears from the gauge fixed action. Hence, the action under this superconformal gauge is
\begin{equation}\label{RNS2}
S_{\text{sconf}}=\frac{1}{2}\int d^{2}\xi \left(\delta^{mn}\partial_{m}x^{\mu}\partial_{n}x_{\mu}-i\psi_{\mu}\gamma^{m}\partial_{m}\psi^{\mu} \right). 
\end{equation}
We should realize that there are still some symmetries. Such a remnant part of local supersymmetry is called {\it global} supersymmetry
\begin{equation}\label{globalSS}
\delta x^{\mu}=\varepsilon^{\alpha}\psi^{\mu}_{\alpha},\hspace{0.3cm}\delta\psi^{\mu}_{\alpha}=-i(\gamma^{m})_{\alpha}^{\beta}\varepsilon_{\beta}\partial_{m}x^{\mu},
\end{equation}
where $\varepsilon^{\alpha}$ is a constant spinor.

\section{Super Riemann Surfaces} 

It is not obvious that the actions given by eq.(\ref{RNS2}) and eq.(\ref{RNS1}) are globally and locally supersymmetric, respectively. Therefore, we want to rewrite these actions in a way which makes supersymmetry manifest. In order to do this, we first introduce the superspace formalism and later the supersurface language.

\subsection{Superspace}

The first attempts to introduce superspace were in 1974 by Salam and Strathdee \cite{salam1, salam2}. From its introduction its applications has been increasing. We adopt this formalism in the next lines.

Basically, when superspace is considered, we deal with an extension of the usual bosonic (commuting) coordinates to fermionic ones. In this way, our space is not only described by the coordinates $(z, \overline{z})$, it is now given by $(z, \overline{z}; \theta, \overline{\theta})$ \footnote{Recall we are working in two dimensions.} instead. Where $\theta$ and $\overline{\theta}$ are called: fermionic, Grassmann or simply {\it odd} variables. Their main properties are
\begin{eqnarray}
\theta \overline{\theta} = - \overline{\theta} \theta,\label{PropertieTHETA1}\\
\theta \theta = 0,\label{PropertieTHETA2}\\
\overline{\theta} \overline{\theta} = 0.\label{PropertieTHETA3}
\end{eqnarray}
Such properties have deep  consequences to this theory. For instance, if we define a function depending on $\theta$ only, say $F(\theta)$, we have that due to eq.(\ref{PropertieTHETA2}) only the first two terms do not vanish,
\begin{eqnarray}\label{superfieldexpansion}
F(\theta)= \sum_{n=0}^{\infty} F_{n}(\theta)^{n} = F_{0}+F_{1}\theta + \sum_{k=0}^{\infty} F_{k+2}\,(\theta)^{2} (\theta)^{k}= F_{0}+F_{1}\theta .
\end{eqnarray}
If there are no boundary terms the following properties are satisfied: 
\begin{eqnarray}
\int d\theta = 0,\label{0propertie}\\
\int d\theta \theta = 1.\label{1propertie} 
\end{eqnarray}
A direct consequence of eq.(\ref{1propertie}) is that $\delta(\theta)= \theta$. Another consequence is due to eqs.(\ref{superfieldexpansion}), (\ref{0propertie}) and (\ref{1propertie}), thus in odd variables we have an equivalence between integrals and partial derivatives  
\begin{eqnarray}
\int d\theta F(\theta)=\frac{\partial F}{\partial \theta}.
\end{eqnarray}
If a function as eq.(\ref{superfieldexpansion}) is extended to bosonic variables as well, it becomes a special function which is called a superfield. For instance, the superfield $X(z,\theta)$ is expanded as
\begin{eqnarray}\label{superfield}
X(z,\theta)= x(z) +  \theta \psi + \overline{\theta}\,\overline{\psi} +\overline{\theta}\theta f. 
\end{eqnarray}
From eq.(\ref{superfield}) it is not difficult to see that the first term is its bosonic part and the rest are constrained because of Grassmman variables.

In case of two variables, eq.(\ref{1propertie}) should be replaced to be  
\begin{eqnarray}
\int d\theta^{1}\int d\theta^{2} \theta^{2}\theta^{1}=1.
\end{eqnarray}

\subsection*{Why to introduce superspace?}

The main point of introducing superspace is because supersymmetry becomes a geometric transformation. Because odd variables are now points within the superspace, its transformations are a type of superdiffeomorphism. Hence, we have
\begin{eqnarray}
\delta z = -i \varepsilon^{\theta} \gamma_{\theta\theta}^{z}\theta = -\varepsilon^{\theta}\theta &&, \hspace{0.5cm}\delta\overline{z}=-i\varepsilon^{\overline{\theta}}
\gamma_{\overline{\theta}\,\overline{\theta}}^{\overline{z}}\overline{\theta}
=\varepsilon^{\overline{\theta}}\overline{\theta},\\
\delta\theta = \varepsilon^{\theta}&&, \hspace{0.5cm}\delta\overline{\theta} = \varepsilon^{\overline{\theta}},
\end{eqnarray}
where $\varepsilon$ is a Grassmann constant spinor. 

Under an infinitesimal supersymmetry transformation, the superfield $X$ such as eq.(\ref{superfield}) transforms as
\begin{eqnarray}
\delta X &=& \delta z \partial_{z}X + \delta \overline{z} \partial_{\overline{z}}X + \delta \theta \partial_{\theta}X +  \delta \overline{\theta} \partial_{\overline{\theta}}X  \\
&=& \varepsilon^{\theta}Q_{\theta}X + \varepsilon^{\overline{\theta}}Q_{\overline{\theta}}X,
\end{eqnarray}
where 
\begin{eqnarray}
Q_{\theta}&=&\partial_{\theta}-i\theta\gamma^{z}_{\theta\theta}\partial_{z}=\partial_{\theta}-\theta\partial_{z},\\
Q_{\overline{\theta}}&=&\partial_{\overline{\theta}}-i\overline{\theta}\gamma^{z}_{\overline{\theta}\,\overline{\theta}}\partial_{\overline{z}}=\partial_{\overline{\theta}} + \overline{\theta}\partial_{\overline{z}}.
\end{eqnarray}
On a flat supersurface, the supercovariant derivative $D_{\theta}$ is easy to construct. In order to satisfy the anticommutation relations $\{D_{\theta},Q_{\theta}\}=\{D_{\overline{\theta}},Q_{\overline{\theta}}\}=\{D_{\theta},Q_{\overline{\theta}}\}=0 $, we have
\begin{equation}
D_{\theta}=\partial_{\theta}+\theta\partial_{z}, \hspace{0.5cm}D_{\overline{\theta}}=\partial_{\overline{\theta}} - \overline{\theta}\partial_{\overline{z}}.
\end{equation}
Therefore, the quantity 
\begin{equation}\label{IFlat}
I=\int d^{2}z d^{2}\theta D_{\overline{\theta}}X D_{\theta}X,
\end{equation}
is supersymmetric invariant.

\subsection{Supersurface notation and flat metric}

We will follow the notation given in \cite{hatfield} \cite{hoker2} \cite{Martinec}. For $\mathcal{N}=1$ superspace, supercoordinates are labelled by $z^{M} = (z, \overline{z}; \theta, \overline{\theta})$ and superdifferentials by $\partial_{M} = (\partial_{z}, \partial_{\overline{z}}; \partial_{\theta}, \partial_{\overline{\theta}})$. Furthermore, tangent space indices are written as $A = (\eta, \overline{\eta}; +, -)$. On the other hand, local Lorentz transformation is restricted to the $SO(1,1)$ subgroup of the more general super-rotation. As a result, tensors transform with the $U(1)$ charge under local Lorentz transformation.

Flat supersurface has the metric
\begin{equation}\label{Flatsupermetric}
g_{MN}=
\begin{pmatrix}
\delta_{mn} & 0 \\
0 & \delta_{\alpha\beta}\\
\end{pmatrix},
\end{equation}
where,
\begin{equation}
\delta_{mn}=
\begin{pmatrix}
0 & 1 \\
1 & 0\\
\end{pmatrix},\hspace{1cm}
\delta_{\alpha\beta}=
\begin{pmatrix}
0 & 1 \\
-1 & 0\\
\end{pmatrix}. 
\end{equation}
The latter metrics are recognized as the flat two dimensional bosonic and fermionic metric, respectively.

\subsection{Global RNS action in supersurface language}

Now we want to show the explicit form  of the action given by eq.(\ref{RNS2}), also called global RNS action, in supersurface language. In eq.(\ref{superfield}) we defined a superfield for one degree of freedom. In order to make contact with global RNS action, it is defined the collection of $d$ superfields, $X^{I} (I=1,...,d)$ as
\begin{equation}\label{superfieldXd}
X^{I}=x^{I}+\theta \psi^{I}_{\theta}+\overline{\theta}\psi_{\overline{\theta}}^{I}+\overline{\theta}\theta F^{I}.
\end{equation}
Furthermore, we already saw that a supersymmetric invariant way to express an action in a non-trivial way in supersurface language is given by eq.(\ref{IFlat}). Therefore, for a superfield of $d$ degrees  of freedom, we have that the action 
\begin{equation}\label{RNS3}
S=\frac{1}{2}\int d^{2}z d^{2}\theta D_{\overline{\theta}}X^{I}D_{\theta}X_{I},
\end{equation}
is supersymmetric invariant. In fact, the action given by eq.(\ref{RNS3}) when is expressed in components is equivalent to the one given by eq.(\ref{RNS2}) \cite{hatfield} .

On the other hand, the superspace formulation of the full locally supersymmetric RNS action requires the introduction of curved supersurfaces. That requirement is equivalent to introducing a generalization of the already known zweibein which from now on will be called the superzweibein.

\section{Curved supersurfaces}

\subsection{The superzweibein $E_{A}^{M}$}

The supersurface extension of the ordinary zweibein is the superzweibein $E_{A}^{M}$ which consists of $16$ superfield components. The $M$ index transforms as an ordinary supersurface coordinate under the superdiffeormorphism group, whilst the $A$ index transforms as living in a flat tangent space. Hatfield \cite{hatfield} indicates that $6$ superfield components can be fixed. These superfields are: one superfield which is necessary to fix the flat supersurface action eq.(\ref{RNS3}), four superfields to describe superdiffeomorphisms and finally one superfield which parametrize rotations in the Tangent space. 

However, what happens to the remaining $10$ superfields which have been not fixed? As pointed out by Eric D'Hoker and D.H. Phong \cite{hoker2}, it is possible to choose the following constraints over the torsion field $T_{\alpha\beta}^{m}$, in order to fix those remaining $10$ superfields:
\begin{eqnarray}\label{torsiongauge}
T_{\alpha\beta}^{m} &=& 2(\gamma^{m})_{\alpha \beta},\\
T^{\alpha}_{\beta\gamma}&=&T^{m}_{np}=0.
\end{eqnarray}
Nonetheless, in a equivalent and more useful way, we can replace such constraints by another gauge where the curvature $R_{\alpha\beta}$ be proportional to $(\gamma_{5})_{\alpha\beta}$. This constraint replaces the another one given by $T^{m}_{np}=0$. We adopt this choice as our gauge \cite{hoker2} \cite{nakayama}. Therefore, we rewrite the constraints given by eq.(\ref{torsiongauge}) as
\begin{eqnarray}
R_{\alpha\beta}&=&-2i(\gamma_{5})_{\alpha\beta}Y\label{gauge1},\\
T^{\alpha}_{\beta\gamma}&=&0\label{gauge2},\\
T^{m}_{\alpha\beta}&=&2(\gamma^{m})_{\alpha\beta},\label{gauge3}
\end{eqnarray}
where $Y$ is understood as the supercurvature scalar.

\section{$\mathcal{N}=1$ Supersymmetric Liouville Theory}

In the same way as we did in the bosonic case, it is instructive to analyse how Super Liouville theory arose in the context of $2D$ supergravity.

The generalization of eq.(\ref{covariantderivative}) is given by the definition of the covariant superderivative which is given as follows:
\begin{eqnarray}
\mathcal{D}_{M}V_{A} &=& \partial_{M}V_{A} + \Omega_{M}E_{A}^{B}V_{B},\\
\mathcal{D}_{M}V^{A} &=& \partial_{M}V^{A} + V^{B}E^{A}_{B}\Omega_{M}.
\end{eqnarray}
In this case, the generator of the Lorentz transformation is defined as $E_{a}^{b} = \epsilon_{a}^{b}$, $E_{a}^{\beta} = E_{\alpha}^{b} = 0$, $E_{\alpha}^{\beta} = \frac{1}{2}(\gamma_{5})_{\alpha}^{\beta} $ \cite{hatfield} \cite{DistlerKawai} \cite{nakayama} \cite{hoker2}.

By using eq.(\ref{RNS3}), the partition function of the string theory with local supersymmetry on the world sheet is given by:
\begin{equation}\label{Z1}
\mathcal{Z} = \int[\mathcal{D}\,E_{M}^{A}][\mathcal{D} X^{I}]\mathit{e}^{-S[X,E]}.
\end{equation}
The matter part, for our purposes, is taken to be
\begin{eqnarray}\label{matteraction1}
S = \frac{1}{8\pi}\int d^{2}Z\,E\,\mathcal{D}^{\alpha}X^{I}\mathcal{D}_{\alpha}X_{I}.
\end{eqnarray}
The super-field $X^{I}$ was defined in eq.(\ref{superfieldXd}) as   
\begin{equation}
X^{I} = x^{I} + \theta^{\alpha}\Psi_{\alpha}^{I} + i\theta\overline{\theta}F^{I},
\end{equation}
where $F^{I}$ is an auxiliary field, $\theta^{\alpha}\Psi_{\alpha}^{I} = \theta^{+}\Psi_{+}^{I} + \theta^{-}\Psi_{-}^{I}$ and $\Psi^{I}$ is a Majorana spinor. The integration measure is given by:
\begin{equation}
d^{2}Z\,E = d^{2}zd\theta d\overline{\theta}\,\text{Sdet}(E_{M}^{A}).
\end{equation}
On the other hand, the Super Gauss-Bonnet theorem for a supersurface $\mathcal{M}$ is defined as:
\begin{equation}\label{superGB}
\chi(\mathcal{M}) = \frac{1}{2\pi}\int d^{2}Z\,E\, Y. 
\end{equation}
In the present gauge, given by eqs.(\ref{gauge1}), (\ref{gauge2}) and (\ref{gauge3}), the supersurface version of Gauss-Bonnet theorem eq.(\ref{superGB}) becomes the usual Gauss-Bonnet theorem, thus we obtain:
\begin{equation}\label{XM1}
\chi(\mathcal{M}) = \frac{1}{2\pi}\int d^{2}Z\,E\, Y = \frac{1}{4\pi}\int d^{2}z\sqrt{g} R. 
\end{equation}
At this point, it is important to point out that the surface $\mathcal{M}$ will be taken topologically equivalent to the sphere, thus $\chi(\mathcal{M}) = 2$. From eq.(\ref{XM1}), we have:
\begin{equation}
\int d^{2}Z\,E\,Y = 4 \pi.
\end{equation}
Since the matter field part eq.(\ref{matteraction1}) into the partition function eq.(\ref{Z1}) is Gaussian, the path integration give an straightforward result, 
\begin{equation}
\mathit{e}^{-S_{m}[E]} = \Omega \left[\frac{8\pi^{2}\,\text{Sdet}'(\square)_{E}}{\int d^{2}Z\,E}\right]^{-d/2}, 
\end{equation}
where $\square=\mathcal{D}_{+}\mathcal{D}_{-}$ \cite{DistlerKawai} and we recall $d$ is the number of superfields ($I = 1, 2,\dots ,d$).
 
The effective action $S_{m}[E]$ is not invariant under the following super Weyl transformation \cite{hatfield} \cite{hoker1} \cite{hoker2} \cite{nakayama} \cite{Martinec} \cite{DistlerKawai} :
\begin{eqnarray}
E_{M}^{a} &=& \mathit{e}^{\Phi}\widehat{E}_{M}^{a},\\
E_{M}^{\alpha} &=& \mathit{e}^{\Phi/2}[\widehat{E}_{M}^{\alpha} + \widehat{E}_{M}^{a}(\gamma_{a})^{\alpha\beta}\widehat{\mathcal{D}}_{\beta}\Phi].
\end{eqnarray} 
It is common in the literature \cite{DistlerKawai} to abbreviate this transformation as $E_{M}^{A} = \mathit{e}^{\Phi}\widehat{E}_{M}^{A}$. On the other hand, Nakayama \cite{nakayama} points out that the variation of the effective action under super Weyl transformations is as follows:
\begin{equation}
\log\left(\frac{\text{Sdet}\square}{\text{Sdet}'\widehat{\square}}\right) = c - S_{SL}(\Phi). 
\end{equation}
As a consequence, the effective action has the following transformation,
\begin{equation}\label{mattertransf1}
S_{m}[E] \rightarrow S_{m}[\mathit{e}^{\Phi}E] = S_{m}[E] - \frac{2}{3}c_{m}S_{SL}[\Phi,E].  
\end{equation}
where $c_{m}= 3d/2$.

As in the bosonic case, we now proceed to fix the gauge of $E_{M}^{A}$. This gauge is called the superconformal gauge and is given as follows,
\begin{equation}\label{superconformalgauge}
f*(E)=\mathit{e}^{\Phi}\widehat{E}(\Gamma).
\end{equation}
We explain briefly the meaning of eq.(\ref{superconformalgauge}). The measure $\mathcal{D}E_{M}^{A}$ will be decomposed into the superdiffeomorphism volume, super local Lorentz transformations, super Weyl transformations and the moduli $\Gamma$. 

After dividing it by the volume of the gauge group, the supervielbein $\mathcal{D}E_{M}^{A}$ is decomposed as,
\begin{equation}\label{BC}
\int[\mathcal{D}E_{M}^{A}] = \int \{d\Gamma\}[\mathcal{D}_{E}\Phi]\mathcal{D}_{E}B\mathcal{D}_{E}C\,\mathit{e}^{-S_{gh}[B,C;E]},
\end{equation}
where $B(\mathbf{z}) = \beta(z) + \theta b(z)$, $C(\mathbf{z}) = c(z) + \theta\gamma(z)$. The super ghost path integral is also Gaussian (as in eq.(\ref{mattertransf1})). Hence, the following equation can be obtained \cite{nakayama} 
\begin{equation}\label{ghosttransf1}
S_{gh}[E] \rightarrow S_{gh}[\mathit{e}^{\Phi}E] = S_{gh}[E] - \frac{2}{3}c_{gh}S_{SL}[\Phi,E],  
\end{equation}
where $c_{gh}=-15$.
From Eqs.(\ref{Z1}) and (\ref{BC}), in this superconformal gauge the partition function is given by:
\begin{equation}\label{Z2}
\mathcal{Z} = \int d\Gamma [\mathcal{D}_{E}\Phi]\mathit{e}^{-S_{m}[\mathit{e}^{\Phi}\widehat{E}]-S_{gh}[\mathit{e}^{\Phi}\widehat{E}]}.
\end{equation}
Combining Eqs.(\ref{mattertransf1}), (\ref{ghosttransf1}) and (\ref{Z2}) we have,
\begin{equation}
\mathcal{Z} = \int d\Gamma [\mathcal{D}_{\widehat{E}}\Phi]\mathit{e}^{-S_{m}[\widehat{E}]-S_{gh}[\widehat{E}]} \underbrace{\mathit{e}^{-(10-d)S_{SL}[\Phi,E]}}_{\text{Super Liouville part}}.
\end{equation} 
For critical dimension $d=10$, the $\Phi$ integration yields a constant. If $d\neq 10$, we have to evaluate this integral.

As in the bosonic case, the measure is defined by the norm:
\begin{equation}
||\delta\Phi||^{2}_{\widehat{E}} = \int d^{2}Z\,E(\delta\Phi)^{2} = \int d^{2}Z\,\widehat{E}\mathit{e}^{\Phi}(\delta\Phi)^{2},
\end{equation}
Therefore, the measure $\mathcal{D}_{\widehat{E}}\Phi$ is defined through
\begin{equation}
\int \mathcal{D}_{\widehat{E}}\Phi\,\mathit{e}^{-||\delta\Phi||^{2}_{\widehat{E}}} = 1.
\end{equation}
However, the norm $||\delta\Phi||^{2}_{\widehat{E}}$ is neither Gaussian nor invariant under the translation in the functional space. In order to linearise this norm we follow the supersymmetric version of the Distler-Kawai hypothesis \footnote{This hypothesis was already mentioned in Chapter \ref{Chapter3}.} \cite{hoker2} \cite{Martinec}. Therefore, the linearised norm is now given by:
\begin{equation}
||\delta\Phi||^{2}_{\widehat{E}} = \int d^{2}Z\,\widehat{E}(\delta\Phi)^{2}.
\end{equation}
From all of these contributions, the form of $S_{SL}[\Phi,E]$ is explicitly given as
\begin{equation}\label{SLaction}
S_{SL}[\Phi,E] = \frac{1}{4\pi}\int d^{2}Z\,\widehat{E} \left( \frac{1}{2} \widehat{\mathcal{D}}^{\alpha}\Phi\widehat{\mathcal{D}}_{\alpha}\Phi + Q\widehat{Y}\Phi + \mu\mathit{e}^{b\Phi}\right).  
\end{equation}
In the next section, by following E. Abdalla et. al \cite{3} and R. Rashkov and M. Stanishkov \cite{Rashkov}, we will study the $N$ - super point function. Our starting point will be the action described by eq.(\ref{SLaction}). 

\section{$N$ - super point function}

Current research on Liouville theory and its supersymmetric generalization have been focused on finding the exact form of its $N$-point functions. Nonetheless, this is a difficult task. Therefore, researchers have concentrated all their effort to first compute the basic case when $N=3$. As stated at the beginning of this work, the focus of present research has been to find the exact form of the $3$-point function in the bosonic and supersymmetric case. Literature on this topic shows a variety of approaches \cite{zamo} \cite{dotto1} \cite{3} \cite{Rashkov}. In order to obtain that point function in the supersymmetric case, we need to first find the exact integral expression of the $N$-point function. 

The super Liouville action is given by \cite{3} \cite{Rashkov},
\begin{equation}
S_{SL}[\Phi,E] = \frac{1}{4\pi}\int d^{2}Z\,E \left( \frac{1}{2} \mathcal{D}^{\alpha}\Phi\mathcal{D}_{\alpha}\Phi + Q\,Y\Phi + \mu\,e^{b\Phi}\right),
\label{superliouvilleaction}
\end{equation}
where $d^{2}Z\,E=d^{2}zd^{2}\theta \,\text{Sdet} E^{A}_{M}$ and $\Phi = \phi + \theta \psi + \overline{\theta}\,\overline{\psi} + \theta\,\overline{\theta} F$. Now, vertex operators in the Neveu-Schwarz sector are represented by \cite{3} \cite{Rashkov}:
\begin{equation}\label{potentialV1}
V_{\alpha}= e^{\alpha \Phi(Z,\overline{Z})}.
\end{equation}
Its superconformal dimension is given by \cite{nakayama}:
\begin{equation}
\Delta_{(\alpha)}=\frac{1}{2}\alpha(Q-\alpha).
\end{equation}
We should notice that $\Delta_{(\alpha)}=\Delta_{(Q-\alpha)}$. As in the bosonic case, it reflects a duality in our supersymmetric theory. 

Hence, the $N$ - point function of vertex operators is given by eq.(\ref{potentialV1}) it is then defined as:
\begin{equation}\label{NsuperpointF}
\left\langle \prod_{I=1}^{N}\,e^{\alpha_{I}\Phi(Z_{I})}\right\rangle = \int [\mathcal{D}_{E}\Phi]\,\,e^{-S_{SL}}\,\prod_{I=1}^{N}\,e^{\alpha_{I}\Phi(Z_{I})}.
\end{equation}
    
\subsection{Computing the N - super point function}

In this subsection we will compute a suitable expression for the $N$ - point function. From eq.(\ref{NsuperpointF}) we have 
\begin{equation}\label{GNpoint1}
\mathcal{G}_{N}=\left\langle \prod_{I=1}^{N}\mathit{e}^{\alpha_{I}\Phi_{SL}(Z_{I})}\right\rangle = \int [\mathcal{D}_{E}\Phi]\,\mathit{e}^{-S_{SL}}\,\prod_{I=1}^{N}\mathit{e}^{\alpha_{I}\Phi_{SL}(Z_{I})}, 
\end{equation}
where the $N$- point function is represented by $\mathcal{G}_{N}$. In the same way we did in the bosonic case, integration over the zero mode will be performed. As usual we separate in two pieces the superfield $\Phi_{SL}$. Therefore, the Liouville superfield $\Phi_{SL}$ decomposes as follows,
\begin{equation}\label{Phidescomposes}
\Phi_{SL}(Z)=\Phi_{0} + \Phi(Z),
\end{equation}
where $D_{\alpha} D^{\alpha}\Phi_{0} = 0$ and $\int d^{2}Z\,\widehat{E}\Phi(Z)=0$. Replacing eq.(\ref{Phidescomposes}) into eq.(\ref{superliouvilleaction}) and also using eq.(\ref{superGB}) gives rise to
\begin{equation}\label{EQ1HOY}
S_{SL} = Q\Phi_{0} + \frac{1}{4\pi}\int\,d^{2}Z\,E\left[ \frac{1}{2}D_{\alpha}\Phi D^{\alpha}\Phi + Q Y\Phi + \mu\,e^{b(\Phi_{0} + \Phi)}\right]. 
\end{equation}
The next step will be to use the fact that $\Phi_{0}$ is the constant mode of the super field $\Phi_{SL}$. Hence, we have 
\begin{eqnarray}
 \frac{\mu}{4\pi}\int\,d^{2}Z\,E\, e^{b(\Phi_{0} + \Phi)}=\frac{\mu }{4\pi}e^{b\Phi_{0}}\int\, d^{2}Z\,E\, e^{b\Phi}.
\end{eqnarray}
This means that eq.(\ref{EQ1HOY}) is written as
\begin{equation}
S_{SL} = Q\Phi_{0} + \frac{\mu}{4\pi}\,e^{b\Phi_{0}}\int\,d^{2}Z\,E\,e^{b \Phi} + \frac{1}{4\pi}\int\,d^{2}Z\,E\,\left[ \frac{1}{2}D_{\alpha}\Phi D^{\alpha}\Phi + Q \,Y\,\Phi \right],
\end{equation}
where we recognize the last term as the super Liouville action without the exponential term $e^{b\Phi}$. Therefore, it is useful to denote such a term as the free super Liouville action $S_{0\, SL}$: 
\begin{equation}
S_{0\,SL}= \frac{1}{4\pi}\int\,d^{2}Z\,E\,\left[ \frac{1}{2}D_{\alpha}\Phi D^{\alpha}\Phi + Q \,Y\,\Phi \right].
\end{equation}
Taking into account eq.(\ref{Phidescomposes}), the measure $\mathcal{D}_{E}\Phi_{SL}$ splits as \cite{Rashkov} 
\begin{equation}\label{measure}
\mathcal{D}_{E}\Phi_{SL}=\mathcal{D}_{E}\Phi_{0}\,\mathcal{D}_{E}\Phi.
\end{equation}
From eqs.(\ref{superliouvilleaction}), (\ref{Phidescomposes}) and (\ref{measure}), we have
\begin{equation}
\mathcal{G}_{N}=\int[\mathcal{D}_{E}\Phi_{0}][\mathcal{D}_{E}\Phi]\prod_{I=1}^{N}\,e^{\alpha_{I}\Phi_{0}}\,e^{\alpha_{I}\Phi(Z_{I})}\,e^{-S_{SL}}.
\end{equation}
Thus, $\mathcal{G}_{N}$ becomes:
\begin{eqnarray}
 \mathcal{G}_{N}&=&\int [\mathcal{D}_{E}\Phi_{0}]\,\sum_{s=0}^{\infty}\,e^{(\sum_{I}\alpha_{I}-Q+sb)\Phi_{0} }\,\frac{(-1)^{s}}{s!}\left( \frac{\mu}{4\pi}\right)^{s}\times\nonumber\\&& \int [\mathcal{D}_{E}\Phi]\,\prod_{I=1}^{N} e^{\alpha_{I}\Phi(Z_{I})}\left(\int\,d^{2}Z\,E\,e^{b\Phi(Z)} \right)^{s}\,e^{-S_{0\,SL}}. 
\end{eqnarray}
Following E. Abdalla et. al \cite{3} we choose the ``on-shell" condition:
\begin{equation}\label{ONSHELL}
s=\frac{1}{b}\left[ Q- \sum_{I=1}^{N}\alpha_{I} \right]. 
\end{equation}
Reabsorbing the constant mode contribution into normalization of $\mathcal{G}_{N}$, we arrive to 
\begin{equation}\label{GN1}
\mathcal{G}_{N} = \sum_{s=0}^{\infty}\frac{(-1)^{s}}{s!}\left( \frac{\mu}{4\pi}\right)^{s} \left\langle \left( \int d^{2}Z\,E\,e^{b\Phi(Z)}\right)^{s}\prod_{I=1}^{N}e^{\alpha_{I}\Phi(Z_{I})} \right\rangle_{S_{0\,SL}}.  
\end{equation}
We can rewrite $\mathcal{G}_{N}$ as an expansion in terms of $\mathcal{G}_{N}^{s}$,
\begin{equation}\label{GNGNS}
\mathcal{G}_{N} = \sum_{s=0}^{\infty} \mathcal{G}_{N}^{s}.
\end{equation}
Comparing eqs.(\ref{GN1}) and (\ref{GNGNS}), we realize that $\mathcal{G}_{N}^{s}$ is given as 
\begin{equation}\label{GN2S}
\mathcal{G}_{N}^{s} = \frac{(-1)^{s}}{s!}\left( \frac{\mu}{4\pi}\right)^{s} \left\langle \left( \int d^{2}Z\,E\,e^{b\Phi(Z)}\right)^{s}\prod_{I=1}^{N}\,e^{\alpha_{I}\Phi(Z_{I})} \right\rangle_{S_{0\,SL}}.  
\end{equation}
It is also well known that the residue of the Gamma function $\Gamma(z)$ is
\begin{equation}
 \operatorname{Res}(\Gamma,-s)=\frac{(-1)^{s}}{s!}, 
\end{equation}
where $s \in \mathbb{N}$. In addition we rewrite the residue of Gamma function for negative integer values $\operatorname{Res}(\Gamma,-s)$ as follows  
\begin{equation}
\Gamma_{R}(-s)= \frac{(-1)^{s}}{s!}.
\end{equation}
In this way, we are able to rewrite $\mathcal{G}_{N}^{s}$ as
\begin{equation}
\mathcal{G}_{N}^{s} = \Gamma_{R}(-s)\left( \frac{\mu}{4\pi}\right)^{s} \left\langle \left( \int d^{2}Z\,E\,e^{b\Phi(Z)}\right)^{s}\prod_{I=1}^{N}e^{\alpha_{I}\Phi(Z_{I})} \right\rangle_{S_{0\,SL}}.
\end{equation}
Despite the factor which is independent of $\Phi(Z)$, we just focus in to compute the expectation value $\left\langle \dots \right\rangle_{S_{0\,SL}} $. Moreover, if we realize that $\alpha_{I}\Phi(Z_{I})=\int d^{2}Z\alpha_{I}\Phi(Z)\delta^{2}(Z-Z_{I})$ and an analogous relation for $b\Phi(Z_{I})$, it is not difficult to obtain 
\begin{eqnarray}\label{superpointfunction1}
\left\langle \left( \int d^{2}Z\,E\,e^{b\Phi(Z)}\right)^{s}\prod_{I=1}^{N}e^{\alpha_{I}\Phi(Z_{I})} \right\rangle_{S_{0\,SL}}  \hspace{-0.5cm}&=& \prod_{I=1}^{N}\int[\mathcal{D}_{E}\Phi]\\&&\prod_{i=1}^{s}
\int d^{2}Z_{i}\,E_{i}\, e^{-\frac{1}{4\pi}\int d^{2}Z\,E\,\left[ \frac{1}{2} D_{\alpha}\Phi D^{\alpha}\Phi - j(Z)\Phi\right] }.\nonumber
\end{eqnarray}
Note that in eq.(\ref{superpointfunction1}), we put together the linear contributions on $\Phi$ in order to build the current $j(Z)$. Hence, such a current in eq.(\ref{superpointfunction1}) is given as
\begin{equation}\label{SLcurrent}
j(Z)= 4\pi \sum_{I}\alpha_{I}\,E^{-1}\delta^{2}(Z-Z_{I}) + 4\pi b  \sum_{I}\,E^{-1}\delta^{2}(Z-Z_{i}) - Q\,Y.
\end{equation}
From eq.(\ref{superpointfunction1}), it is not difficult to realize that the following expression is a Gaussian integral:
\begin{equation}
\int[\mathcal{D}_{E}\Phi]\,e^{-\frac{1}{4\pi}\int d^{2}Z\,E\left[ \frac{1}{2} D_{\alpha}\Phi D^{\alpha}\Phi - j(Z)\Phi\right] }.
\end{equation} 
Additionally, we define the inner product
\begin{eqnarray}
\langle f, g\rangle = \int d^{2}Z\,E [f(Z).g(Z)].
\end{eqnarray}
Hence, by neglecting the surface terms due to the kinetic term we are now able to rewrite eq.(\ref{superpointfunction1}) as the Gaussian functional integral
\begin{eqnarray}
\prod_{i=1}^{s}\int d^{2}Z_{i}\,E_{i}\,\left\lbrace  \int [\mathcal{D}_{E}\Phi] \,e^{-\langle \Phi, (\frac{\square}{4\pi}) \Phi\rangle }\right\rbrace \,e^{\langle \frac{j}{4\pi}, (\frac{\square}{4\pi})^{-1} \frac{j}{4\pi}\rangle}, 
\end{eqnarray}
where we should recall $\square=D_{+}D_{-}$. In addition, Polchinski \cite{Polchinski} gives an expression for the generalized super Green function which satisfy the next kind of Poisson equation:
\begin{equation}\label{SGreenEq}
D_{+}D_{-}G(Z_{i},Z_{j})=\pi \delta^{2}(Z_{i}-Z_{j}).
\end{equation}
Therefore, the super Green function is written in supersurface notation as  
\begin{equation}\label{SGreenFunct}
G(Z_{i},Z_{j})=-\frac{1}{2}\ln|Z_{i}-Z_{j}|^{2},
\end{equation}
where $|Z_{i}-Z_{j}|=|z_{i}-z_{j}-\theta_{i}\theta_{j}|$ \cite{Polchinski} \cite{Martinec} \cite{Alvarez1, Alvarez2}. From eq.(\ref{SGreenEq}) it is possible to identify
\begin{eqnarray}\label{Gfunction}
\left( \frac{\square}{4\pi}\right)^{-1}\rightarrow 4\,G(Z_{i}, Z_{j}), 
\end{eqnarray}
as in the bosonic case. Substituting eqs.(\ref{SLcurrent}), (\ref{SGreenFunct}), (\ref{Gfunction}) into (\ref{superpointfunction1}), we obtain
\begin{eqnarray}\label{Expansion}
\frac{1}{\left[ \text{Sdet}(\frac{\square}{4\pi})\right]^{1/2}}\prod_{i=1}^{s}\int d^{2}Z_{i} \,E_{i}\,e^{\langle \frac{j}{4\pi}, (\frac{\square}{4\pi})^{-1} \frac{j}{4\pi}\rangle}&&=
\frac{1}{\left[ \text{Sdet}(\frac{\square}{4\pi})\right]^{1/2}}\prod_{i=1}^{s}\int d^{2}Z_{i} \,E_{i}\nonumber\\
&&\times e^{\sum_{I,J}\alpha_{I}\alpha_{J}G(Z_{I}, Z_{J})}\times e^{\sum_{J}\sum_{i}\alpha_{J}\,b\,G(Z_{J}, Z_{i})}\nonumber\\
&&\times  e^{Q\sum_{J}\alpha_{J}\int d^{2}W \frac{E_{W}}{4\pi} G(Z_{J}, W)\,Y}\times e^{b\sum_{j}\sum_{I}\alpha_{I}G(Z_{j}, Z_{I})}\nonumber\\
&&\times e^{b^{2}\sum_{j}\sum_{i}G(Z_{i}, Z_{j})}\times e^{bQ\sum_{j}\int d^{2}W \frac{E_{W}}{4\pi}G(Z_{j}, W)\,Y(W)}\nonumber\\
&&\times e^{Q\sum_{I}\alpha_{I}\int d^{2}X \frac{E_{X}}{4\pi}G(X, Z_{I})\,Y(X)}\nonumber\\
&&\times e^{Qb\int d^{2}X \frac{E_{X}}{4\pi} \sum_{i} G(X, Z_{i})\,Y(X)}\nonumber\\
&&\times e^{\frac{Q^{2}}{16\pi^{2}}\int d^{2}X \,E(X)\,\int d^{2}W \,E(W)\,Y(X)G(X, W)\,Y(W)}.\nonumber\\
\end{eqnarray}
The last term on the right hand side is recognized again (see Chapter \ref{Chapter4}) as the Polyakov term \cite{dual} which contributes to the central charge with the factor proportional to $Q^{2}$. As in the bosonic case that term will not have further implication for our purposes \footnote{For a more complete treatment about superconformal currents and charges we refer to S. Ketov \cite{K}, R. Blumenhagen and E. Plauschinn \cite{B} and E. Martinec \cite{Martinec}.}. Therefore, the integrand of eq.(\ref{Expansion}) can be rewritten as
\begin{eqnarray}\label{Expansion1}
e^{\sum_{I,J}\alpha_{I}\alpha_{J}G(Z_{I}, Z_{J})}\times e^{Q\sum_{J}\alpha_{J} P(Z_{J})}\times e^{Q\sum_{I}\alpha_{I} P(Z_{I})}\times e^{\mathcal{F}^{b}(Z_{j})}\times e^{bQ\sum_{j}P(Z_{j})}\times e^{bQ\sum_{i}P(Z_{i})},\nonumber\\ 
\end{eqnarray}
where we have defined the functions $P(Z)$ and $\mathcal{F}^{b}(Z)$ as:
\begin{eqnarray}
P(X)&=&\frac{1}{4\pi}\int d^{2}W E_{W}\,G(X, W)\,Y,\label{Psuperfunction}\\
\mathcal{F}^{b}(Z_{j})&=&2b\sum_{j}\sum_{I}\alpha_{I}\,G(Z_{j}, Z_{I}) + b^{2}\sum_{i}\sum_{j}\,G(Z_{i}, Z_{j}).\label{Ffunction}
\end{eqnarray}
In the definition of $P(X)$, $X$ and $W$ are supercoordinates. Replacing eq.(\ref{SGreenFunct}) into eq.(\ref{Expansion1}), we arrive to
\begin{eqnarray}\label{Ersa}
\prod_{I<J}^{N}|Z_{I}-Z_{J}|^{-2\alpha_{I}\alpha_{J}}\times \lim_{Z_{I}\rightarrow Z_{J}}e^{\sum_{I,J}\alpha_{I}\alpha_{J}G(Z_{I}, Z_{J})}\times e^{2Q\sum_{I}\alpha_{I}P(Z_{I})}\times \mathcal{H}_{N}
\end{eqnarray}
where $\mathcal{H}_{N}$ is given as 
\begin{eqnarray}\label{HI}
\mathcal{H}_{N}=\prod_{i}^{s}\int d^{2}Z_{i}E_{i}\,e^{2bQ P(Z_{i})}\times \prod_{I=1}^{N}|Z_{I}-Z_{i}|^{-2b\alpha_{I}}\prod_{i<j}|Z_{i}-Z_{j}|^{-2b^{2}}\times \lim_{Z_{i}\rightarrow Z_{j}}e^{b^{2}G(Z_{i}, Z_{j})}.\nonumber\\
\end{eqnarray} 
In order to write $\mathcal{H}_{N}$, the integral $P(X)$ has been defined as follows 
\begin{eqnarray}\label{ultimita}
P(X)=\frac{1}{4\pi}\int d^{2}W\, E\,G(X, W)\,Y(W).
\end{eqnarray} 
At this point, Distler et. al \cite{DistlerKawai} apply a flat super conformal transformation, which in terms of the super determinant of $E_{M}^{A}$ is  $E\rightarrow e^{\Omega}\widehat{E}$, thus we obtain 
\begin{eqnarray}\label{ULTIMASA}
P(X)=\frac{1}{4\pi}\int d^{2}W\, \widehat{E}\,e^{\Omega}\,G(X, W)\,Y,
\end{eqnarray} 
where $\widehat{E}$ is a reference superzweibein. Following \cite{hoker2} \cite{DistlerKawai} we use the transformation between $Y$ and $\widehat{Y}$ ,
\begin{eqnarray}
Y=e^{-\Omega}\left(\widehat{Y}-2D_{+}D_{-}\Omega \right).
\end{eqnarray}
Therefore, by substituting the latter relation into eq.(\ref{ULTIMASA}), we have 
\begin{equation}
P(X)=-\frac{1}{2\pi}\int d^{2}W\,G(X, W)\,\square\Omega(W),
\end{equation}
where we have also used the flat conformal gauge which implies $\,\text{Sdet} \widehat{E}_{M}^{A}= \widehat{E}=1$ and thus $\widehat{Y}=0$. After integrating by parts we obtain
\begin{equation}
P(X)=-\frac{1}{2\pi}\int d^{2}W\,\left[ \square G(X, W)\right] \,\Omega(W) + \text{surface terms}.
\end{equation}
Disregarding the surface terms and taking into account eq.(\ref{SGreenEq}),
\begin{eqnarray}
P(X)=-\frac{1}{2}\int d^{2}W\,\delta^{2}(X-W)\,\Omega(W).
\end{eqnarray}
Finally, we arrive at a compact expression for $P(X)$ in the flat conformal gauge:
\begin{eqnarray}\label{Pfunctionchapter5}
P(X)=-\frac{1}{2}\ln E_X.
\end{eqnarray}
Additionally, we consider the following relations \cite{hatfield} \cite{Alvarez1}
\begin{eqnarray}
E^{A}_{M}&=&\frac{\partial Z'^{N}}{\partial Z^{M}}\widehat{E}^{A}_{N},\label{DF1A}\\
dZ^{M}&=&\frac{\partial Z^{M}}{\partial Z'_{N}}\, dZ^{N}\label{DF1B}.
\end{eqnarray}
From eqs.(\ref{DF1A}) and (\ref{DF1B}) we define $d\mathcal{S}^{A}=E^{A}_{M}\,dZ^{M}=\widehat{E}^{A}_{N}dZ'^{N}$ and the scalar invariant $d\mathcal{S}^{2}=d\mathcal{S}^{A}d\mathcal{S}_{A}$. Introducing a renormalization scale $L$ in eq.(\ref{SGreenFunct}) and using the definition of $d\mathcal{S}$, we have
\begin{eqnarray}
\lim_{Z_{i}\rightarrow Z_{j}}G(Z_{i}, Z_{j})&=&-\lim_{Z_{i}\rightarrow Z_{j}} \ln \left|\frac{Z_{i}-Z_{j}}{L}\right|\nonumber\\
&=&-\ln \left|\frac{dZ_{i}}{L}\right|\nonumber\\
&=&\ln E_{i}-\frac{1}{2}\ln\left|\frac{d\mathcal{S}}{L}\right|^{2}\label{WSP}.
\end{eqnarray}
As in the bosonic case the second term on the right hand side of eq.(\ref{WSP}) is used to renormalize the cosmological constant $\mu$. Therefore, we can disregard this term and only define eq.(\ref{WSP}) as
\begin{eqnarray}\label{GMzrt}
G^{R}(Z_{i})=\ln E_{i}.
\end{eqnarray}
Using eqs.(\ref{Pfunctionchapter5}) and (\ref{GMzrt}) into eq.(\ref{HI}) it is possible to obtain
\begin{eqnarray}\label{HI1}
\mathcal{H}_{N}=\prod_{i=1}^{s}\int d^{2}Z_{i}\,(E_{i})^{1-bQ+b^{2}}\,\prod_{I=1}^{N}|Z_{I}-Z_{i}|^{-2b\alpha_{I}}\prod_{i<j}^{s}|Z_{i}-Z_{j}|^{-2b^{2}}.
\end{eqnarray}
By recalling the relation $Q=b+\frac{1}{b}$, we have that $1-bQ+Q^{2}=0$. Hence, the exponent of the super zweibein vanishes, and eq.(\ref{HI1}) is simplified to 
\begin{eqnarray}\label{HI2}
\mathcal{H}_{N}=\prod_{i=1}^{s}\int d^{2}Z_{i}\,\prod_{I=1}^{N}|Z_{I}-Z_{i}|^{-2b\alpha_{I}}\prod_{i<j}^{s}|Z_{i}-Z_{j}|^{-2b^{2}}.
\end{eqnarray}
We can easily recognize the latter equation as the supersymmetric extension of the Dotsenko-Fateev integral \cite{dofa} \cite{Alvarez1, Alvarez2}. Much research on this topic obtained an exact expression for eq.(\ref{HI2}). In particular, we will use the results of L. Alvarez-Gaume and PH. Zaugg \cite{Alvarez1, Alvarez2} for this integral. Carrying on with the computation, from eq.(\ref{Ersa}) and the result given above, we have
\begin{eqnarray}\label{superpointfunction2}
&& \lim_{Z_{I}\rightarrow Z_{J}}e^{\sum_{I,J}\alpha_{I}\alpha_{J}G(Z_{I}, Z_{J})}\times e^{2Q\sum_{I}\alpha_{I}P(Z_{I})}\times \prod_{I<J}^{N}|Z_{I}-Z_{J}|^{-2\alpha_{I}\alpha_{J}}\\
&&\prod_{i=1}^{s}\int d^{2}Z_{i}\prod_{I=1}^{N}|Z_{i}-Z_{J}|^{-2b\alpha_{I}}\prod_{i<j}^{s}|Z_{i}-Z_{j}|^{-2b^{2}}.\nonumber
\end{eqnarray}
Hence,
we must now compute this expression 
for the particular case $N=3$. This is the task for the next section.

\section{$N = 3$ superpoint function}

So far, we have discussed a general expression for the $N$ point function in $\mathcal{N}=1$ Super Liouville theory. Now we will focus particularly on the special case $N=3$. Following L. Alvarez-Gaume and Ph. Zaugg \cite{Alvarez1, Alvarez2} and E. Abdalla et. al \cite{3}, the usual procedure to perform the present computation is to fix the following points  $z'_{1}=1, z'_{2}=0, z'_{3}= R$, $\theta'_{1}= \theta'_{2}=0$ and $\theta'_{3}$ is chosen to be $R\eta$ \footnote{Because $z'_{3}$ and $\theta'_{3}$ dependent on $R$, they are fixing only one degree of freedom.}. The invariant parameter $\eta$ will be defined soon. Thereby, the holomorphic part of the superpoints considered are $Z'_{1}=(1,0)$, $Z'_{2}=(0,0)$ and $Z'_{3}=(R, R\eta)$. After replacing such points in eq.(\ref{superpointfunction2}) the parameter $R$ will be taken as going to infinity. It means that the pair $(z'_{3}, \bar{z}'_{3})$ is identified again as the infinity of the Riemann sphere as $R\rightarrow \infty$. 

Before we proceed with such a computation, it is important to point out some definitions. First of all, the difference between super coordinates will be written as
\begin{eqnarray}\label{DiferenciaKO}
Z_{IJ}=Z_{I}-Z_{J}.
\end{eqnarray}
In addition, the parameter $\eta$ used to write $\theta'_{3}=R\eta$ is defined as a $SL(2|1)$  invariant quantity given as \cite{Alvarez1, Alvarez2}
\begin{eqnarray}
\eta=\frac{\theta_{1}Z_{23}+\theta_{2}Z_{31}+\theta_{3}Z_{12}+\theta_{1}\theta_{2}\theta_{3}}{(Z_{12}Z_{13}Z_{23})^{1/2}}.
\end{eqnarray}
To perform a $SL(2|1)$ transformation from $Z \rightarrow Z'$, we use the following relations \cite{Alvarez1, Alvarez2}
\begin{eqnarray}\label{SL21}
z'(Z)=\frac{a z+ b + \alpha\theta}{cz + d + \beta \theta}, \hspace{1cm}\theta'(Z)=\frac{\overline{\alpha} z+ \overline{\beta} + \overline{A} \theta}{cz + d + \beta \theta},
\end{eqnarray} 
along with their conjugate expressions. We also have that $\overline{A}=\sqrt{ad-bc-3\alpha\beta}$, $\bar{\alpha}=(a\beta-c\alpha)/(\sqrt{ad-bc})$ and $\bar{\beta}=(b\beta-d\alpha)/(\sqrt{ad-bc})$ in eq.(\ref{SL21}). On the other hand, Alvarez-Gaume and Zaugg \cite{Alvarez1} express the superdifferential $dZ$ transforming as
\begin{equation}
dZ\,=\,D\theta \,dZ',
\end{equation} 
where according to the chain rule we have $|\partial Z/\partial Z'|= D\theta$. Furthermore, they pointed out that
\begin{eqnarray}
D\theta &=&\frac{\text{Sdet} g}{|cz' + d + \beta\theta'|^{2}},\label{Music}\\
|Z_{IJ}| &=& \frac{\text{Sdet} g\,|Z'_{IJ}|}{(cz'_{I} + d + \beta\theta'_{I})(cz'_{J} + d + \beta\theta'_{J})},\label{Music1}
\end{eqnarray}
where $\text{Sdet} g = ad-bc-\alpha\beta $. Now we rewrite eq.(\ref{superpointfunction2}) as
\begin{eqnarray}\label{Tower1}
\mathcal{F}_{3}(Z')&=&\prod_{I<J}^{N}|Z_{I}-Z_{J}|^{-2\alpha_{I}\alpha_{J}}\,\prod_{i=1}^{s}\int d^{2}Z_{i}\prod_{I=1}^{N}|Z_{i}-Z'_{J}|^{-2b\alpha_{I}}\prod_{i<j}^{s}|Z_{i}-Z_{j}|^{-2b^{2}}.\nonumber\\
\end{eqnarray}
In this equation we must realize that the supercoordinates $Z_{I}$ were just relabel as $Z'_{I}$ ($I= 1, 2, 3$). Moreover, by using eq.(\ref{Music1}) the first factor $\prod_{I<J}^{N}|Z'_{I}-Z'_{J}|^{-2\alpha_{I}\alpha_{J}}$ was transformed to $\prod_{I<J}^{N}|Z_{I}-Z_{J}|^{-2\alpha_{I}\alpha_{J}}$. Hence, we obtain an extra contribution coming from the factor $|cz+d+\beta\theta|$ which acts as a Jacobian in eq.(\ref{Music1}). Putting together such contributions we obtain that $\mathcal{F}_{3}$ transform under the $SL(2|1)$ supergroup as
\begin{eqnarray}\label{LLLL}
\mathcal{F}_{3}(Z')=\prod_{I=1}^{3}|cz_{I} + d + \beta\theta_{I}|^{2(Q-\sum_{I}\alpha_{I})\alpha_{I}}\,\mathcal{F}_{3}(Z).
\end{eqnarray}
As studied in the bosonic case (Chapter \ref{Chapter4}) to find the coefficients of the $SL(2|1)$ transformations given in eq.(\ref{SL21}) we must must solve the system of equations
\begin{eqnarray}
z'_{1}&=&\frac{a z_{1}+b+\alpha \theta_{1}}{cz_{1} + d+ \beta\theta_{1}}=1,\\
z'_{2}&=&\frac{a z_{2}+b+\alpha \theta_{2}}{cz_{2} + d+ \beta\theta_{2}}=0,\\
z'_{3}&=&\frac{a z_{3}+b+\alpha \theta_{3}}{cz_{3} + d+ \beta\theta_{3}}=R,\\
\theta'_{1}&=&\frac{\bar{\alpha}z_{1}+\bar{\beta}+\bar{A}\theta_{1}}{cz_{1} + d + \beta\theta_{1}}=0,\\
\theta'_{2}&=&\frac{\bar{\alpha}z_{2}+\bar{\beta}+\bar{A}\theta_{2}}{cz_{2} + d + \beta\theta_{2}}=0,
\end{eqnarray}
which becomes solvable by adding the extra condition $ad-bc-3\alpha\beta=1$. From the latter equations the really important quantities are $c$, $d$ and $\beta$, since $\prod_{I=1}^{3}|cz_{I} + d + \beta\theta_{I}|^{2(Q-\alpha_{I})\alpha_{I}}$ is the factor which acts as a Jacobian in eq.(\ref{LLLL}). We compute this coefficients using {\it Mathematica} \cite{Mathematica} and write them in Appendix \ref{AppendixD}. As we are interested in how such a Jacobian behaves as $R\rightarrow \infty$, we just write this contribution directly \footnote{In this expression we disregard factors which are proportionals to $R^{-1}$.} \cite{Alvarez1}
\begin{eqnarray}\label{MMMM}
R^{2(Q-\sum_{I}\alpha_{I})\alpha_{3}}\times \left|\frac{Z_{23}}{Z_{12}Z_{31}}\right|^{(Q-\sum_{I}\alpha_{I})\alpha_{1}}\times \left|\frac{Z_{31}}{Z_{12}Z_{23}}\right|^{(Q-\sum_{I}\alpha_{I})\alpha_{2}}\times \left|\frac{Z_{12}}{Z_{13}Z_{23}}\right|^{(Q-\sum_{I}\alpha_{I})\alpha_{3}}.\nonumber\\
\end{eqnarray}
Putting together eqs.(\ref{LLLL}) and (\ref{MMMM}), we have
\begin{eqnarray}
\mathcal{H}_{3}(Z)&=&\lim_{R\rightarrow \infty}R^{2(Q-\sum_{I}\alpha_{I})\alpha_{3}} \left| \frac{Z_{23}}{Z_{12}Z_{31}}\right|^{(Q-\sum_{I}\alpha_{I})\alpha_{1}}\times \left|
\frac{Z_{31}}{Z_{12}Z_{23}}\right|^{(Q-\sum_{I}\alpha_{I})\alpha_{2}}\times \nonumber\\
&&\left|\frac{Z_{12}}{Z_{13}Z_{23}}\right|^{(Q-\sum_{I}\alpha_{I})\alpha_{3}}\times
|Z_{12}|^{-2\alpha_{1}\alpha_{2}}|Z_{23}|^{-2\alpha_{2}\alpha_{3}}|Z_{13}|^{-2\alpha_{1}\alpha_{3}}\,\nonumber\\
&&\prod_{i=1}^{s}\int d^{2}Z_{i}\,|Z'_{1}-Z_{i}|^{-2b\alpha_{1}}|Z'_{2}-Z_{i}|^{-2b\alpha_{2}}\,|Z'_{3}-Z_{i}|^{-2b\alpha_{3}}\prod_{i<j}^{s}|Z_{i}-Z_{j}|^{-2b^{2}}.\nonumber\\
\end{eqnarray}
Using the fixed points $z'_{1}=1$, $z'_{2}=0$, $z'_{3}=R$, $\theta_{1}=0$, $\theta_{2}=0$, $\theta'_{3}=R\eta$ and eq.(\ref{DiferenciaKO}), we have
\begin{eqnarray}
\lim_{R\rightarrow \infty}&&R^{2(Q-\sum_{I}\alpha_{I})\alpha_{3}}|Z_{12}|^{-\delta_{12}}|Z_{23}|^{-\delta_{23}}|Z_{13}|^{-\delta_{13}}\,\prod_{i=1}^{s}\int d^{2}Z_{i}\,|z_{i}|^{-2b\alpha_{1}}\,|1-z_{i}|^{-2b\alpha_{2}}\nonumber\\
&&|R-z_{i}-R\eta\theta_{i}|^{-2b\alpha_{3}}\,\prod_{i<j}^{s}|Z_{i}-Z_{j}|^{-2b^{2}},
\end{eqnarray}
where $\delta_{ij}= \Delta_{i}+\Delta_{j}-\Delta_{k}$, $i\neq j \neq k$. Additionally, taking into account that $\prod_{i=1}^{s}|R-z_{i}-R\eta\theta_{i}|^{-2b\alpha_{3}}=R^{-2bs\alpha_{3}}\prod_{i=1}^{s}|1-\frac{z_{i}}{R}-\eta\theta_{i}|^{-2b\alpha_{3}}$, we obtain
\begin{eqnarray}\label{Usual}
\lim_{R\rightarrow \infty}&&R^{2(Q-\sum_{I}\alpha_{I})\alpha_{3}}\times R^{-2bs\alpha_{3}}|Z_{12}|^{-\delta_{12}}|Z_{23}|^{-\delta_{23}}|Z_{13}|^{-\delta_{13}}\,\prod_{i=1}^{s}\int d^{2}Z_{i}\,|z_{i}|^{-2b\alpha_{1}}\,\nonumber\\
&&|1-z_{i}|^{-2b\alpha_{2}}|1-\frac{z_{i}}{R}-\eta\theta_{i}|^{-2b\alpha_{3}}\,\prod_{i<j}^{s}|Z_{i}-Z_{j}|^{-2b^{2}}.
\end{eqnarray}
Based on the on-shell condition eq.(\ref{ONSHELL}) we realize that there is no dependence on $R$ in eq.(\ref{Usual}), this fact is in accordance with the results from \cite{Alvarez1, Alvarez2} \cite{3}. Hence, the next step is to deal with the remaining integral. In order to do this, it is usual in the literature \cite{3} \cite{Alvarez1} \cite{Rashkov} to take $\eta$ small, thus eq.(\ref{Usual}) becomes
\begin{eqnarray}\label{ORA1}
|Z_{12}|^{-\delta_{12}}|Z_{23}|^{-\delta_{23}}|Z_{13}|^{-\delta_{13}}&&\,\prod_{i=1}^{s}\int d^{2}Z_{i}\,|z_{i}|^{-2b\alpha_{1}}\,|1-z_{i}|^{-2b\alpha_{2}}\nonumber\\
&&|1+2b\alpha_{3}\eta\sum_{i=1}^{s}\theta_{i}|\,\prod_{i<j}^{s}|Z_{i}-Z_{j}|^{-2b^{2}}.
\end{eqnarray} 
Following E. Abdalla et. al \cite{3}, R.C. Rashkov and M. Stanishkov \cite{Rashkov} and L. Alvarez-Gaume and Ph. Zaugg \cite{Alvarez1, Alvarez2}, we rewrite the latter equation in a general form as
\begin{equation}\label{ORA2}
\left\langle \prod_{I=1}^{3}\,e^{\alpha_{I}\Phi(Z_{I})}\right\rangle = |Z_{12}|^{-\delta_{12}}|Z_{13}|^{-\delta_{13}}|Z_{23}|^{-\delta_{23}}\left(D_{even}+ D_{odd}\,\eta\overline{\eta} \right), 
\end{equation}
where $D_{even}$ represents the contribution due to an even number of fermionic variables $\theta$ ($s$ even), while $D_{odd}$ comes from an odd number of these variables ($s$ odd). In this way, the integral in eq.(\ref{ORA1}) is identified with the factor $D_{even}+ D_{odd}\,\eta\overline{\eta}$ in eq.(\ref{ORA2}). Such an integral is written as 
\begin{eqnarray}\label{Rultimasa}
J^{s+1}(\alpha_{1},\alpha_{2},b; Z)&=&\int \prod_{i=1}^{s}d^{2}Z_{i}\, |z_{i}|^{-2b\alpha_{1}}\times |z_{i}-1|^{-2b\alpha_{2}}\times |1+2b\eta\alpha_{3}\sum_{i=1}^{s}\theta_{i}|\nonumber\\
&\times &\prod_{i<j}^{s}|Z_{i}-Z_{j}|^{-2b^{2}},
\end{eqnarray}    
where we have labelled this integral by $J^{s+1}(\alpha_{1},\alpha_{2},b; Z)$. L. Alvarez-Gaumé and Ph. Zaugg \cite{Alvarez2} writes that integral as
\begin{eqnarray}\label{Rultimasasa}
\int \prod_{i=1}^{k}d^{2}Z_{i}\,z_{i}^{2a}\times |z_{i}-1|^{2b}\times |1-2c\eta\sum_{i=1}^{k}\theta_{i} | \times \prod_{i<j}^{k}|z_{i}-z_{j}-\theta_{i}\theta_{j}|^{-2\rho}.
\end{eqnarray}
Hence, from eqs.(\ref{Rultimasa}) and (\ref{Rultimasasa}) it is no difficult to identify that
\begin{eqnarray}
k &=& s,\label{H1a}\\
a &\rightarrow & -b\alpha_{1},\label{H1b}\\
b &\rightarrow & -b\alpha_{2},\label{H1c}\\
c &\rightarrow & -b\alpha_{3},\label{H1d}\\
\rho &=& b^{2}\label{H1e}.
\end{eqnarray}
In this way, by taking into account the result of the supersymmetric extension of the Dotsenko-Fateev integral obtained in \cite{Alvarez1, Alvarez2}, we have
\begin{eqnarray}\label{Equation}
\pi^{k}k!\left( \Delta(\frac{\rho+1}{2})\right)^{k } \left(\frac{-\rho}{2} \right)^{2\left[\frac{ k}{2}\right] }&&\prod_{i=1}^{k}\Delta(\frac{-\rho + 1}{2}i-\left[ \frac{i}{2}\right] )\nonumber\\
&&\prod_{i=0}^{k-1}\Delta(-a-c+\rho(k-1)+\frac{-\rho - 1}{2}i+\left[\frac{i}{2} \right] )\nonumber\\
&&\Delta (1+a+\frac{-\rho -1}{2}i+\left[\frac{i}{2} \right] ) \Delta(1+c+\frac{-\rho -1}{2}i+\left[\frac{i}{2} \right] ).\nonumber\\ 
\end{eqnarray}    
By using eqs.(\ref{H1a}), (\ref{H1b}), (\ref{H1c}), (\ref{H1d}) and (\ref{H1e}), eq.(\ref{Equation}) becomes
\begin{eqnarray}\label{LatterR}
\pi^{s}\,s!\left( \Delta(\frac{b^{2}}{2}+\frac{1}{2})\right)^{s}\left(\frac{-b^{2}}{2} \right)^{2\left[\frac{ s}{2}\right] }&&\prod_{j=1}^{s}\Delta(\frac{-b^{2} + 1}{2}j-\left[ \frac{j}{2}\right] )\nonumber\\
&&\prod_{j=0}^{s-1}\Delta(b\alpha_{1}+b\alpha_{2}+b^{2}(s-1)+\frac{-b^{2} - 1}{2}j+\left[\frac{j}{2} \right] )\nonumber\\
&&\Delta (1-b\alpha_{1}+\frac{-b^{2} -1}{2}j+\left[\frac{j}{2} \right] )\nonumber\\
&&\Delta(1-b\alpha_{2}+\frac{-b^{2} -1}{2}j+\left[\frac{j}{2} \right] ).
\end{eqnarray}      
Then we analyse the argument $\Delta(b\alpha_{1}+b\alpha_{2}+b^{2}(s-1)+\frac{-b^{2} - 1}{2}j+\left[\frac{j}{2} \right] )$,
\begin{eqnarray}
b\alpha_{1}+b\alpha_{2}+b^{2}(s-1)+\frac{-b^{2} - 1}{2}j+\left[\frac{j}{2} \right]&=&b\alpha_{1}+b\alpha_{2}+b^{2}s -b^{2}-\frac{b^{2}}{2}j-\frac{j}{2}+\left[ \frac{j}{2}\right] \nonumber\\
&=& b(\alpha_{1}+\alpha_{2}+ bs)-b^{2}-\frac{b^{2}}{2}j-\frac{j}{2}+\left[ \frac{j}{2}\right]\nonumber\\
&=&\underbrace{b(Q-b)}-\alpha_{3}b-\frac{j}{2}+\left[\frac{j}{2} \right]-\frac{b^{2}}{2} j\nonumber\\
&=&\hspace{0.6cm} 1\hspace{0.5cm} -\alpha_{3}b-\frac{j}{2}+\left[\frac{j}{2} \right]-\frac{b^{2}}{2} j.
\end{eqnarray}
As such, eq.(\ref{LatterR}) can be written as:
\begin{eqnarray}\label{GNS3}
\pi^{s}\, s!\left(\Delta(\frac{b^{2}+1}{2})\right)^{s}\left(\frac{-b^{2}}{2} \right)^{2\left[\frac{ s}{2}\right] }&&\prod_{j=1}^{s}\Delta(\frac{-b^{2} + 1}{2}j-\left[ \frac{j}{2}\right] )\nonumber\\
&&\prod_{j=0}^{s-1}\prod_{j=1}^{3}\Delta (1-b\alpha_{j}-\frac{b^{2}}{2}j- \frac{1}{2}j+\left[\frac{j}{2} \right] ).\nonumber\\
\end{eqnarray}
The result for the $s^{th}$ term in the expansion of $\mathcal{G}_{N}$ is obtained by combining eqs.(\ref{GN2S}) and (\ref{GNS3}) \cite{3} \cite{Rashkov}:
\begin{eqnarray}
\mathcal{G}_{N=3}^{s}= \left(- \frac{\mu}{4}\Delta(\frac{b^{2}+1}{2})\right)^{s}\left(\frac{-b^{2}}{2} \right)^{2\left[\frac{ s}{2}\right] }&&\prod_{I<J}^{3} |Z_{I}-Z_{J}|^{-\delta_{ij}}\prod_{j=1}^{s}\Delta(\frac{-b^{2} + 1}{2}j-\left[ \frac{j}{2}\right] )\nonumber\\
&&\prod_{j=0}^{s-1}\prod_{j=1}^{3}\Delta (1-b\alpha_{j}-\frac{b^{2}}{2}j- \frac{1}{2}j+\left[\frac{j}{2} \right] ).\nonumber
\end{eqnarray} 
We realize that $\mathcal{G}_{N=3}^{s}$ changes by depending on $s$. In this way, we have
\begin{equation*}
\mathcal{G}_{N=3}^{s}= 
|Z_{12}|^{-\delta_{12}}|Z_{13}|^{-\delta_{13}}|Z_{23}|^{-\delta_{23}} \left\{
\begin{aligned}
&D_{\text{even}}\,; \hspace{1.3cm} s\in 2\mathbb{N},\\
&D_{\text{odd}}\eta\overline{\eta}\,;\hspace{1cm} s\in 2\mathbb{N}+1.
\end{aligned}\right .
\end{equation*}
Hence, by applying the results of eq.(\ref{GNS3}) and the ``on-shell'' condition $s=(Q-\sum_{I}\alpha_{I})/b$, $D_{\text{even}}$ and $D_{\text{odd}}$ are given as
\begin{eqnarray}
D_{\text{even}}=\left(\frac{\mu b^{2}}{8}\Delta(\frac{b^{2}+1}{2})\right)^{(Q-\sum_{I}\alpha_{I})/b}&&\hspace{-0.7cm}\prod_{I<J}^{3} |Z_{I}-Z_{J}|^{-\delta_{ij}}\prod_{j=1}^{(Q-\sum_{I}\alpha_{I})/b}\Delta(\frac{-b^{2} + 1}{2}j-\left[ \frac{j}{2}\right] )\nonumber\\
&&\hspace{-1.5cm}\prod_{j=0}^{(Q-\sum_{I}\alpha_{I})/b-1}\prod_{j=1}^{3}\Delta (1-b\alpha_{j}-\frac{b^{2}}{2}j- \frac{1}{2}j+\left[\frac{j}{2} \right] ),\nonumber\\
\label{Deven}\\
D_{\text{odd}}=-\frac{b^{2}}{2}\left(\frac{\mu b^{2}}{8}\Delta(\frac{b^{2}+1}{2})\right)^{(Q-\sum_{I}\alpha_{I})/b}&&\hspace{-0.7cm}\prod_{I<J}^{3} |Z_{I}-Z_{J}|^{-\delta_{ij}}\prod_{j=1}^{(Q-\sum_{I}\alpha_{I})/b}\Delta(\frac{-b^{2} + 1}{2}j-\left[ \frac{j}{2}\right] )\nonumber\\
&&\hspace{-1.5cm}\prod_{j=0}^{(Q-\sum_{I}\alpha_{I})/b-1}\prod_{j=1}^{3}\Delta (1-b\alpha_{j}-\frac{b^{2}}{2}j- \frac{1}{2}j+\left[\frac{j}{2} \right] ),\nonumber\\
\label{Dodd}
\end{eqnarray}
where $\Delta(x)=\frac{\Gamma(x)}{\Gamma(1-x)}$.

Additionally, following the usual interpretation of the $s^{\text{th}}$ term under the on-shell condition \cite{Goulian} \cite{Polyakov2} \cite{Difra} \cite{Rashkov} \cite{3} \cite{zamo2} \cite{dotto1},
\begin{eqnarray}
\text{Res}(\sum_{s=(Q-\sum_{I}\alpha_{I})/b}\mathcal{G}_{3})=\mathcal{G}_{3}^{s}.
\end{eqnarray}
Finally, such prescription allows us to give an exact result for the three point function into $\mathcal{N}=1$ Super Liouville theory, which by using eqs.(\ref{Deven}) and (\ref{Dodd}) is
\begin{eqnarray}
\mathcal{G}_{3}= |Z_{12}|^{-\delta_{12}}|Z_{13}|^{-\delta_{13}}|Z_{23}|^{-\delta_{23}}(D_{\text{even}} + D_{\text{odd}}\,\eta\overline{\eta}).
\end{eqnarray} 

\chapter{Final Remarks} 

\label{Final Remarks} 

\lhead{Chapter 6. \emph{Final Remarks}} 
\vspace{-1cm}
One of the main purposes of this dissertation has been to give a pedagogical introduction to Liouville field theory. Another important purpose was to demonstrate one of the more remarkable aspects of such a theory; this is, the exact computation of the three point function in the bosonic and supersymmetric case. 

For computing Liouville correlators, it is usual to use the Coulomb Gas method \cite{zamo} \cite{nakayama} \cite{ginsparg} \cite{1} \cite{K}. In contrast, we used an alternative method for our first case, which was called the OPS method \cite{dual, dual2}. That method showed to be a good alternative for obtaining the DOZZ formula, because the lattice of poles appeared in a natural way due to the introduction of the Watson-Sommerfeld transform \cite{WATSON} \cite{dual}.

However, for computing the supersymmetric version of the DOZZ formula we applied the usual Coulomb Gas method \cite{3} \cite{Rashkov} \cite{Alvarez1, Alvarez2}. Therefore, further research should be done in order to check if the OPS method also works in the supersymetric case.  

As stated in the introduction, there is a potentially interesting use of the Liouville theory which lies in the field of the $AdS/CFT$ correspondence. We recall a very interesting result given by O. Coussaert, M. Henneaux and P. van Driel \cite{AdS3/CFT2}. In that paper they related a three dimensional Einstein theory of gravity with negative cosmological constant ($AdS_{3}$), and a two dimensional conformal field theory, which was identified to be the Liouville field theory ($CFT_{2}$). Hence, they stated a correspondence of the kind $AdS_{3}/CFT_{2}$. They achieved such correspondence in a straightforward manner. In this way, it is natural to ask ourselves if the minimal surface created by fixing three points at the boundary, as on the Liouville side, will correspond to the exact form of the DOZZ formula. As far as we know, this is an unsolved question. Another interesting issue is the existence of a duality Gravity/SuperLiouville. Based on a lack of research on this issue, it is viewed as an open problem which is subject to future investigation.      

Finally, we hope this dissertation provides a starting point for future research in this field.




\addtocontents{toc}{\vspace{2em}} 

\appendix


\chapter{Appendix A} 

\label{AppendixA} 

\lhead{\emph{Appendix A.}} 

\section{Conformal Coordinates}

Here, we obtain a convenient expression for eq.(\ref{Pfunction}) in conformal coordinates. In order to do this, we take into account the conformal transformation between a general coordinate $y$ and a euclidean flat one $y'$:
\begin{eqnarray}
ds^{2}(y)&=& \mathit{e}^{2\sigma}ds^{2}_{\text{Euclidean}},\\
ds^{2}(y)&=& \mathit{e}^{2\sigma}({dy'}_{1}^{2}+{dy'}_{2}^{2}).
\end{eqnarray}
From the latter relation we get, 
\begin{equation}\label{matrixmetric}
g_{\mu\nu}(y)=
\begin{pmatrix}
\mathit{e}^{2\sigma} & 0\\
0 & \mathit{e}^{2\sigma} \\
\end{pmatrix}.
\end{equation}
Therefore, it is straightforward to obtain:
\begin{equation}\label{1equation}
\sqrt{\det(g)_{(y)}}= e^{2\sigma}.
\end{equation}
The relation between the physical scalar of curvature $R$ and the reference one $\widehat{R}$ is given by:
\begin{equation}\label{2equation}
R = e^{-2\sigma}(\widehat{R}-\nabla^{2}\sigma).
\end{equation}
Additionally, we recall 
\begin{equation}
\Delta =-\frac{1}{\sqrt{g}}\partial_{\mu}\,\sqrt{g}\,g^{\mu\nu}\,\partial_{\nu}.
\end{equation}
By using this definition and eq.(\ref{2equation}) we get,
\begin{equation}\label{3equation}
e^{2\sigma} \, R = \widehat{R} + e^{2\sigma} \Delta \sigma .
\end{equation}
If we now set the reference metric as $\widehat{g}_{\mu\nu}=\delta_{\mu\nu}$, it is possible to simplify eq.(\ref{3equation}) because of $\widehat{R} = 0$. Therefore, we obtain a simple form for the physical scalar of curvature $ R = \Delta \sigma $. By combining eq.(\ref{Deltadefinition}) and the latter equation of $R$, we are able to write (we use $x$ and $y$ as arbitrary variables):
\begin{eqnarray}
P(x) &=& \frac{1}{\pi}\int d^{2}y\sqrt{g(y)} G(x, y)R_{[g]}(y),\nonumber\\
P(x) &=& \frac{1}{\pi}\int d^{2}y\,\mathit{e}^{2\sigma} G(x, y)\Delta_{g}\sigma(y),\nonumber\\
P(x) &=& \frac{1}{\pi}\int d^{2}y\, \mathit{e}^{2\sigma} (\Delta_{g}G(x, y))\,\sigma(y)\hspace{0.5cm} + \text{surface terms},\nonumber\\
P(x) &=& \frac{1}{\pi}\int d^{2}y\,\pi\,\delta^{2}(x,y) \frac{1}{2}\ln\sqrt{\det g_{(y)}} + \text{surface terms},\nonumber\\
P(x) &=& \frac{1}{2}\int d^{2}y\,\delta^{2}(x,y)\,\ln\sqrt{\det g_{(y)}} \hspace{0,6cm}+ \text{surface terms}.\nonumber\\
\end{eqnarray}
Disregarding the surface terms, such as $\int d^{2}y\,\Delta (G(x, y) \,\ln\sqrt{\det g_{(y)}})\,$,
\begin{equation*}
\int d^{2}y\Delta (G(x, y) \,\ln\sqrt{\det g_{(y)}}\,) \longrightarrow 0.
\end{equation*}
Finally, we obtain the desired expression for $P(x)$:
\begin{equation}
P(x) = \frac{1}{2} \,\ln\sqrt{\det g_{(x)}}.
\end{equation}


\chapter{Appendix B} 

\label{AppendixB} 

\lhead{\emph{Appendix B.}} 
\section{UV behaviour of 2D Green functions}

Starting by introducing the conformal relation in complex coordinates:
\begin{eqnarray}
ds^{2}=\sqrt{g} dz d\overline{z}.
\end{eqnarray}
We want to analyse the special limit of $G(x,y)$ as $x \rightarrow y$,
\begin{eqnarray}\label{LIMIT1}
\lim_{x\rightarrow y}G(x, y).
\end{eqnarray}
From eq.(\ref{Bus3}) we have,
\begin{eqnarray}
G(x, y)=-\frac{1}{2}\ln\frac{|x-y|}{\Lambda}.
\end{eqnarray}
Hence, the limit given by eq.(\ref{LIMIT1}) becomes:
\begin{eqnarray}
\lim_{x\rightarrow y}G(x, y)&=& -\frac{1}{2}\ln\frac{|dz|}{\Lambda}\\
&=& -\frac{1}{2}\ln\frac{(g)^{-1/4}|ds|}{\Lambda}\\
&=& -\frac{1}{2}\ln\frac{|ds|}{\Lambda}+\frac{1}{4}\ln(\sqrt{g}).
\end{eqnarray}
Therefore, by reabsorbing the geodesic invariant distance $|ds|$ into the renormalization scale $\Lambda$ and also by defining $G^{R}(x,x)= \lim_{x\rightarrow y}G(x, y)$, we have
\begin{eqnarray}
G^{R}_{0}(x,x)=\frac{1}{4}\ln(\sqrt{g}).
\end{eqnarray}
This computation along with the one given in Appendix \ref{AppendixA} are also valid in the Supersymmetric case. We performed analogous computations in supersymmetric language in Chapter \ref{Chapter5}. 

\chapter{Appendix C} 

\label{AppendixC} 

\lhead{\emph{Appendix C.}} 
\section{The Dotsenko-Fateev integral}

In this last Appendix we derive eq.(\ref{I31}) by starting from the result obtained by Dotsenko and Fateev \cite{dofa}. They solve the integral
\begin{eqnarray}\label{ApIntegral}
\mathcal{I}_{3}(m,n)&=&\prod_{i<j}^{m}\prod_{r<s}^{n}\int\,d^{2}x_{i}d^{2}y_{r}|x_{i}-x_{j}|^{4\rho}|y_{r}-y_{s}|^{4\rho'}\nonumber\\&&|x_{i}-y_{r}|^{-4}|x_{i}|^{2\alpha}|x_{i}-1|^{2\beta}|y_{r}|^{2\alpha'}|y_{r}-1|^{2\beta'}.\nonumber\\
\end{eqnarray}
By identifying the parameters $\rho, \rho', \alpha, \alpha', \beta$ and $\beta'$ as
\begin{eqnarray}\label{dictionary}
\rho &=& -b^{2},\hspace{0.3cm} \rho' =-\frac{1}{b^{2}},\hspace{0.3cm} \alpha =-2b\alpha_{1},\nonumber\\
\beta &=&-2b\alpha_{2},\hspace{0.3cm}  \alpha' =-\frac{2\alpha_{1}}{b},\hspace{0.3cm} \beta' =-\frac{2\alpha_{2}}{b},
\end{eqnarray}
we realize that the latter expression is equal to eq.(\ref{I3}). Applying the dictionary given by eq.(\ref{dictionary}) and using the result by Dotsenko and Fateev \cite{dofa}, into eq.(\ref{ApIntegral}). We have
\begin{eqnarray}\label{EqNoV}
\mathcal{I}_{3}(m,n)&=&\left[m!n! \pi^{m+n} b^{-8mn} (\gamma(-b^2))^{-m}(\gamma(-b^{-2}))^{-n} \right]\left(\prod_{I=1}^{3}\prod_{t=0}^{m-1}\gamma(2b\alpha_{I}+tb^{2}) \right)^{-1}\nonumber\\
&&\left(\prod_{I=1}^{3}\prod_{t=0}^{n-1}\gamma(2b\alpha_{I}+m+tb^{-2}) \right)^{-1}\left(\prod_{t=1}^{m}\gamma(1+tb^{2})\prod_{t=1}^{n}\gamma(1+m+tb^{-2}) \right)^{-1}.\nonumber\\
\end{eqnarray}
In order to extrapolate the latter result to non-integer values of $m$ and $n$, we will transform the latter three terms in eq.(\ref{EqNoV}) and express them by using the Zamolodchikov's function $\Upsilon(x)$. From eq.(\ref{upsilonAhhh}), we have \cite{witten2012}
\begin{eqnarray}
\gamma(bx)=\frac{\Upsilon(x+b)}{\Upsilon(x)}\,b^{2bx-1}.
\end{eqnarray}
Hence, by replacing $x \rightarrow x+tb$ it is possible to obtain the following recursion relation
\begin{eqnarray}\label{CCCRECU}
\gamma(bx+tb^{2})=\frac{\Upsilon(x+(t+1)b)}{\Upsilon(x+tb)}b^{2bx+2tb^{2}-1}.
\end{eqnarray}
Additionally, from eq.(\ref{CCCRECU}) it is possible to obtain 
\begin{eqnarray}
\prod_{t=0}^{m-1}\gamma(bx+tb^{2})&=&\frac{\Upsilon(x+mb)}{\Upsilon(x)}\,b^{m[(2bx-1)+(m-1)b^{2}]},\label{CCC1}\\
\prod_{t=1}^{m}\gamma(bx+tb^{2})&=&\frac{\Upsilon(x+(m+1)b)}{\Upsilon(x+b)}\,b^{m[(2bx-1)+(m+1)b^{2}]}\label{CCC2}.
\end{eqnarray}
Replacing eqs.(\ref{CCC1}) and (\ref{CCC2}) into the first and second term of eq.(\ref{EqNoV})
\begin{eqnarray}
\left(\prod_{I=1}^{3}\prod_{t=0}^{m-1}\gamma(2b\alpha_{I}+tb^{2}) \right)^{-1}&=&(b)^{-m[4b\sum_{I}\alpha_{I}-3+3(m-1)b^{2}]}\prod_{I=1}^{3}\frac{\Upsilon(2\alpha_{I})}{\Upsilon(2\alpha_{I}+mb)},\label{C11}\\
\left(\prod_{I=1}^{3}\prod_{t=0}^{n-1}\gamma(2b\alpha_{I}+m+tb^{-2}) \right)^{-1}&=&(\frac{1}{b})^{-n[\frac{4}{b}\sum_{I}\alpha_{I}-3+6m+\frac{3(n-1)}{b^{2}}]}\prod_{I=1}^{3}\frac{\Upsilon(2\alpha_{I}+mb)}{\Upsilon(2\alpha_{I}+mb+\frac{n}{b})}.\nonumber\\\label{C12}
\end{eqnarray}
For the remaining term in eq.(\ref{EqNoV}) we must be careful, because as we mentioned in Chapter \ref{Chapter4} the Zamolodchikov's function $\Upsilon(x)$ has poles at $x=-mb-n/b$ and $x=(m+1)b+(n+1)/b$. To avoid this issue, we introduce a non-zero parameter $\epsilon$ and compute what we want. As usual at the end of the calculation we will take the limit $\epsilon \rightarrow 0$. Hence, we have 
\begin{eqnarray}\label{EQLastC}
\left(\prod_{t=1}^{m}\gamma(1+tb^{2}+\epsilon)\prod_{t=1}^{n}\gamma(1+m+tb^{-2}+\epsilon) \right)^{-1}&=&(b)^{n[(1+2m)+\frac{(n+1)}{b^{2}}]-m[1+(m+1)b^{2}]}\times\nonumber\\
&&\left\lbrace\frac{\Upsilon(b+\frac{1}{b}+\epsilon)}{\Upsilon((1+m)b+\frac{(n+1)}{b}+\epsilon)} \right\rbrace.\nonumber\\
\end{eqnarray}
Now we use the relation $Q=b+1/b$ in the latter equation. Thus, it is easy to obtain
\begin{eqnarray}\label{ReflectQX}
\lim_{\epsilon\rightarrow 0}\,b^{n[(1+2m)+\frac{(n+1)}{b^{2}}]-m[1+(m+1)b^{2}]}\left\lbrace\frac{\Upsilon(Q+\epsilon)}{\Upsilon(Q+mb+\frac{n}{b}+\epsilon)} \right\rbrace.
\end{eqnarray}
Additionally, we use the reflection property $\Upsilon(x)=\Upsilon(Q-x)$ into eq.(\ref{ReflectQX}). In this way, the last equation involving $\Upsilon(x)$ functions becomes
\begin{eqnarray}
\frac{\Upsilon(Q+\epsilon)}{\Upsilon(Q+mb+\frac{n}{b}+\epsilon)}=\frac{\Upsilon(\epsilon)}{\Upsilon((-mb-\frac{n}{b})-\epsilon)}.
\end{eqnarray}
Applying the well-known L'Hospital's rule, we easily find
\begin{eqnarray}
\lim_{\epsilon\rightarrow 0}\left\lbrace\frac{\Upsilon(\epsilon)}{\Upsilon((-mb-\frac{n}{b})-\epsilon)}\right\rbrace=-\frac{\Upsilon'(0)}{\Upsilon'(-mb-\frac{n}{b})}.
\end{eqnarray}
Therefore, eq.(\ref{EQLastC}) becomes 
\begin{eqnarray}\label{C13}
-(b)^{(\frac{n}{b}-mb)(2Q-\sum_{I}\alpha_{I})-2mn}\frac{\Upsilon'(0)}{\Upsilon'(-mb-\frac{n}{b})}.
\end{eqnarray}
Finally, combining eqs.(\ref{C11}), (\ref{C12}) and (\ref{C13}), we have
\begin{eqnarray}
\mathcal{I}_{3}(m, n)= - m! n! \phi^{m}_{b}\phi^{n}_{1/b}\frac{\Upsilon'_{(0)}}{\Upsilon'_{(-bm-(\frac{1}{b})n)}}\prod_{I=1}^{N=3}\frac{\Upsilon_{(2\alpha_{I})}}{\Upsilon_{(\sum_{J}\alpha_{J}-2\alpha_{I})}},
\end{eqnarray}
where $\phi_{b}$ and $\phi_{1/b}$  are defined by eqs.(\ref{phib}) and (\ref{phi1b}).

\chapter{Appendix D} 

\label{AppendixD} 

\lhead{\emph{Appendix D.}}  

\section{$c$, $d$ and $\beta$}
In this last appendix we show the result for the variables $c$, $d$ and $\beta$ which were computed by using {\it Mathematica}. Recalling that the system of equations to be solved is
\begin{eqnarray}
z'_{1}=\frac{a z_{1}+b+\alpha \theta_{1}}{cz_{1} + d+ \beta\theta_{1}}&=&1,\\
z'_{2}=\frac{a z_{2}+b+\alpha \theta_{2}}{cz_{2} + d+ \beta\theta_{2}}&=&0,\\
z'_{3}=\frac{a z_{3}+b+\alpha \theta_{3}}{cz_{3} + d+ \beta\theta_{3}}&=&R,\\
\theta'_{1}=\frac{\bar{\alpha}z_{1}+\bar{\beta}+\bar{A}\theta_{1}}{cz_{1} + d + \beta\theta_{1}}&=&0,\\
\theta'_{2}=\frac{\bar{\alpha}z_{2}+\bar{\beta}+\bar{A}\theta_{2}}{cz_{2} + d + \beta\theta_{2}}&=&0,\\
ad-bc-3\alpha\beta &=&1.
\end{eqnarray}
Therefore, we have:

\begin{figure}[htbp]
  \vspace{-1cm} \hspace{-3cm} \includegraphics[scale=0.9]{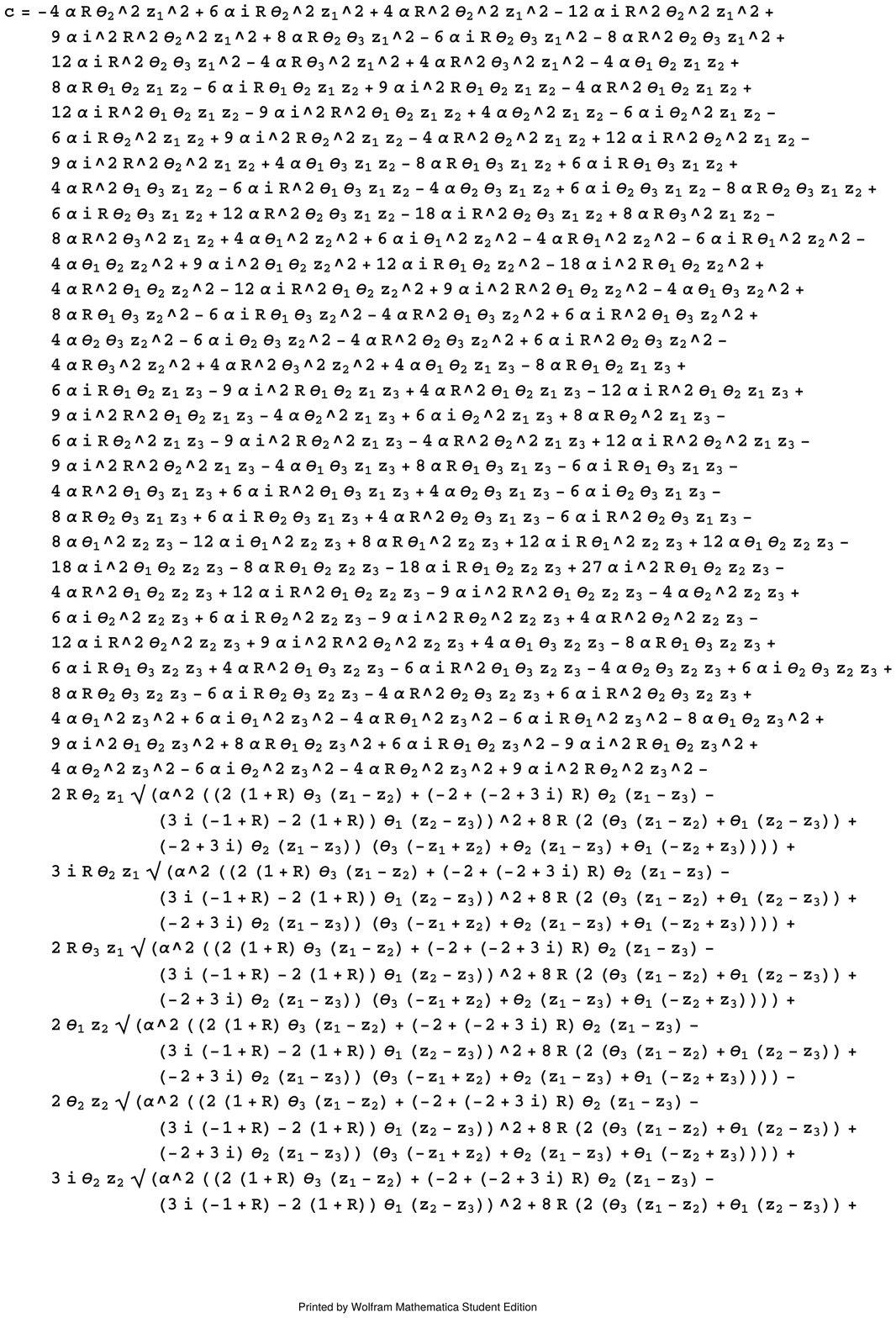}
  \label{fig:A1}
\end{figure}

\begin{figure}[htbp]
  \vspace{-1cm} \hspace{-3cm} \includegraphics[scale=0.9]{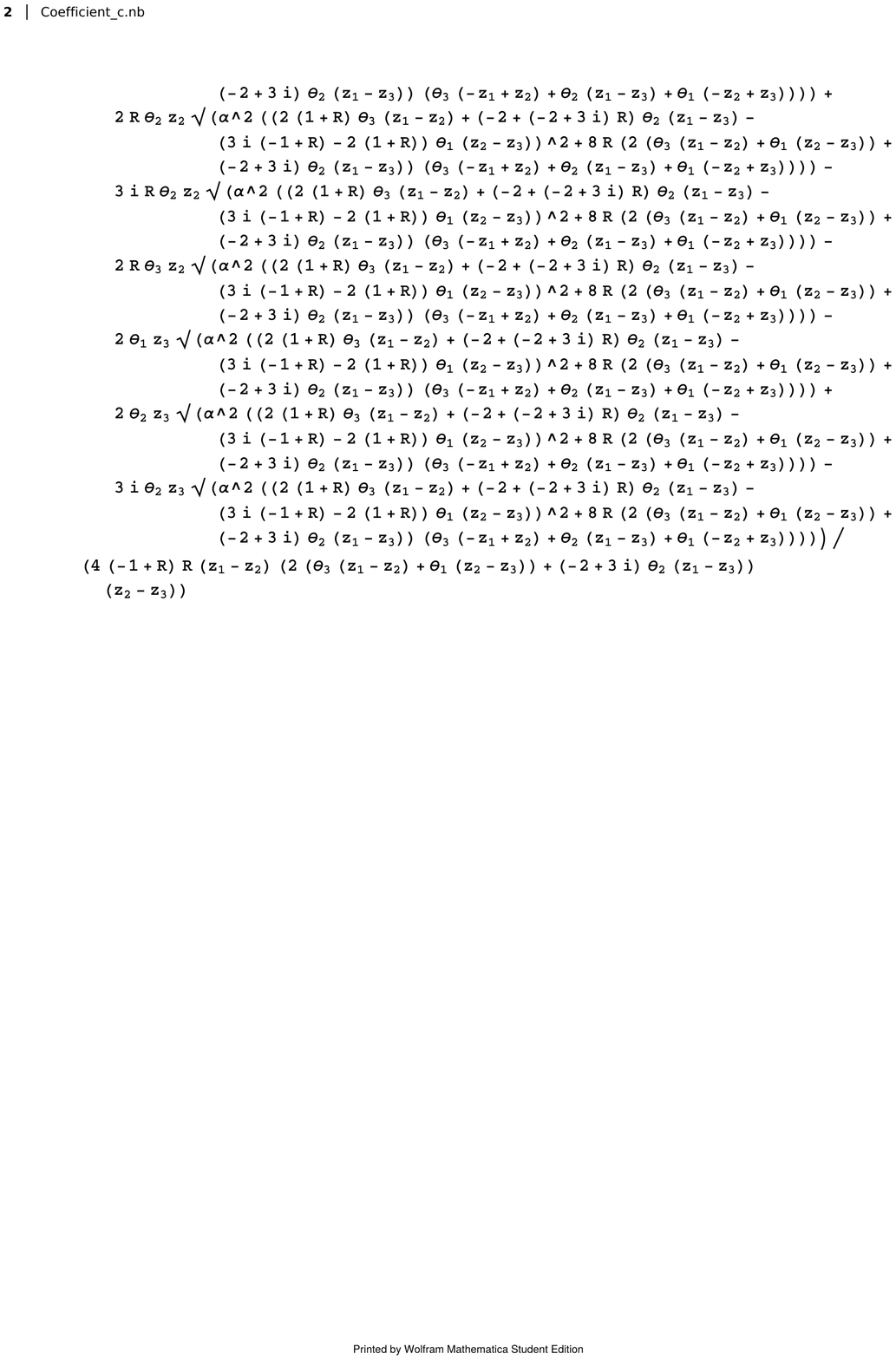}
  \label{fig:A2}
\end{figure}

\begin{figure}[htbp]
  \vspace{-1cm} \hspace{-3cm} \includegraphics[scale=0.9]{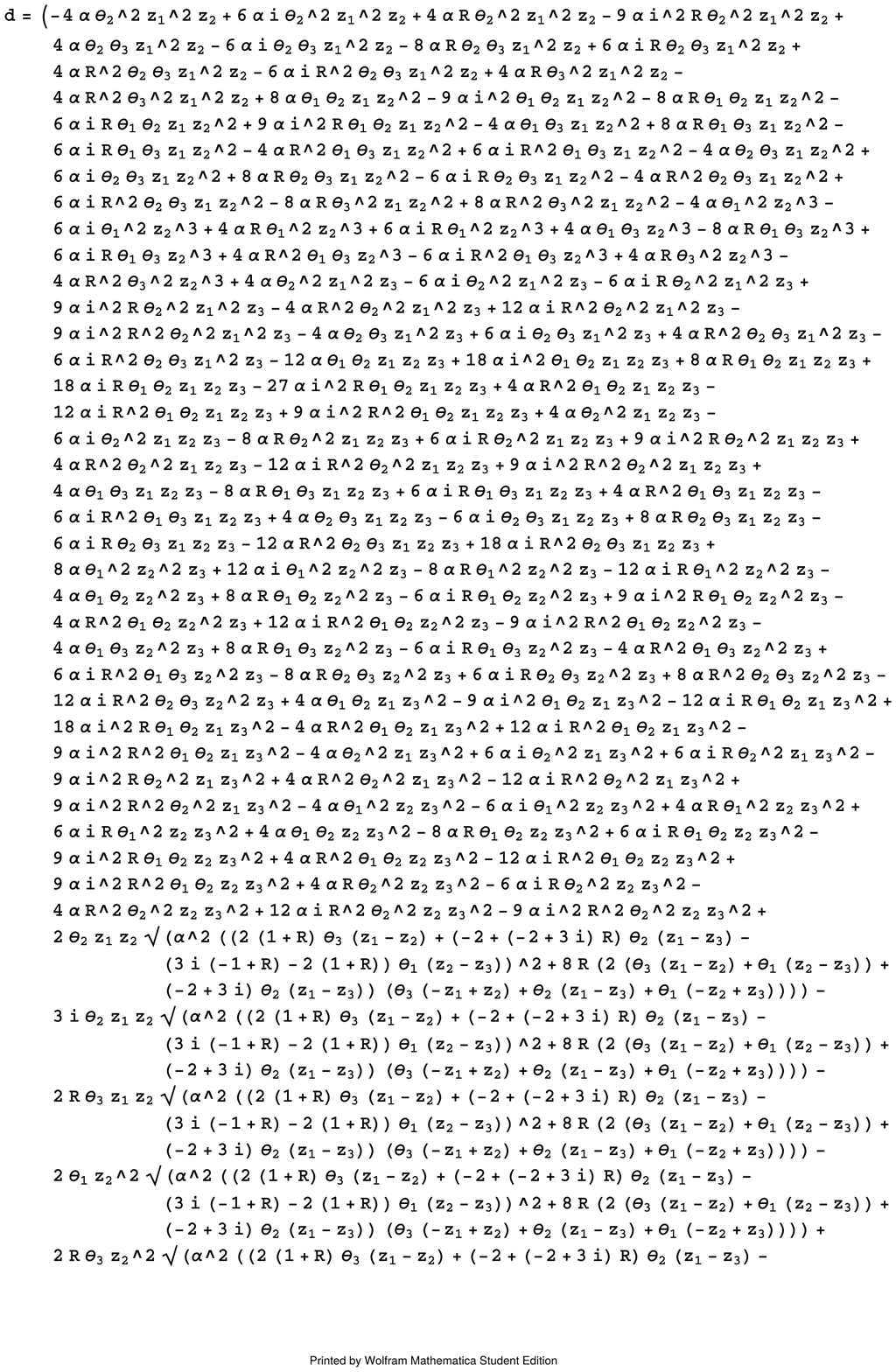}
  \label{fig:A3}
\end{figure}

\begin{figure}[htbp]
  \vspace{-1cm} \hspace{-3cm} \includegraphics[scale=0.9]{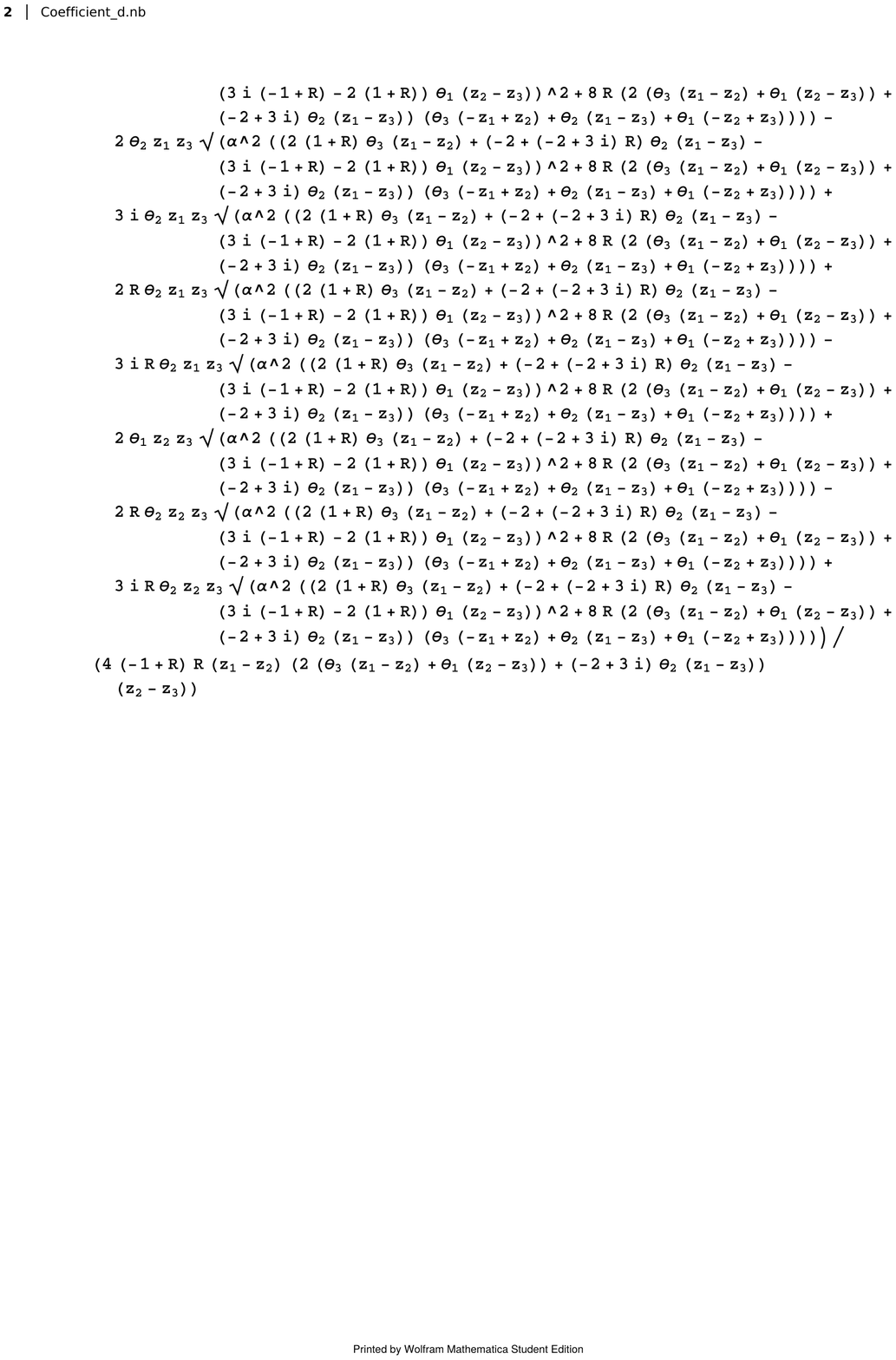}
  \label{fig:A4}
\end{figure}

\begin{figure}[htbp]
  \vspace{-1cm} \hspace{-3cm} \includegraphics[scale=0.9]{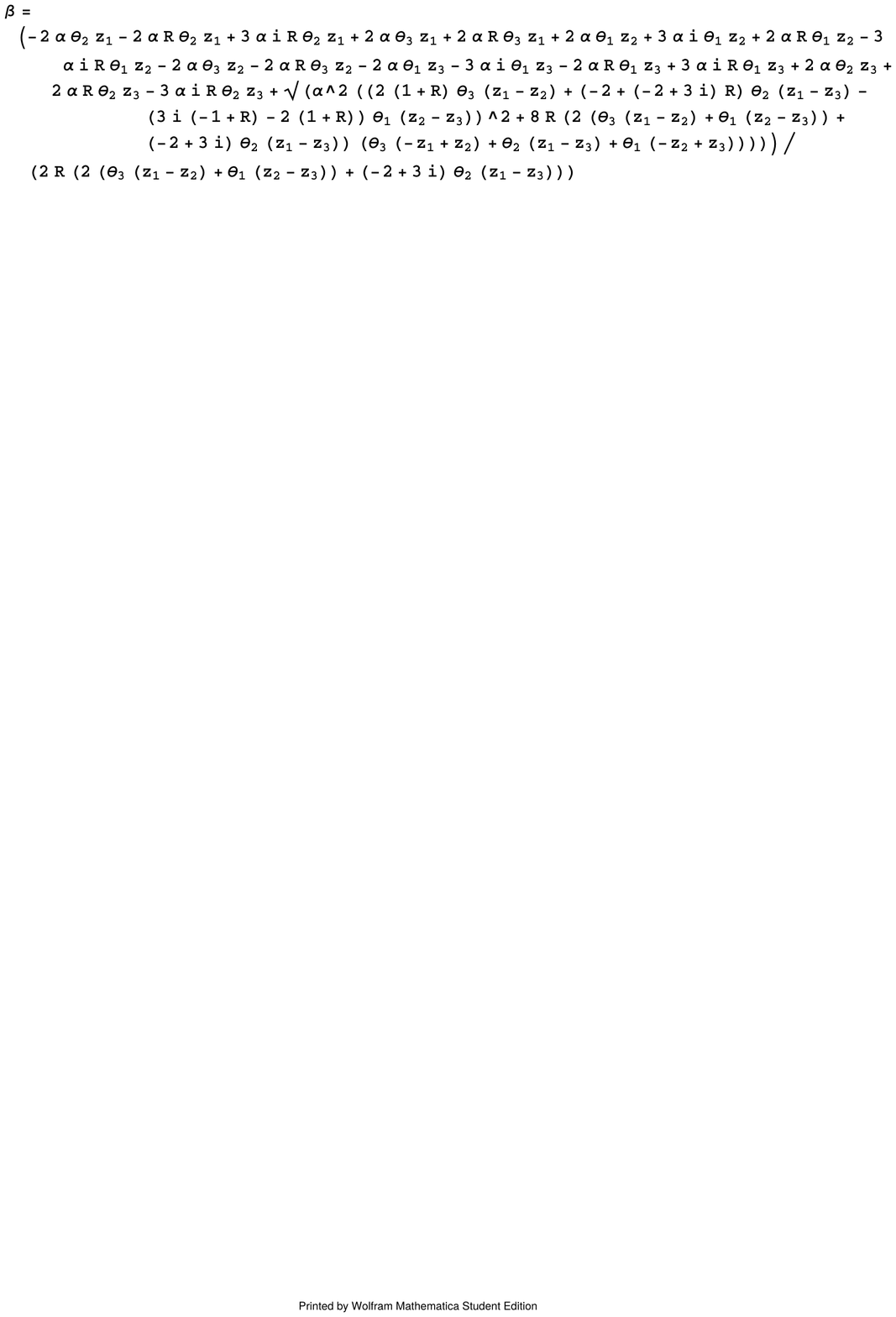}
  \label{fig:A5}
\end{figure}

\addtocontents{toc}{\vspace{2em}}

\backmatter



\lhead{\emph{Bibliography}} 

\end{document}